# Synaptic placement reflects shared input in *Drosophila* descending neurons

Xizhe Zhang

ORCID: https://orcid.org/0000-0002-8684-4591

## Abstract

Network topology describes connections between neurons, whereas synaptic placement specifies how those connections are arranged within individual cells. How these levels of organization correspond remains incompletely understood. Here we show that connectivity between presynaptic neurons is reflected in relative input placement within *Drosophila* descending neurons (DNs). Across thousands of one-way DN connections in the independently reconstructed MaleCNS and FlyWire brains, inputs from sources also contacting the partner DN lay on average 9.4 and 8.0 µm nearer that partner's inputs along the receiver's neurites than other non-DN inputs. The same ordering held when source-input positions were fixed and partners with and without recorded source input were compared after standardizing reference-set size. In BANC and MaleCNS, shared input and measured spatial overlap provided complementary predictive information about DN interconnection. Two interconnected DNs sharing an ascending source differed in their identified cord outputs. Together, these findings link a three-neuron topological relationship to relative input placement, revealing a correspondence between network connectivity and the internal spatial organization of a receiving neuron.

## Introduction

Descending pathways translate brain computations into motor control. In *Drosophila*, descending neurons (DNs) form a major interface between the brain and circuits in the ventral nerve cord. Individual DN types can initiate backward walking or distinct forward-walking programs[1,2], while descending population activity accompanies locomotion and contributes to graded wing movements[3,4]. Their effects depend on ongoing behavioral state[5], and connections among DNs allow descending commands to recruit broader neuronal populations[6,7]. Ascending neurons also convey behavioral-state information to brain regions involved in sensory integration and action selection[8]. Local DN connections can therefore create converging routes through which an upstream source reaches a receiving DN directly and through a DN partner. How the inputs of these routes are arranged within the receiving neuron remains an open cellular question.

Common-neighbor relationships and nonrandom connectivity are established features of cortical neuronal groups[9,10], and convergent and recurrent connections contribute to the organization of insect connectomes[11,12]. The configuration examined here consists of a source contacting two DNs and a connection from one DN, the partner, to the other, the receiver. These three directed edges form a feed-forward configuration[13]. A synapse-count-weighted adjacency matrix records the connections and their contact counts, but leaves the positions of those contacts unspecified. The same connectivity can thus be realized by different arrangements of inputs along a receiving arbor. Physical-network models that represent the internal structure of spatially extended nodes provide a way to retain this spatial information alongside connectivity[14]. For a neuronal circuit, this raises a concrete question: is a connection between presynaptic cells associated with the relative placement of their inputs within a common receiver?

Synaptic location is relevant to this question because input interactions can depend on dendritic position[15]. Reconstructions have revealed inputs concentrated on fine neurites in *Drosophila*[16] and clustered contacts from individual excitatory axons onto CA1 pyramidal-cell dendrites[17]. In *Caenorhabditis elegans*, local synaptic organization has been associated with three-neuron configurations and shared-neighbor relationships[18], and synaptic clustering has been examined across development[19]. In identified fly DNs, reconstruction and biophysical modeling have linked input topography and morphology to integration[20]. We asked whether source–partner connectivity corresponds to relative input placement in adult descending circuits, using two complementary comparisons. For a fixed partner-to-receiver connection, are shared-source inputs nearer the partner's input sites than other inputs on that receiver? Conversely, with the source, receiver and source-input positions fixed, are those inputs nearer sites supplied by partners that also receive source input than sites supplied by partners with none recorded? We refer to these as the between-source comparison and the within-source partner comparison.

Anatomical atlases and comparative brain–cord reconstructions identify DN arbors, projections and cellular partners[21–24]. Dense electron-microscopy reconstructions retain individual contact locations[25,26], and adult fly connectomes combine these positions with systematic cell identities[27–29]. Together, these resources allow an upstream source to be followed from its contacts onto two DNs to the positions at which source and partner inputs enter the receiver. They also distinguish cellular relationships that coarser descriptions can merge: two DNs may share a sensory category or project to the same cord region while receiving input from different source cells or contacting different downstream neurons. Source sharing, input placement and cord-target connectivity can therefore be examined within the same identified circuit.

Here, we analyzed synaptic placement in the male MaleCNS connectome and the independently reconstructed female FlyWire brain, and used MaleCNS and the female Brain And Nerve Cord (BANC) connectome to examine network and cord-output organization[24,27–29]. In both placement datasets, shared-source inputs lay nearer partner-DN inputs along receiving neurites. With source, receiver and source-input positions fixed, the same inputs were also nearer sites supplied by partners that received source input. An identified ascending–descending circuit localized these converging routes to named cells and receiving branches. The network analyses linked shared input to DN interconnection beyond measured spatial overlap, while the identified DNs differed in their cord-target composition and anatomical reach. Exploratory comparisons in interneurons from two adult *C. elegans* reconstructions extended the placement question to another species and neuronal population[30]. These analyses connect a relationship between cells to the arrangement of their inputs within a common receiver.

## Results

### Source–partner connectivity is associated with relative input placement along receiving neurites

We first asked whether shared-source connectivity in the network was reflected in input placement within the receiving DN. For each one-way partner-to-receiver DN connection, we divided the receiver's non-DN inputs according to whether their source also innervated the partner DN. We then measured, along the receiving neurite, the distance from each source input to the nearest input site supplied by the partner DN, restricting comparisons to the same receiver and endpoint brain region (Fig. 1a).

Across 13,318 eligible one-way MaleCNS connections, inputs from shared sources lay closer to partner-DN inputs than other non-DN inputs on the same receiver: mean path distances were 60.77 and 70.18 µm, respectively, a difference of −9.42 µm (95% type-endpoint interval, −10.52 to −8.31 µm; Fig. 1b and Supplementary Table 1). The same ordering was present in 74.7% of directed DN pairs and, after averaging within receivers, in 82.7% of 1,079 receiving DNs (Fig. 1c,d).

We next asked the complementary question while holding source and receiver fixed: were the same source inputs closer to sites supplied by DN partners that also received input from that source? We compared partners receiving at least five synapses from the source with partners having no recorded source input, while holding the source-input positions, endpoint region and skeleton component fixed. Because nearest-neighbor distance depends on the number of available partner input sites, we standardized the reference-set size by calculating the expected nearest distance to three uniformly sampled distinct partner input sites. With receivers weighted equally, mean distances were 64.40 µm for source-connected partners and 73.54 µm for partners with no recorded source input, a difference of −9.14 µm (conditional 95% interval, −11.21 to −7.07 µm). The difference was negative in 749 of 922 receivers (81.2%; Fig. 1b,e and Supplementary Table 2).

Both comparisons showed the same ordering in the independent female FlyWire brain. Across 5,807 eligible one-way DN connections, shared-source inputs lay 7.98 µm nearer partner-DN inputs than other inputs (means, 51.37 and 59.36 µm; 95% type-endpoint interval for the difference, −9.27 to −6.70 µm). The difference was negative in 581 of 720 receivers (80.7%). In the within-source partner comparison, receiver-equal means were 62.79 and 72.93 µm, a difference of −10.15 µm (conditional 95% interval, −12.15 to −8.14 µm). This difference was negative in 483 of 589 receivers (82.0%; Fig. 1b, Supplementary Fig. 10 and Tables 23–24).

The between-source relationship retained its direction under equal type-pair and receiver weighting in both MaleCNS and FlyWire, and Euclidean distances on the same retained records likewise placed shared-source inputs nearer partner-DN inputs (Supplementary Figs. 1 and 10; Tables 1 and 23). In MaleCNS, all 17 sufficiently sampled regional groups showed the same ordering. Individual connections varied, including examples with the opposite ordering (Supplementary Table 3). Across the two fly brains, source–partner connectivity was therefore associated with relative input placement both when comparing sources for a fixed DN connection and when comparing DN partners for fixed source inputs on a receiver.

An exploratory extension examined author-classified interneurons in two adult *C. elegans* reconstructions, using the whole reconstructed brain arbor as the spatial domain. The within-source partner differences were −9.88 µm in Dataset7 (conditional 95% interval, −15.03 to −4.73 µm; 10 receivers) and −5.39 µm in Dataset8 (−9.83 to −0.94 µm; 13 receivers), with negative differences in 10 of 10 and 11 of 13 receivers, respectively. Between-source estimates were also negative in both adults, although the Dataset8 interval crossed zero (Supplementary Fig. 11 and Table 25). These exploratory intervals describe within-animal dependence, with few receiver-type clusters. A head-to-trunk population in a *Platynereis dumerilii* larva[31] yielded only four eligible between-source pairs with mixed directions and one within-source receiver, so it provided descriptive evidence rather than an estimable population-level comparison (Supplementary Note 8 and Table 26).

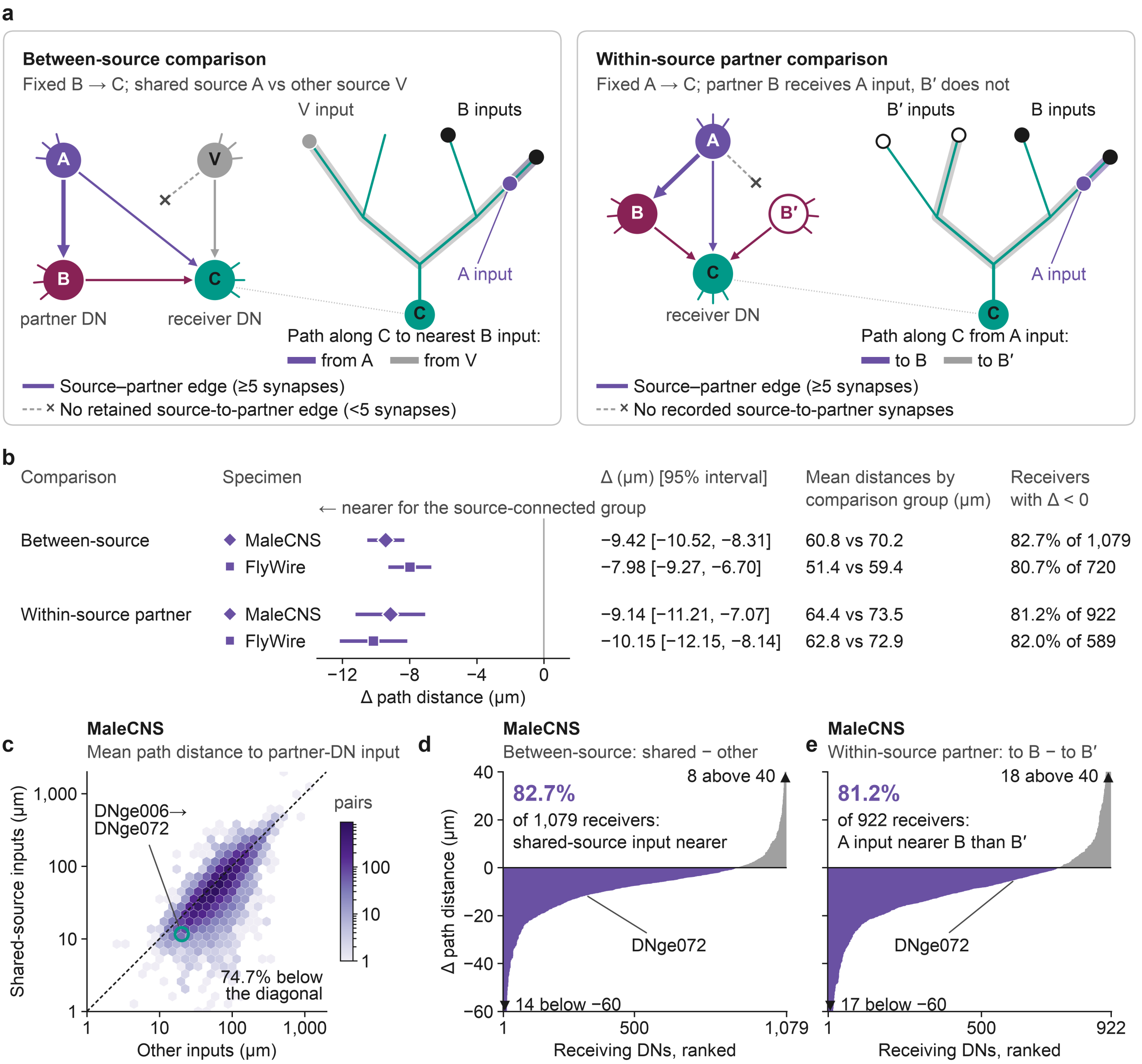

**Figure 1 | Source–partner connectivity is associated with relative input placement along receiving neurites. a,** The two comparisons. Left, with partner-to-receiver connection B→C fixed, non-DN sources A and V contact C, but only A contacts B in the retained input graph (at least five synapses). Right, with source A and receiver C fixed, distances from the same A inputs to sites supplied by B and B′ are compared: B receives at least five A synapses; B′ has none recorded. Dashed stubs denote fewer than five source-to-partner synapses (left) or zero recorded synapses (right). Both endpoints share the same brain-region label; path distances follow C's skeleton, and the intervening path may leave that region; the within-source statistic uses the same connected component and the expected nearest distance to three uniformly sampled distinct partner input sites. **b,** Mean differences and 95% intervals for MaleCNS (diamonds) and FlyWire (squares). Negative values indicate nearer inputs for the source-connected group. Between-source estimates weight directed connections equally (13,318 MaleCNS; 5,807 FlyWire) and use type-endpoint intervals. Within-source estimates weight receivers equally (922; 589) and use conditional intervals clustered by receiver type and source cell. Columns give group means and the fraction of receivers with a negative difference. **c–e,** MaleCNS distributions; corresponding FlyWire distributions are in Supplementary Fig. 10. **c,** Between-source mean path distances in hexagonal bins, colored by connection count on a logarithmic scale; dashed line, equality; ring, DNge006→DNge072. **d,e,** Receiver-level differences for the between-source and within-source comparisons, ranked on a common vertical scale. DNge072 is marked; triangles and adjacent counts indicate values outside the displayed range. Receivers and connections are observations within specimens, not independent animals.

**An identified ascending–descending circuit localizes converging input routes**

Within the BANC and MaleCNS brain–cord reconstructions, AN19B014 contacted both DNge006 and DNge072, while DNge006 supplied an additional input to DNge072, forming direct and partner-mediated anatomical routes from the same source to the receiving DN (Fig. 2a,b). In MaleCNS, AN19B014 extended from the ventral nerve cord into the brain, whereas both DNs projected to the cord. Of the 260 brain synaptic records among these cells, 223 were located in the gnathal ganglion (GNG).

Cells of the same three types formed this arrangement in both BANC and MaleCNS (Fig. 2b; Supplementary Fig. 2). AN19B014 was the leading shared upstream source of the selected DN pair in each specimen, accounting for 100% of retained shared-input cosine overlap in BANC and 81.7% in MaleCNS (Supplementary Table 4). In the selected MaleCNS circuit, this source contacted the two DNs predominantly through distinct presynaptic sites: of 159 AN19B014 sites contacting either DN, 21 contacted both, whereas the remaining sites contacted one DN or the other (Supplementary Fig. 3a). Thus, the shared-source relationship was realized largely through separate contacts from the same upstream neuron rather than through presynaptic sites jointly contacting both DNs.

On the receiving side, DNge072 exhibited the same distance ordering as the population-level between-source comparison. The MaleCNS pair shared 12 upstream source cells in the retained input graph, including AN19B014. For input sites in the GNG, mean distances along DNge072's arbor to the nearest DNge006 input site in the same region were 10.7 μm for the pooled inputs from these 12 shared sources and 19.0 μm for other non-DN inputs (Fig. 2c,d; Supplementary Table 3). Thus, the shared-source configuration of this identified DN pair was accompanied by closer placement of shared-source inputs to partner-DN inputs along the receiving arbor.

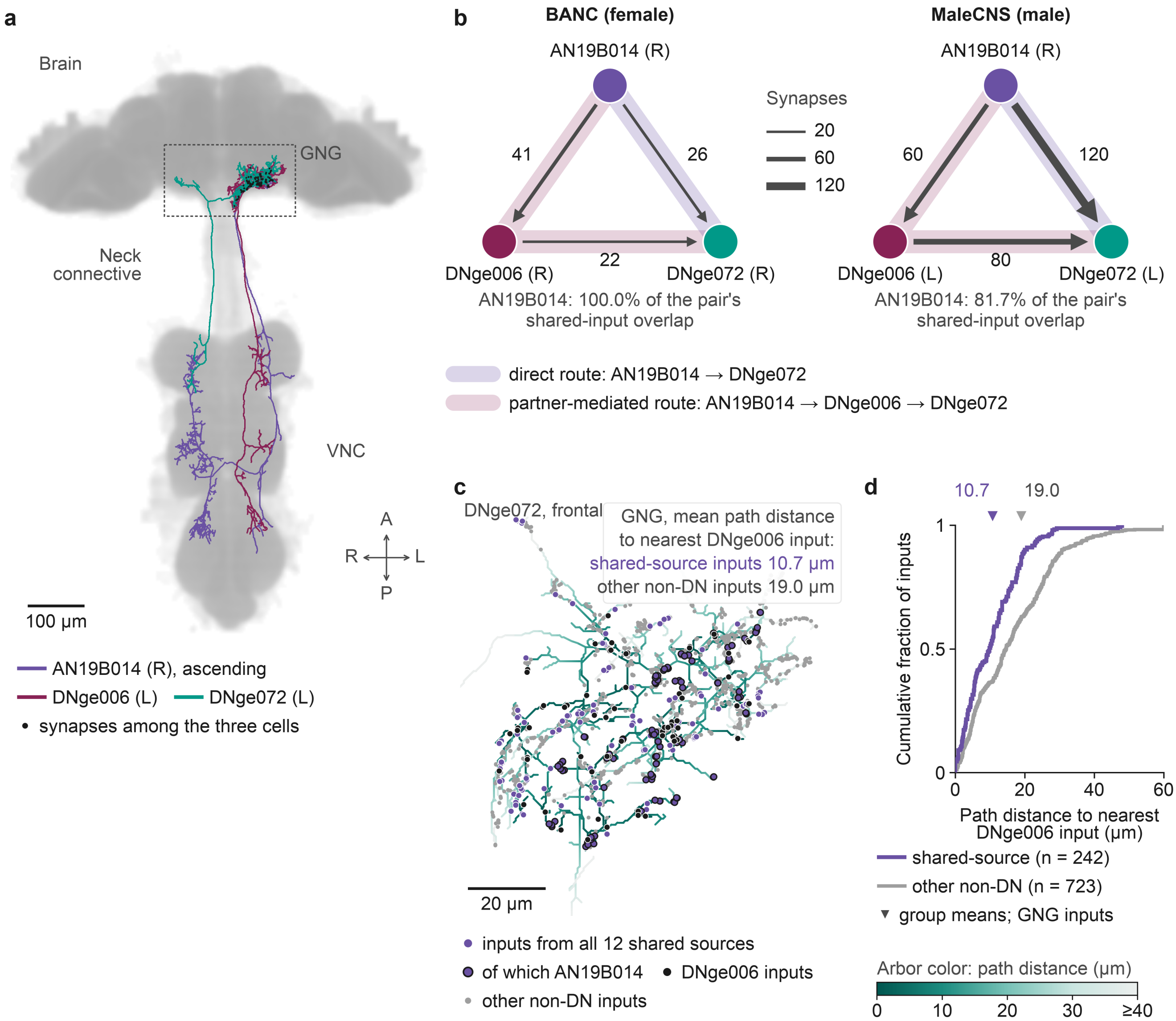

**Figure 2 | Shared-source inputs lie closer to partner-DN inputs within an identified circuit. a,** Horizontal projection of AN19B014, DNge006 and DNge072 in MaleCNS over a synapse-density background. Black points mark brain synaptic records among the three cells, with 223 of 260 in the gnathal ganglion (GNG; box). VNC, ventral nerve cord. **b,** Selected individual-cell circuits in BANC and MaleCNS, each containing a one-way DNge006→DNge072 connection. Arrow labels and widths indicate brain synapse counts; L and R denote annotated side. Shading distinguishes the direct and partner-mediated routes from AN19B014 to DNge072. Percentages give AN19B014's contribution to the selected DN pair's retained shared-input cosine overlap. **c,** Frontal projection of DNge072's brain arbor in MaleCNS, colored by path distance to the nearest DNge006 input site in the GNG. Purple points show inputs pooled across all 12 shared sources, with AN19B014 inputs outlined. Black points mark DNge006 inputs; gray points mark other non-DN inputs. Displayed points represent retained brain inputs; annotated means use the eligible GNG subset. **d,** Cumulative distributions of path distance along DNge072 to the nearest DNge006 input site in the GNG, comparing shared-source and other non-DN inputs in the eligible GNG subset. Triangles mark group means; 13 other non-DN inputs lie beyond the displayed range. An example measured path is shown in Supplementary Fig. 1d.

## Shared input and spatial overlap provide complementary information about DN interconnection

We next examined how shared input related to interconnection across the descending network. Among pairs of DNs of different types, at least one shared non-DN source was present in 65.0% of one-way connected pairs in BANC and 85.6% in MaleCNS, compared with 9.3% and 27.4% of unconnected pairs, respectively. Shared sources were most frequent among reciprocal pairs in both connectomes (Fig. 3a; Supplementary Fig. 4d).

Within one-way connected pairs with shared input, shared-input overlap was concentrated in a small number of upstream cells. After sources were ranked by their contribution within each pair, the five leading sources accounted on average for 91.3% of shared-input cosine overlap in BANC and 87.7% in MaleCNS (Fig. 3b; Supplementary Fig. 4a and Table 6). A median of two leading sources in BANC and three in MaleCNS accounted for 80% of shared-input overlap in one-way pairs with shared input (Supplementary Table 6).

Shared input also distinguished connection rates among DN pairs with comparable measured voxel overlap (Fig. 3c). Spatial-overlap measures summarized the distributions of the two DNs across brain voxels and named regions, whereas shared-input measures retained the identities and synaptic weights of individual non-DN sources. Within comparable ranges of 40-μm voxel overlap, pairs with greater shared-input overlap were more often connected. In logistic models with the same baseline covariates, each set of measures added predictive information beyond the other. For test pairs whose two DN types were both withheld from model fitting, adding shared-input measures to models containing spatial overlap reduced mean held-out log loss by 2.9% in BANC and 3.4% in MaleCNS. Conversely, adding spatial-overlap measures to models containing shared input reduced log loss by 10.0% and 10.5%, respectively. Both contributions remained positive when type pairs were weighted equally (Fig. 3d; Supplementary Fig. 5 and Tables 8–10).

Coarser similarities described multisynaptic access from eight annotated upstream source groups and output distributions across seven cord regions. Their joint contribution to prediction was smaller once spatial overlap and shared input were included (Supplementary Fig. 8 and Tables 8–10). Together, measured spatial overlap and source-resolved shared input provided complementary predictive information about DN interconnection in both reconstructions.

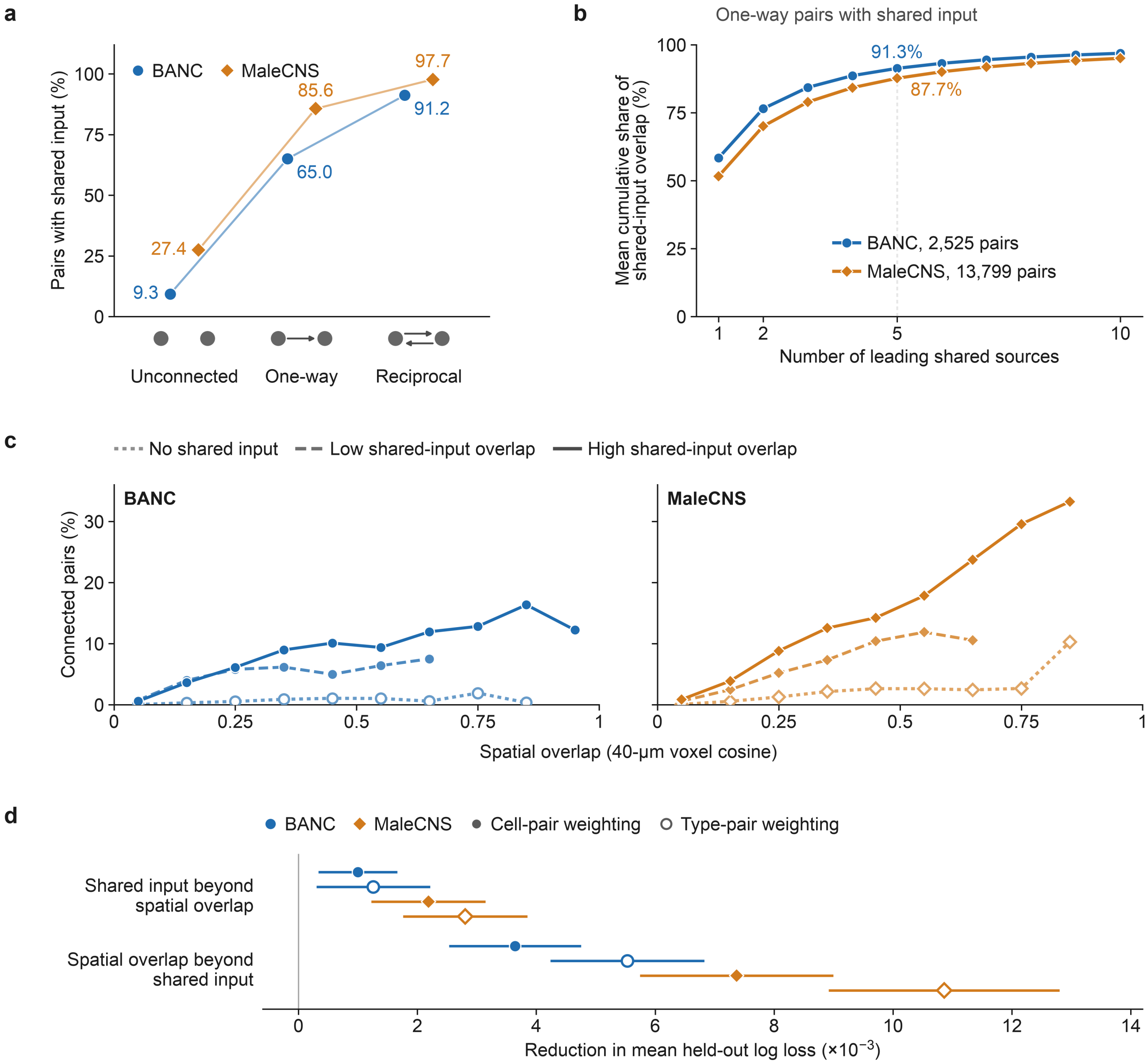

**Figure 3 | Shared input and spatial overlap provide complementary information about DN interconnection. a,** Percentage of pairs of different DN types with at least one shared non-DN source, grouped by connection class. Connections require at least five brain synapses in a direction; shared sources supply at least five synapses to each DN. **b,** Mean cumulative share of shared-input cosine overlap contributed by the leading sources in one-way connected pairs with shared input. Sources are ranked by contribution within each pair; the vertical dotted line marks five sources. **c,** Connection rates across 40-μm voxel-overlap bins, stratified by shared-input cosine. No shared input denotes zero; low and high strata contain nonzero values at or below and above the specimen median, respectively. Bins with fewer than 100 evaluated pairs are omitted. **d,** Reduction in mean held-out log loss from adding each set of measures to models already containing the other. All models include the same baseline covariates. Both DN types in each test pair were excluded from model fitting. Filled and open symbols denote equal weighting of cell pairs and type pairs, respectively; bars show unadjusted 95% type-endpoint intervals. Loss reductions are displayed in units of $10^{-3}$. Panels **c,d** use 93,550 BANC and 162,638 MaleCNS evaluation pairs. DN–DN edges were excluded from shared-input and spatial-overlap measurements.

### Interconnected DNs share an ascending source but distribute output differently

DNge006 and DNge072 shared input from the ascending neuron AN19B014 and were interconnected in the brain, but differed in their identified cord outputs. In both BANC and MaleCNS, cord interneurons received most of DNge006's retained identified output. Direct motor-neuron contacts accounted for nearly half of DNge072's retained identified output, compared with about 13–17% for DNge006 in the pooled type profiles. This contrast was present in both cells of each type in each reconstruction (Fig. 4a,b; Supplementary Table 18).

The difference extended from target classes to individual downstream cells. The selected DNge006–DNge072 pair had no common retained identified cord target in either reconstruction (Fig. 4c, inset; Supplementary Fig. 9c). Across one-way connected DN pairs with shared input and retained output from both neurons, the median fraction of their combined retained output directed to common targets was approximately 1% in each reconstruction, although target overlap spanned a broad range (Fig. 4c; Supplementary Table 19). Connected pairs with shared input nevertheless showed slightly greater overlap in identified cord output than unconnected pairs matched for measured anatomy and shared input (Supplementary Fig. 9b and Table 20).

Output allocation and anatomical reach distinguished the two DNs in different ways. Despite its smaller motor-neuron output fraction, DNge006 had more retained identified output synapses overall, contacted more distinct motor neurons directly, and reached a larger motor-neuron set through one identified cord interneuron (Fig. 4d). Thus, DNge072 allocated a larger proportion of its identified output directly to motor neurons, whereas DNge006's connections extended to more distinct motor-neuron targets. Shared ascending input and local DN interconnection therefore coexisted with different downstream target compositions and anatomical reach.

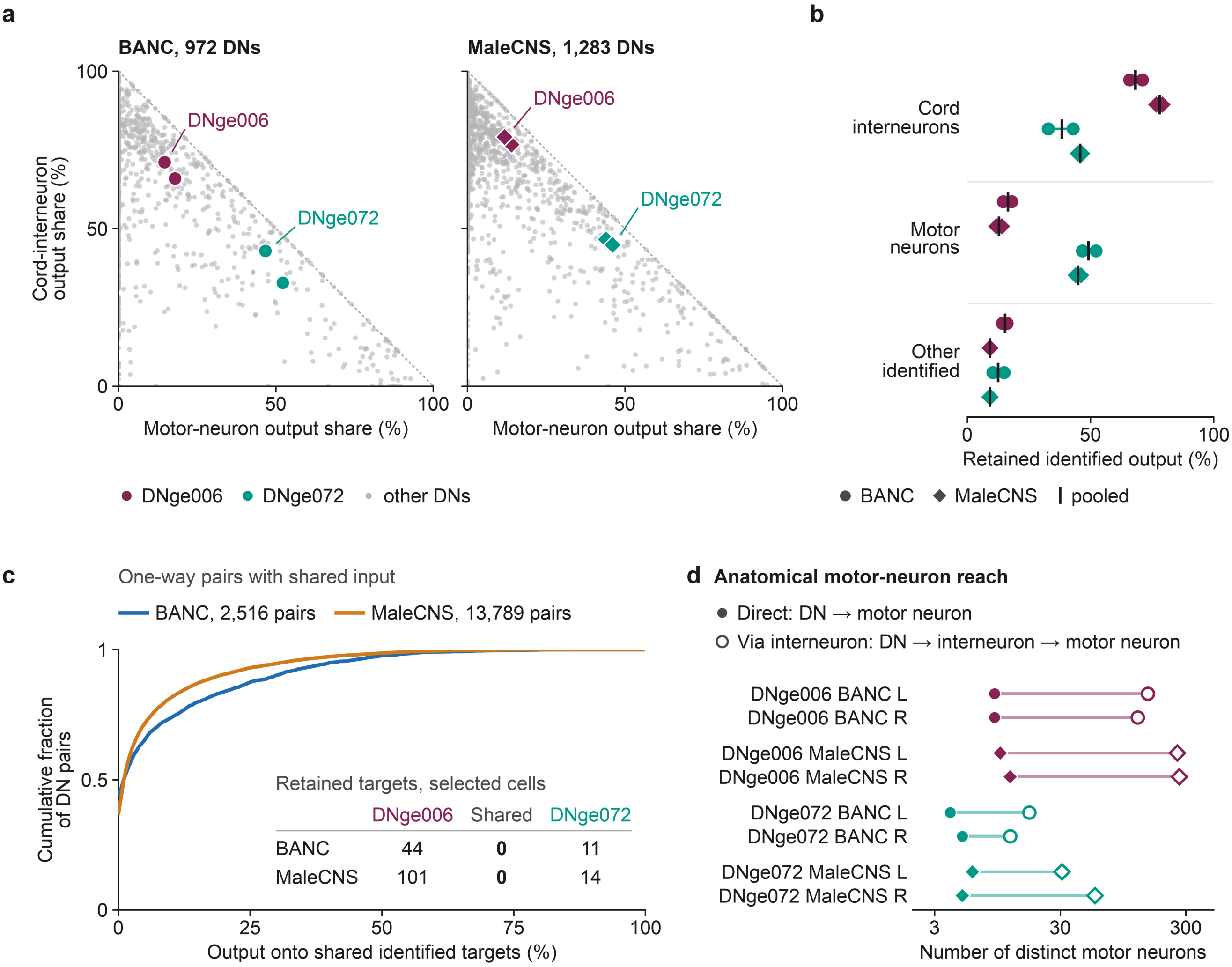

| | DNge006 | Shared | DNge072 |
|---|---|---|---|
| BANC | 44 | **0** | 11 |
| MaleCNS | 101 | **0** | 14 |

**Figure 4 | Interconnected DNs share an ascending source but distribute output differently.** Retained targets are identified neurons receiving at least five cord synapses from a DN; output shares are calculated from synapse counts onto these targets. **a,** Motor-neuron and cord-interneuron output shares across DNs with retained targets, highlighting both cells of DNge006 and DNge072 in each reconstruction. Gray points show other DNs; the diagonal denotes output entirely allocated to these two target classes. **b,** Output shares for individual cells. Circles denote BANC and diamonds MaleCNS; lines join the two cells of each type within a reconstruction, and vertical ticks mark their pooled synapse-weighted shares. Other identified targets include ascending, efferent, descending and other accepted neuronal classes. **c,** Cumulative distributions of the fraction of each pair's combined retained output directed to targets retained for both DNs, for one-way connected pairs with shared input and retained output from both cells. The inset lists retained target counts and the number shared by the selected cells shown in Fig. 2. **d,** Numbers of distinct identified motor neurons contacted directly or reached through one identified cord interneuron, shown by filled and open symbols, respectively, on a logarithmic axis. Lines connect the two counts for each DN. Each edge requires at least five synapses; a motor neuron may belong to both sets and is counted once within each. Annotation coverage and alternative output summaries are reported in Supplementary Tables 18–19.

## Discussion

The identified ascending–descending circuit illustrates how a relationship among neurons can be reflected in input organization within one of them. AN19B014 contacted both DNge006 and DNge072, while DNge006 supplied an additional input to DNge072, creating direct and partner-mediated anatomical routes to a common receiver. Within DNge072, inputs pooled across the pair's 12 shared sources, including AN19B014, lay nearer DNge006 input sites than other non-DN inputs in the same region. This arrangement exemplified a relationship observed across thousands of DN connections in MaleCNS and the independently reconstructed FlyWire brain. Across the two placement comparisons and two brains, mean differences were approximately 8–10 μm, with the corresponding distance ordering present in about four-fifths of eligible receivers. Connectivity between neurons was therefore associated with relative input placement within individual neurons.

This relationship connects two levels of neuronal organization. At the graph level, a source contacting both a partner and a receiver, together with a partner-to-receiver connection, supplies the three edges of a feed-forward configuration[13]. At the cellular level, these edges comprise synaptic contacts distributed along receiving arbors. An adjacency matrix preserves the connections and their contact counts but leaves their spatial arrangement unspecified. The within-source partner comparison related these descriptions while holding the source, receiver and source-input positions fixed: inputs from source-connected partners lay nearer those positions than inputs from partners with none recorded, after standardizing reference-set size. The association thus involved connectivity between two presynaptic neurons and the separation of their contacts within a third neuron.

The network and placement analyses describe complementary aspects of shared-source organization. In BANC and MaleCNS, shared sources were more frequent among interconnected DN pairs than among unconnected pairs, consistent with the broader relevance of common-neighbor relationships to neuronal connectivity[9,10]. Shared input provided predictive information about DN interconnection beyond measured voxel and regional overlap, while spatial overlap contributed information beyond shared input. These descriptions retain different properties: the territories occupied by two cells and the identities of upstream neurons contacting both. Within receiving DNs, source–partner connectivity was also associated with relative input placement. Source sharing was therefore related both to interconnection between DNs and to the arrangement of converging inputs within the receiver. A few upstream cells often contributed most of the shared-input overlap, making this population relationship attributable to specific cellular sources.

Local branch accessibility offers one possible origin of these relationships. Nearby source, partner and receiving neurites could provide opportunities for the three connections and for source and partner inputs to occupy nearby positions on the receiver. Cortical studies have related geometric overlap to functional connectivity[32], and experimentally shifting motoneuron dendrites in *Drosophila* larvae changes their contacts with identified premotor partners[33]. Such local geometry could remain relevant among DN pairs with similar voxel or regional distributions. In the identified circuit, AN19B014 contacted the two DNs predominantly through distinct presynaptic sites. Activity-dependent refinement provides another possible contribution. Nearby inputs can be coactive, and activity-dependent processes influence synaptic clustering and weakening during development[34–36]. Correlated activity in direct and partner-mediated routes could contribute to their relative input placement. Developmental measurements could distinguish placement established as overlapping arbors grow from subsequent changes associated with activity in the converging routes.

For models of neuronal network formation, the observed relationship supplies a joint target across these levels. Generative models of human connectomes have combined geometric and topological factors[37], while neuron-resolved models incorporate topological and spatial or physical-contact constraints[38]. Morphology-based cortical models also show how neuronal geometry contributes to network architecture[39]. A model can reproduce connection probabilities, shared-neighbor relationships and synapse counts while assigning different positions to the same contacts. The present measurements provide an additional observable: whether source–partner connectivity is associated with relative input placement within the common receiver. Physical-network models that represent the internal structure of spatially extended nodes offer a framework for retaining these arrangements alongside connectivity[14].

Earlier synapse-resolved work provides a close precedent. In *C. elegans*, three-neuron configurations and shared-neighbor relationships have been associated with local synaptic organization along neurites[18]. Developmental work has also examined synaptic clustering in the reconstruction series used here, including its adult specimens[19,30]. Our within-source partner comparison asks whether fixed source inputs are nearer sites supplied by partners that also receive source input, after standardizing the number of reference sites. It yielded negative estimates in both adult reconstructions, and the between-source point estimates had the same direction. These observations extend the specified comparison to a different neuronal population; the small samples and the between-source interval crossing zero in Dataset8 limit their precision. The independent fly comparisons establish the relationship in sampled adult descending circuits, while the worm analyses connect it to earlier evidence for local synaptic organization without assuming cell-type homology.

Common upstream organization coexisted with differentiated downstream connectivity. The DNge006–DNge072 example was selected from pairs with similar cord-region output profiles, yet the selected cells had no common retained identified cord target in either reconstruction. Regional similarity therefore accompanied separation at the level of individual targets. DNge072 allocated a larger fraction of retained identified output directly to motor neurons, whereas DNge006 contacted more distinct motor neurons directly and reached a larger set through one cord interneuron. Across one-way connected pairs with shared input, the median fraction of retained output directed to common targets was approximately 1%. A separate matched comparison including one-way and reciprocal pairs nevertheless showed slightly greater output overlap among interconnected pairs than among unconnected controls. Shared premotor-input patterns organize motor-neuron modules in fly motor circuits[40], and descending-to-motor reconstructions distinguish direct contacts from routes through cord interneurons[23]. Here, shared ascending input and local DN interconnection accompanied differences in target composition and anatomical reach.

The physiological question is how input placement affects the joint action of converging routes. DNs participate in state-dependent motor control and interact within descending populations[5–7]. Direct source input and input through a partner could contribute differently to a receiver's activity depending on their positions, strengths and timing. Cortical pyramidal neurons exhibit position-dependent synaptic summation[15]. In passive models of looming-responsive fly DNs, dendritic structure can normalize individual input efficacy, and synapse distributions can support approximately linear encoding of synapse number[20]. Studies of *Drosophila* olfactory circuits have also related wiring and input location to connection strength and subcellular integration[41,42], while cortical work shows sensitivity to input sequence and the relative arrangement of excitation and inhibition[43,44]. Path distance measures geometric separation; its electrical consequences depend on the receiving neurites' biophysical properties. Models preserving connectivity, synapse counts and morphology while varying input positions could test the contribution of placement. Recording DNge072 during AN19B014 stimulation with selective DNge006 perturbation would instead test the partner's contribution to the response. Type-wide perturbations would also involve connections beyond the selected cell pair.

The placement evidence spans one male MaleCNS specimen and one female FlyWire specimen, with intervals describing dependence within specimens rather than variation across animals or a sex difference. FlyWire retained the core thresholds and numerical definitions of the MaleCNS comparisons. Both placement analyses used a fixed five-contact edge threshold; the three- and seven-contact sensitivity analyses concerned DN interconnection prediction. Fine branch accessibility remains a possible contributor even when source-input positions are fixed. The *C. elegans* estimates came from small interneuron samples. In *Platynereis*, three of four eligible between-source pairs had positive differences and all four shared one partner; only one receiver qualified for the within-source comparison. These sparse observations do not establish the fly relationship in that population. Network and cord-output analyses were restricted to BANC and MaleCNS, where accepted target annotations covered approximately one-fifth to two-fifths of recorded cord output for DNge006 and DNge072 before edge thresholding.

The identified circuit provides a concrete example of a relationship supported across two independently reconstructed fly brains: connectivity between presynaptic neurons is associated with the relative placement of their inputs within a common receiver. The network and placement results connect relationships among cells to organization within cells. For physical-network research, this is an empirical example of how topology among spatially extended nodes can be associated with the internal arrangement of their connections. Synapse-resolved connectomes make both levels measurable: which neurons are linked and where their converging inputs lie within the cells that receive them.

## Methods

### Specimens, cohorts and connection definitions

We used public BANC v888 and MaleCNS v1.0 reconstructions from one female and one male *Drosophila*, respectively, for network and cord-output analyses, and MaleCNS plus the independent female FlyWire/FAFB v783 brain for synaptic-placement comparisons.[24,27–29] Two adult *C. elegans* brains and one *Platynereis* larva supplied exploratory extensions described below.[30,31] No new animals were acquired. Released MaleCNS synaptic partner records had confidence scores of at least 0.5. Native identifiers, annotations and source versions were retained. Supplementary Table 21 separates each dataset's cohort from the cells and connections eligible for individual comparisons.

The initial BANC cohort comprised 986 typed DNs with brain input, an exact-ID skeleton containing at least two usable brain vertices, finite geometric measurements and positive brain cable length and at least 50 cord synapses onto the annotated target set used in the initial cohort mask. Removing all DN–DN inputs excluded seven cells with no remaining input at the five-synapse edge threshold. The final 979 DNs, representing 418 exact types, also had nonzero propagated source access and at least 50 synapses in named cord regions. MaleCNS required a known type, nonzero external brain input and source access, at least 50 synapses in named cord regions and at least two external brain synaptic records, yielding 1,287 DNs in 473 types. These cord-output criteria defined the BANC and MaleCNS cohorts; path-distance eligibility was determined separately as described below.

A directed DN connection required at least five brain synapses. Different-type pairs were one-way when only one direction met this threshold, reciprocal when both did, and unconnected when neither did. The unordered-pair prediction response was a connection in either direction. All-pair summaries contained 476,923 BANC and 825,185 MaleCNS pairs. Input graphs retained non-self brain edges with at least five synapses and removed every annotated DN–DN edge, including edges involving DNs outside the cohorts: 7,338 BANC and 19,475 MaleCNS directed edges. Raw response counts and source-to-partner counts were retained separately.

**Shared sources and receiving geometry**

For a one-way connection $i \to j$, $i$ is the partner DN and $j$ the receiver. A shared source is a non-DN neuron with at least five brain synapses onto each DN. Thus $S_{ij}=N^{-}_{\text{nonDN}}(i) \cap N^{-}_{\text{nonDN}}(j)$. Other edges among these cells were unrestricted; the configuration was not an induced-triad census. In the between-source comparison, other inputs came from retained non-DN sources outside $S_{ij}$ and could include sources with one to four synapses onto the partner. The within-source comparison instead required zero recorded source-to-partner synapses for the comparator.

Receiving geometry used official native-coordinate MaleCNS SWC skeletons for 1,079 DNs. Coordinates were multiplied by 0.008 to convert 8-nm units to micrometers. Skeletons were undirected graphs with Euclidean edge lengths; components were not bridged, and skeletons were not resampled or clipped to a brain region. Each postsynaptic input record was projected to its nearest vertex. For partner input sites $R_{ijr}$ in endpoint region $r$, distance from a projected source position $x$ was

$$d_{ijr}(x)= \min_{y \in R_{ijr}} d_{T_j}(x,y).$$

Here $T_j$ is the skeleton graph of receiver $j$, $d_{T_j}(x,y)$ is the shortest path length between vertices $x$ and $y$ along that graph, and $R_{ijr}$ is the set of vertices to which partner-$i$ input records in region $r$ were projected. Both endpoints had to have the same released `primary_post` region label; the intervening path could leave that region. Side-specific labels remained separate during measurement. No maximum projection-distance cutoff was applied. Unreachable records were omitted before checking regional eligibility. A record is a released presynaptic–postsynaptic contact row; a distinct physical site is a unique coordinate triplet within a fixed neuron and stratum; a projected vertex is a skeleton location. These units need not be one-to-one. Supplementary Methods report projection errors, component counts and exclusion denominators. BANC synaptic positions were not inferred from edge totals.

**Between-source comparison along receiving neurites**

Within each partner–receiver connection and endpoint region, we compared shared-source and other retained non-DN input records (Fig. 1a). Each group required at least three reachable records. All eligible records contributed to their group mean. Regional means were weighted by the shared-source record count, imposing the same regional mixture on both groups:

$$\Delta_{ij}=\sum_{r} \frac{n_{ijr}^{\text{shared}}}{\sum_{r'} n_{ijr'}^{\text{shared}}} \left(\overline{d}_{ijr}^{\text{shared}} - \overline{d}_{ijr}^{\text{other}}\right).$$

Here $n_{ijr}^{\text{shared}}$ is the number of eligible shared-source input records of connection $i \to j$ in region $r$, and $\overline{d}_{ijr}^{\text{shared}}$ and $\overline{d}_{ijr}^{\text{other}}$ are the mean distances $d_{ijr}(x)$ over the eligible shared-source and other records in that region. No comparator subsampling was performed. A negative difference indicates nearer shared-source inputs. In MaleCNS, before excluding unreachable projected positions, the Euclidean-distance criteria retained 13,323 of 16,114 one-way connections. Removing unreachable projected positions and reapplying the regional record requirements retained 13,318 connections, 1,079 receivers and 6,260 unordered type pairs. Euclidean distances were recomputed on these same retained records and weights.

The primary mean weighted directed cell pairs equally. Alternative summaries first averaged within unordered type pairs or within receiving cells, then weighted those units equally (Supplementary Fig. 1 and Table 1). Type-endpoint intervals accounted for observations sharing a DN type.[45] Their implementation and the distinct within-source interval are described under Statistical analyses and reproducibility. Individual examples, including opposite orderings, are retained in Supplementary Table 3.

**Within-source partner comparison**

Receivers came from the between-source pool (1,079 MaleCNS; 720 FlyWire). The source, receiver and actual input records were fixed while partners changed (Fig. 1a). Partners were different-type cohort DNs with at least five brain synapses onto the receiver and fewer than five in reverse. Source-to-receiver edges also required five synapses. Within each exact side-specific region and connected skeleton component, at least three source records, three distinct postsynaptic coordinate triplets per partner and both partner classes were required. Partners receiving at least five source synapses were compared with those receiving none recorded in quality-filtered brain records; one-to-four counts were excluded.

Nearest distance depends on reference-set size. We standardized the statistic to three distinct physical partner sites sampled uniformly without replacement. The reference count of three was fixed before this comparison was calculated; no alternative reference counts were screened. For ordered path distances $d_{(1)},\ldots,d_{(n)}$ from a source record to a partner's sites, the exact expectation was

$$\tilde{d}_i(x)=\sum_{\ell=1}^{n\text{-}2}\frac{\binom{n\text{-}\ell}{2}}{\binom{n}{3}}d_{(\ell)}.$$

The calculation averages every three-site subset analytically rather than selecting the three nearest sites. Distinct coordinates projecting to the same vertex remained distinct sampling units; source records retained their multiplicity. Path lengths on skeleton forests were calculated using lowest common ancestors.

Within each source–receiver–region–component stratum, we averaged over the same source records and then equally over partners within each class. Strata were weighted by source-record counts within source–receiver pairs; sources were weighted equally within receivers, and receivers equally in the primary mean. In MaleCNS, the analysis retained 68,609 strata, 49,742 source–receiver pairs, 8,453 sources and 922 receivers in 415 exact types. The other 157 original receivers had no eligible stratum. The 15,110 distinct partner–receiver edges differed from the 13,318 between-source edges because eligibility was defined differently (Supplementary Table 2). Counts of partner-level rows and cross-class combinations describe reused comparisons, not independent biological samples.

**Independent placement comparisons in FlyWire**

FlyWire used the v783 individual-synapse and proofread connectivity releases, the official v2 neuron annotations and the matching released skeletons.[27,28] The endpoint cohort comprised 1,210 typed DNs in 446 exact types with matching skeletons. All 1,303 annotated DNs, including 93 without exact types, were excluded from the source pool. Sources were other proofread neurons. The same five-contact edge threshold, one-way partner-to-receiver criterion, minimum site counts and nearest-three statistic were used. As FlyWire covers the brain, no cord-output criterion was imposed.

Native synapse and skeleton coordinates in nanometres were converted to micrometres by a factor of 0.001. Synapse counts into the cohort were reconciled with the released connectivity table. Postsynaptic coordinates were projected to the nearest receiving-skeleton vertex; ties followed SciPy cKDTree without an additional tie-breaking rule. Published skeletons had already been healed and downsampled by their provider. We added no edges or projection-distance exclusion. Comparisons retained released endpoint neuropil labels, including hemisphere suffixes, and required paths within connected skeleton components. The definitions and weighting of the two comparisons were otherwise unchanged. FlyWire yielded 5,807 between-source pairs and 720 receivers, and 19,246 source–receiver pairs involving 4,140 sources and 589 receivers in the within-source comparison (Supplementary Note 7, Supplementary Fig. 10 and Tables 23–24).

**Exploratory comparisons in other species**

The *C. elegans* analysis used the two adult specimens, Dataset7 and Dataset8, from Witvliet et al.[30] Each endpoint cohort contained 44 author-classified interneurons in 24 classes. All author-classified interneurons were excluded from the source pool; sources were other identified neurons. This population was not treated as homologous to fly DNs. Coordinates and parent edges came from the original skeleton JSON files. Each chemical-contact record connected a native presynaptic active-zone record to a postsynaptic attachment; separate attachments to the same receiving cell were retained. EM-section counts were not used as contact multiplicities. The native postsynaptic treenode identified the receiving position without nearest-vertex projection, and nanometres were converted to micrometres. The entire reconstructed brain arbor was one spatial domain, with skeleton components preserved; it did not constitute a complete descending system. Both adults were analyzed with the five-contact threshold, site-eligibility rules and nearest-three statistic unchanged (Supplementary Note 8, Supplementary Fig. 11 and Table 25).

For *Platynereis*, the published three-day larval reconstruction supplied 138 author-annotated head-to-trunk projection cells, of which 99 had unambiguous types (35 types).[31] All 138 were excluded from upstream sources, which additionally required the author's neuronal annotation. Chemical contacts used native connector-to-postsynaptic-treenode records and unchanged parent graphs, with nanometre coordinates converted to micrometres. The whole receiving arbor was one spatial domain. Applying the same thresholds retained four between-source pairs and one within-source receiver. Their intervals were non-estimable because all between-source type pairs shared a type endpoint and the within-source sample contained only one receiver type (Supplementary Note 8 and Table 26).

**Spatial overlap and shared-input measurements**

Spatial overlap comprised weighted cosine and binary Jaccard overlap of voxel occupancy and external regional synapse counts. BANC voxel weights were skeleton cable lengths in 40-μm native-coordinate voxels: segments with both endpoints at $y<325$ μm were assigned by midpoint, following the public bancr brain separation plane.[46] MaleCNS voxel weights counted the DN's own endpoint coordinates on external brain contacts: presynaptic positions for outgoing contacts and postsynaptic positions for incoming contacts. DN–DN records were excluded. Counts were unthresholded, and separate postsynaptic partners at a polyadic site contributed separate rows. Spatial inputs therefore represented cable distribution in BANC and external synaptic-position distribution in MaleCNS. The retained arrays contained 321 and 594 nonempty voxels and 80 and 79 side-specific regional labels, respectively.

Shared-input cosine and Jaccard used distinct non-DN source IDs and thresholded input counts. The common baseline included centroid distance, axiswise arbor-range overlap, external degrees and strengths, side agreement, direct upstream canonical-class and fine cell-class similarities, first-hop source-category similarity, and source/output diversity. The latter used sums and absolute differences of Shannon entropy. First-hop profiles were obtained by row-normalizing the one-step propagated source matrix $AB$ over the eight annotated source groups, where $A$ is the discounted propagation operator and $B$ the source-group seed matrix (Supplementary Methods), whereas shared-input overlap retained individual source identities. Supplementary Table 22 gives formulas, fields, transformations, zero handling and model membership.

Coarse source access summarized discounted graph walks from eight annotated groups, including ascending neurons, with discount 0.85. Cord-region profiles pooled all recorded outgoing contacts into seven named groups, without requiring identified targets or a per-target edge threshold. They therefore differed from the retained identified-target profiles of Fig. 4. Similarity was 1-JSD($p$,$q$)/log2, using natural-log divergence, not its square root. Exact source and region mappings, the propagation operator and first-arrival alternative are specified in Supplementary Methods.

**Prediction evaluation and sensitivity analyses**

Four core additive logistic models contained the common baseline C0, C0 plus spatial overlap (A0), C0 plus shared input (U0), or both additions (M0). Other models examined coarse source/output similarities (Supplementary Note 3 and Tables 8–10). Predictors used cubic B-splines with four uniformly spaced training-range knots, including the two boundaries, constant extrapolation and no bias column. Logistic fitting minimized mean log loss plus $\lambda\|\beta\|^2/2$, with an unpenalized intercept, using SciPy and scikit-learn. Supplementary Methods specify preprocessing, optimizer settings and model selection.

Five outer folds grouped complete exact cell types. A pair was evaluated only when both endpoint types belonged to the held-out fold; fitting excluded all pairs involving either held-out type. Up to 40,000 training pairs were sampled uniformly without replacement using seed 20260908 + fold, independently of connection labels and identically across models. Three inner complete-type folds selected $\lambda$ from 0.0001, 0.001, 0.01, 0.1 and 1. Both endpoint types were excluded from the corresponding inner training set. Knots and feature standardization were fitted on each applicable training partition. Evaluation retained natural prevalence: 93,550 BANC and 162,638 MaleCNS cell pairs in 17,265 and 22,137 type pairs. Holdout applied to supervised fitting; external anatomy of test cells remained observable.

Log loss used natural logarithms. Conditional contribution was reference-model minus extended-model loss on identical evaluated pairs; positive values indicated improvement. Relative improvement was $100(L_{\text{reference}}\text{-}L_{\text{extended}})/L_{\text{reference}}$, calculated after pooling pair losses, rather than averaging fold-specific percentages. Figure 3d expresses mean loss reductions in units of $10^{-3}$ nats per pair, numerically equal to nats per 1,000 pairs in Supplementary Figs. 5–6 and Tables 9–10 and 14. Type-pair evaluation changed weights without refitting.

Four sensitivity settings changed voxel size to 30 or 50 μm at five synapses, or the DN–DN response threshold to three or seven synapses at 40 μm (Supplementary Fig. 6 and Table 14). Cohorts, folds, regional overlap and non-target input measurements remained fixed. Only the two conditional contributions were evaluated. Held-out predictions supplied Brier scores, AUROC, average precision and ten-bin calibration, without recalibration or threshold optimization (Supplementary Fig. 7 and Tables 15–16).

**Identified circuits, source contributions and presynaptic sites**

DNge006–DNge072 and the two other examples in Supplementary Fig. 2 were selected using BANC development data before inspecting their MaleCNS connections. The candidate and ranking rules are given in Supplementary Methods. Selection did not use receiving-neurite distances. Within each type pair, individual cells were selected by prioritizing a one-way connection, then shared-input cosine and spatial cosine. These are selected cell arrangements, not spatially registered cross-specimen replicas.

A source's contribution to shared-input cosine was its product of input weights onto the two DNs divided by the input-vector norms. Its fraction of total cosine overlap quantified its contribution to overlap, not its fraction of all receiver input. Source concentration and directional asymmetry are reported in Supplementary Note 2 and Supplementary Fig. 4. For the named circuit, AN19B014's overlap contribution, its presynaptic-site allocation and the receiving positions of the pooled 12 shared sources were separate measurements.

A MaleCNS presynaptic site was keyed by upstream ID and the original integer pre-coordinate triplet. Each DN pair counted once per site, regardless of postsynaptic multiplicity. With $s_i$,$s_j$ distinct sites from shared sources supplying each DN and $b$ supplying both, overlap was $2b/(s_i+s_j)$. The 159 AN19B014 sites included 21 contacting both DNs, giving overlap $42/180$ rather than $21/159$. The allocation reference preserved each source's contact counts and site slots (Supplementary Fig. 3 and Supplementary Methods). Transmitter annotations and display rules, including source-specific example paths, are documented separately in Supplementary Methods.

**Cord-output targets and matched comparisons**

Outgoing contacts in the seven named cord groups were aggregated by exact postsynaptic ID. Accepted targets were traced annotated neurons in MaleCNS or proofread/roughly proofread annotated neurons in BANC; non-neuronal classes were excluded. Principal target profiles then retained edges with at least five cord synapses. Seven BANC and four MaleCNS DNs had no retained target and were excluded from target-pair summaries.

We distinguished three denominators. Annotation coverage was identified output before edge thresholding divided by all recorded output in these cord groups. A retained identified-output share was a target class's retained synapses divided by all retained identified-output synapses. An all-contact fraction divided a target class's retained synapses by all recorded cord output, including unaccepted and unresolved targets. Supplementary Table 18 reports the prethreshold identified numerator and retained counts.

For retained target weights $v_{ik}$, shared-target mass was

$$M_{ij}=\frac{\sum_{k:v_{ik}>0,\, v_{jk}>0}(v_{ik}+v_{jk})}{\sum_k v_{ik}+\sum_k v_{jk}}.$$

Cosine and Jaccard measured target-vector overlap separately. Released superclass and region fields defined target classes. Direct motor reach required a retained DN-to-motor edge; two-edge reach required an accepted cord-intrinsic interneuron and at least five cord synapses on each edge. Each motor endpoint counted once per DN. Direct and two-edge sets could overlap and were not added together (Fig. 4d).

Connected pairs with shared input were matched to unconnected pairs with shared input by Euclidean distance after scaling eight anatomical/input covariates by their specimen-wide standard deviations; side agreement was matched exactly. Matching allowed replacement and used no caliper or output measurements. After matching, both pairs had to have retained identified cord output; the same output-availability filter was applied to the presynaptic-site comparison. It accounts for the 27 one-way and one reciprocal MaleCNS pairs omitted there. Supplementary Methods specify the matching metric, post-match exclusions and control reuse. Differences are descriptive.

#### Statistical analyses and reproducibility

The biological units were reconstructed specimens: BANC and MaleCNS for network and output analyses, MaleCNS and FlyWire for DN placement, and two *C. elegans* adults and one *Platynereis* larva for exploratory extensions. Between-source and prediction intervals were unadjusted normal 95% intervals accounting for type-endpoint dependence. Prediction intervals applied that estimator to paired losses, conditional on fitted models. Within-source intervals used crossed clusters by receiver exact type and source cell, preserving receiver-equal weights. These are within-specimen asymptotic intervals, not estimates of between-animal variation or refitting uncertainty. Their finite-sample coverage is not established for the small worm samples. Structurally degenerate or nonpositive signed variances were not reported as valid intervals in the extensions; *Platynereis* point estimates were retained without intervals. Equations, eligibility conditions and alternative weightings are given in Supplementary Methods.

Analyses were exploratory. Spatial-overlap and shared-input contributions were evaluated separately after fitting the combined model. Sensitivity settings and the within-source partner comparison were specified before their respective calculations; the reference count was not chosen by screening alternatives. The FlyWire and cross-species analyses retained these thresholds and numerical definitions. Dataset-specific source populations, coordinate handling and spatial domains were fixed before evaluating their effects; these adaptations were not preregistered. Both adult *C. elegans* specimens and the sparse *Platynereis* results are reported. Supplementary Methods describe the input adaptations, distance calculations, weighting and uncertainty estimation.

#### Interactive visualization with FlyScope

FlyScope v2.1.0 used a Python server and a Three.js interface. Archived results were linked to cell identifiers, receiving-site records and available skeletons. Path views used frozen input-to-vertex mappings for MaleCNS and FlyWire or native receiving treenodes for the other specimens, preserving disconnected skeleton components. Offsets from recorded sites to mapped vertices were displayed separately and excluded from path distances. Overview geometry was simplified for rendering; traced paths used detailed skeletons. The nearest-input tool displays one path; the within-source statistic instead averages the nearest distance over uniform samples of three distinct reference sites, with archived estimates and strata available separately. Pair summaries pool retained shared sources and remain unchanged when the highlighted source is switched. Display filters do not alter the archived study summaries or their eligibility criteria and weights. Dataset coverage is labelled explicitly; BANC provides circuit topology and retained-output summaries.

#### Use of AI tools

OpenAI Codex assisted with analysis code and computational execution. The author is responsible for reviewing and approving the code, results, references and manuscript.

## Data availability

BANC v888 connectivity, cell annotations and skeletons are available from Harvard Dataverse (https://doi.org/10.7910/DVN/7WTH1N) and the public BANC data repository (https://console.cloud.google.com/storage/browser/lee-lab_brain-and-nerve-cord-fly-connectome/).[24] MaleCNS v1.0 synaptic partner records (minimum confidence 0.5), annotations, neurotransmitter predictions and native-coordinate skeletons are available at https://male-cns.janelia.org/download/.[29] FlyWire v783 individual synapses and connectivity are available at https://zenodo.org/records/10676866, matching skeletons at https://zenodo.org/records/10877326, and annotations at https://github.com/flyconnectome/flywire_annotations.[27,28] *C. elegans* Dataset7 and Dataset8 skeletons, synapses and neuron classifications are available at https://github.com/dwitvliet/nature2021.[30] *Platynereis* data are available through https://catmaid.jekelylab.ex.ac.uk/ (publication project 30, April 2024 archive; stack 21) and https://github.com/JekelyLab/*Platynereis*_3D_connectome_2024.[31]

Interactive exploration of the study circuits and archived placement results is available through FlyScope (https://flyscope.xizhe.net/), including a direct view of source and partner inputs on DNge072.

## Author contributions

X.Z. conceived and designed the study, directed the analyses, interpreted the results, and wrote and revised the manuscript.

## Competing interests

The author declares no competing interests.

# Supplementary Information

Synaptic placement reflects shared input in *Drosophila* descending neurons

Xizhe Zhang

ORCID: https://orcid.org/0000-0002-8684-4591

This file contains Supplementary Figures 1-11, Supplementary Notes 1-8, Supplementary Methods and Supplementary Tables 1-26. Notes 1-6 describe the BANC and MaleCNS analyses. Note 7 describes the independent FlyWire placement comparisons; Note 8 describes the exploratory C. elegans and Platynereis comparisons.

## Supplementary Figures

# Supplementary Notes and Methods

# Supplementary Tables

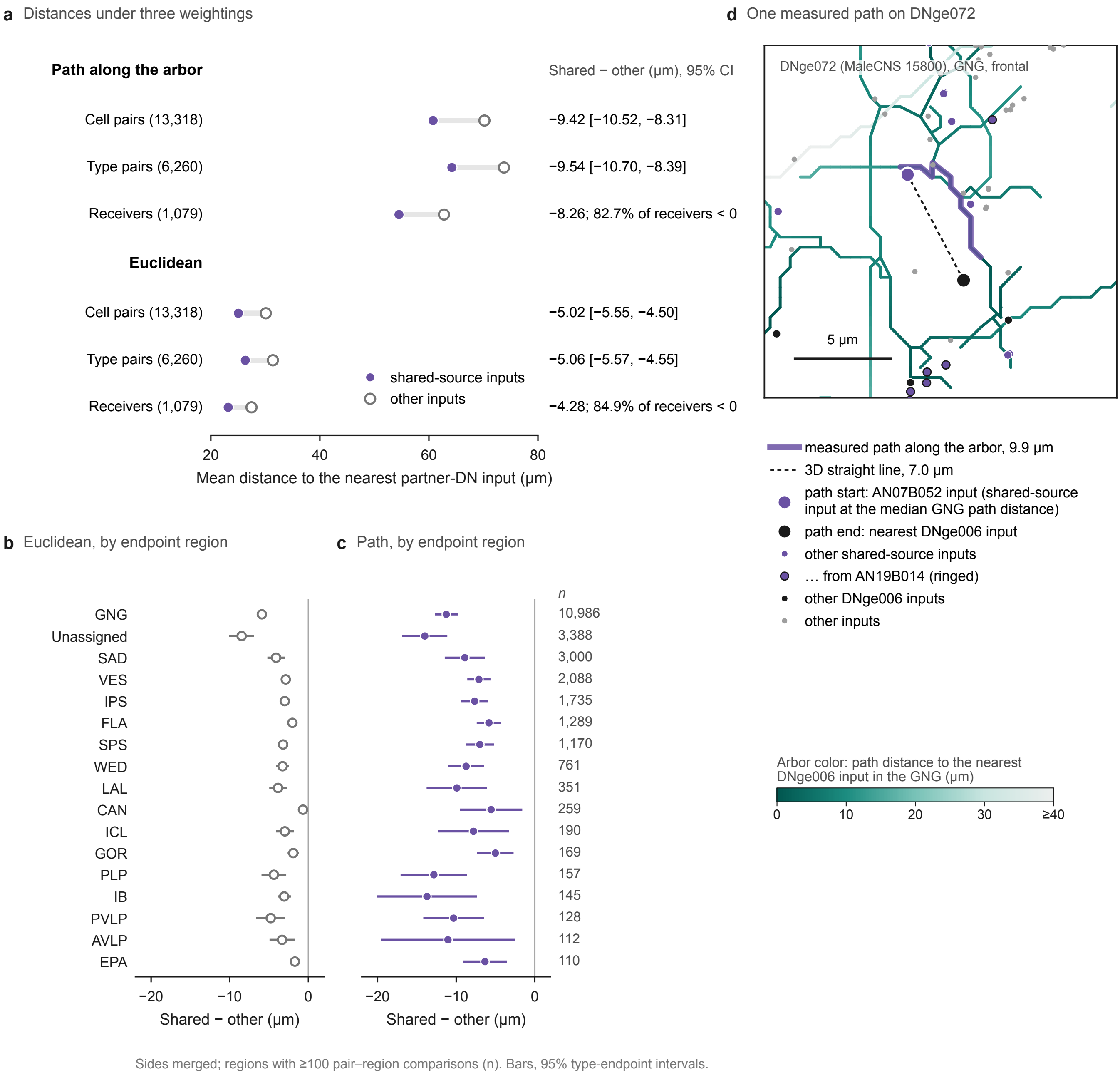

**Supplementary Figure 1. Distances of shared-source and other inputs under alternative weighting schemes.**

a, Mean path (along the reconstructed neurite) and Euclidean distance of shared-source (filled) and other (open) inputs to the nearest partner-DN input in the same receiver and brain region, on identical retained records, weighting 13,318 directed cell pairs, 6,260 type pairs or 1,079 receivers equally. Text gives differences with 95% type-endpoint intervals or, for receiver averages, the percentage of receivers with a negative difference. b, Euclidean differences by endpoint brain region. c, Corresponding path-distance differences; n is the number of pair-region comparisons. Sides are merged; groups require at least 100 comparisons. Unassigned denotes the released CentralBrain-unspecified label (Supplementary Methods, "Record identity and receiving-skeleton exclusions"). d, One measured path on DNge072 (MaleCNS 15800): the GNG shared-source input at the median path distance, from AN07B052, with a purple measured path (9.9 µm) ending at the nearest DNge006 input (large black point) and the three-dimensional straight line (dashed; 7.0 µm). The arbor is colored by path distance to the nearest DNge006 input in the GNG, as in Fig. 2c; purple, shared-source inputs (ringed, AN19B014); black, DNge006 inputs; gray, other inputs. Scale bar, 5 µm.

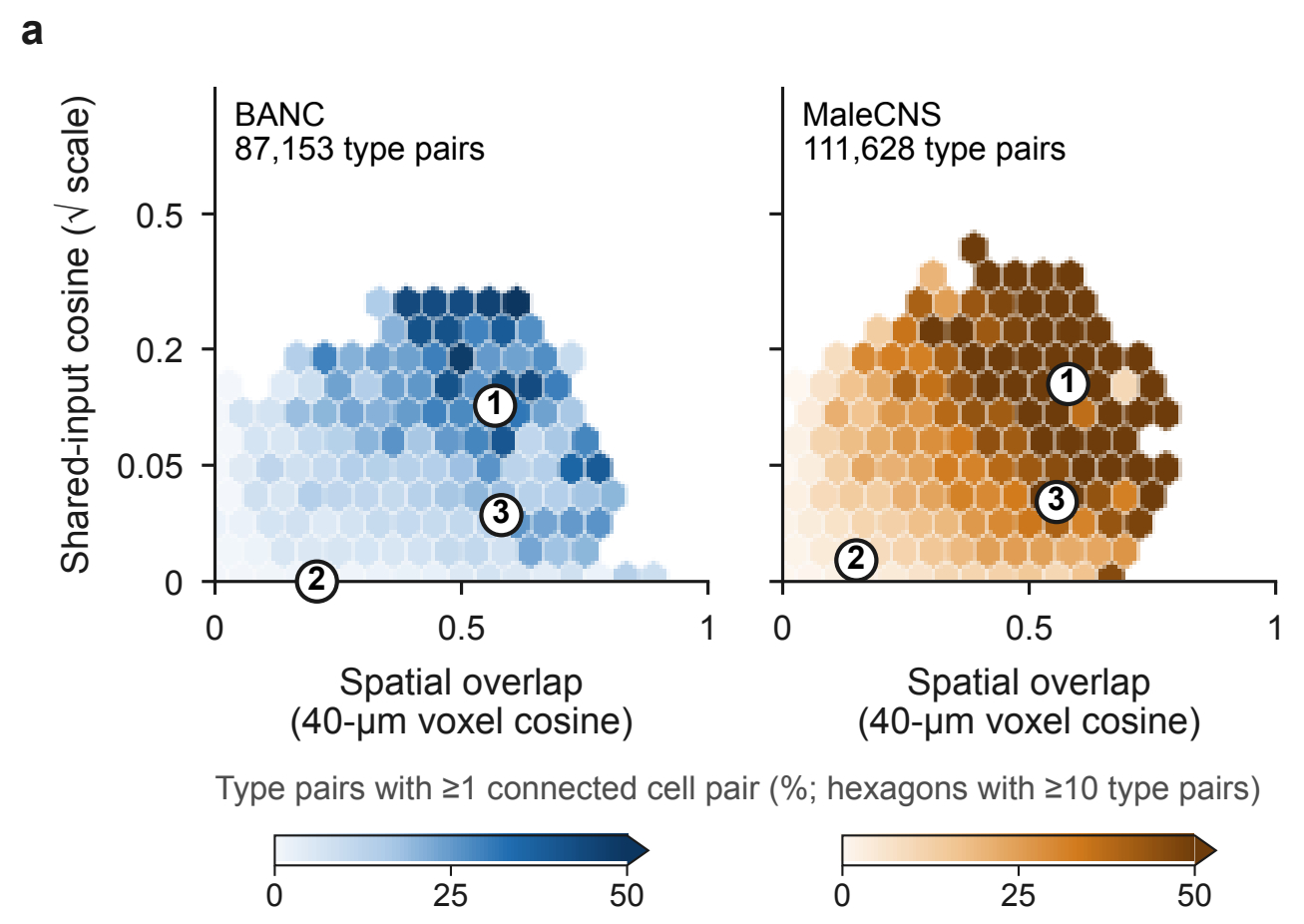

**b**

| | ① B | ① M | ② B | ② M | ③ B | ③ M |
|---|---|---|---|---|---|---|
| Connected cell pairs | 4/4 | 4/4 | 0/4 | 0/4 | 6/6 | 8/8 |
| Cord-region output similarity | 0.66 | 0.68 | 0.94 | 0.82 | 0.68 | 0.73 |
| Sensory-access similarity | 0.97 | 1.00 | 0.86 | 0.91 | 0.95 | 0.93 |
| BANC cable ≤5 µm of other type | 35% | – | 17% | – | 27% | – |
| Spatial overlap (voxel cosine) | 0.57 | 0.58 | 0.21 | 0.15 | 0.58 | 0.56 |
| Shared-input cosine | 0.114 | 0.144 | 0.000 | 0.002 | 0.016 | 0.023 |

B, BANC; M, MaleCNS; shading scaled within each row. Cable: mean of the two types' fractions (BANC only). Output similarity compares distributions over seven cord regions.

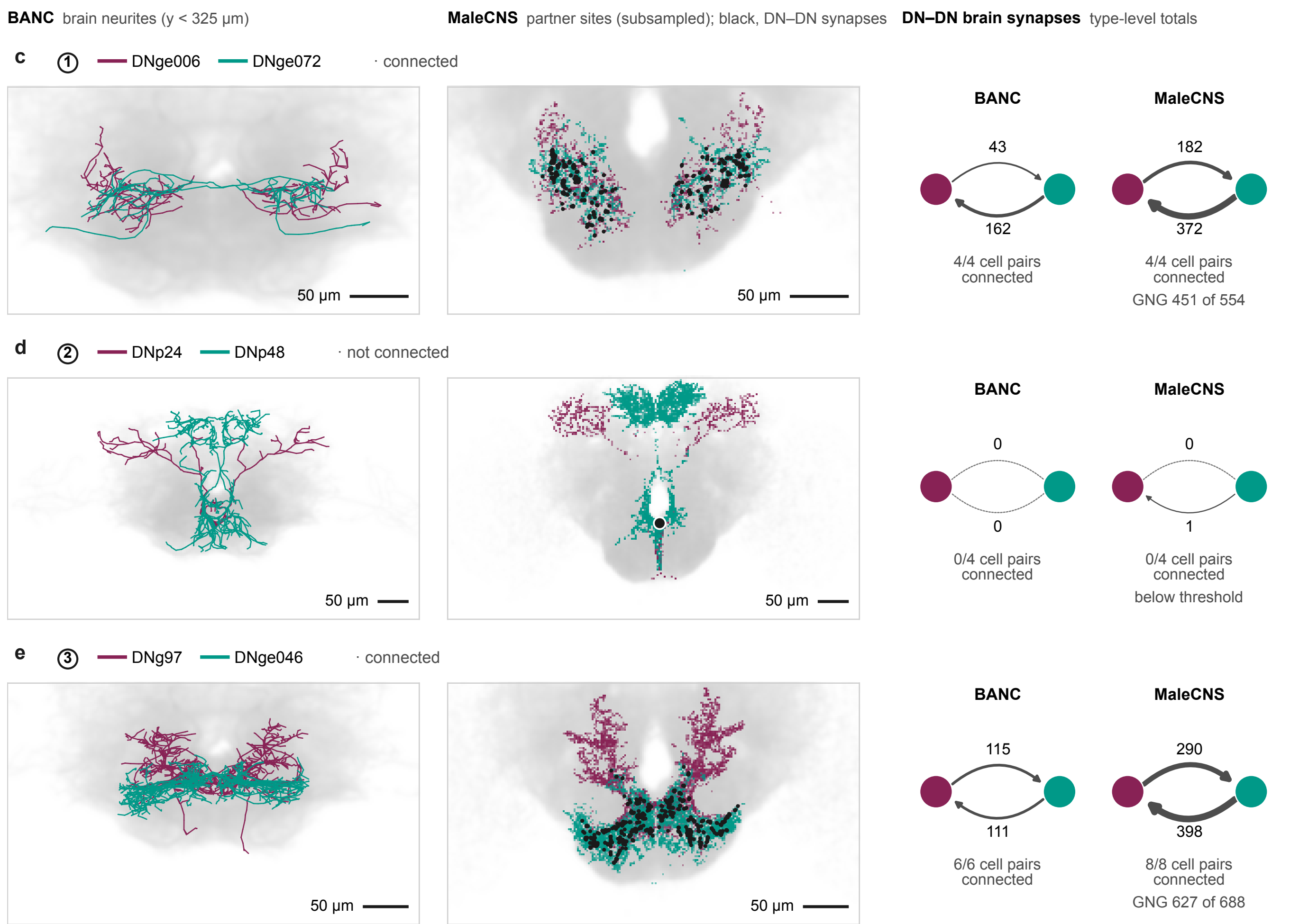

Frontal views, dorsal up, the fly's left to the right. Arrow width is proportional to the synapse count; a connection needs ≥5 synapses in either direction. Type-level totals include cell pairs other than the selected cells of Fig. 2b. Examples were selected from BANC development data.

**Supplementary Figure 2. Three example type pairs in their population context.** **a**, All 87,153 BANC and 111,628 MaleCNS type pairs, positioned by mean voxel cosine and shared-input cosine (square-root scale). Hexagons with at least ten type pairs show the connected fraction, capped at 50%; numbers locate examples. **b**, Connected cell-pair fractions, cord-region output similarity, sensory-access similarity, BANC cable proximity, voxel cosine and shared-input cosine; shading is row-scaled. B/M, BANC/MaleCNS. **c**–**e**, Examples 1–3, first type wine and second teal. Left, BANC brain skeletons; middle, subsampled MaleCNS external partner sites and DN–DN contacts (black); grey, cohort background. Frontal views have dorsal up and fly left to the right; bars, 50 $\mu$m. Right, type-level directional synapse totals over all cross-type cell pairs, connected-pair counts and GNG records. Totals include other cells beyond the selected pair of Fig. 2b. A connection requires five synapses; the single DNp48-to-DNp24 contact is below threshold. Selection and coordinate sampling are specified in Supplementary Methods.

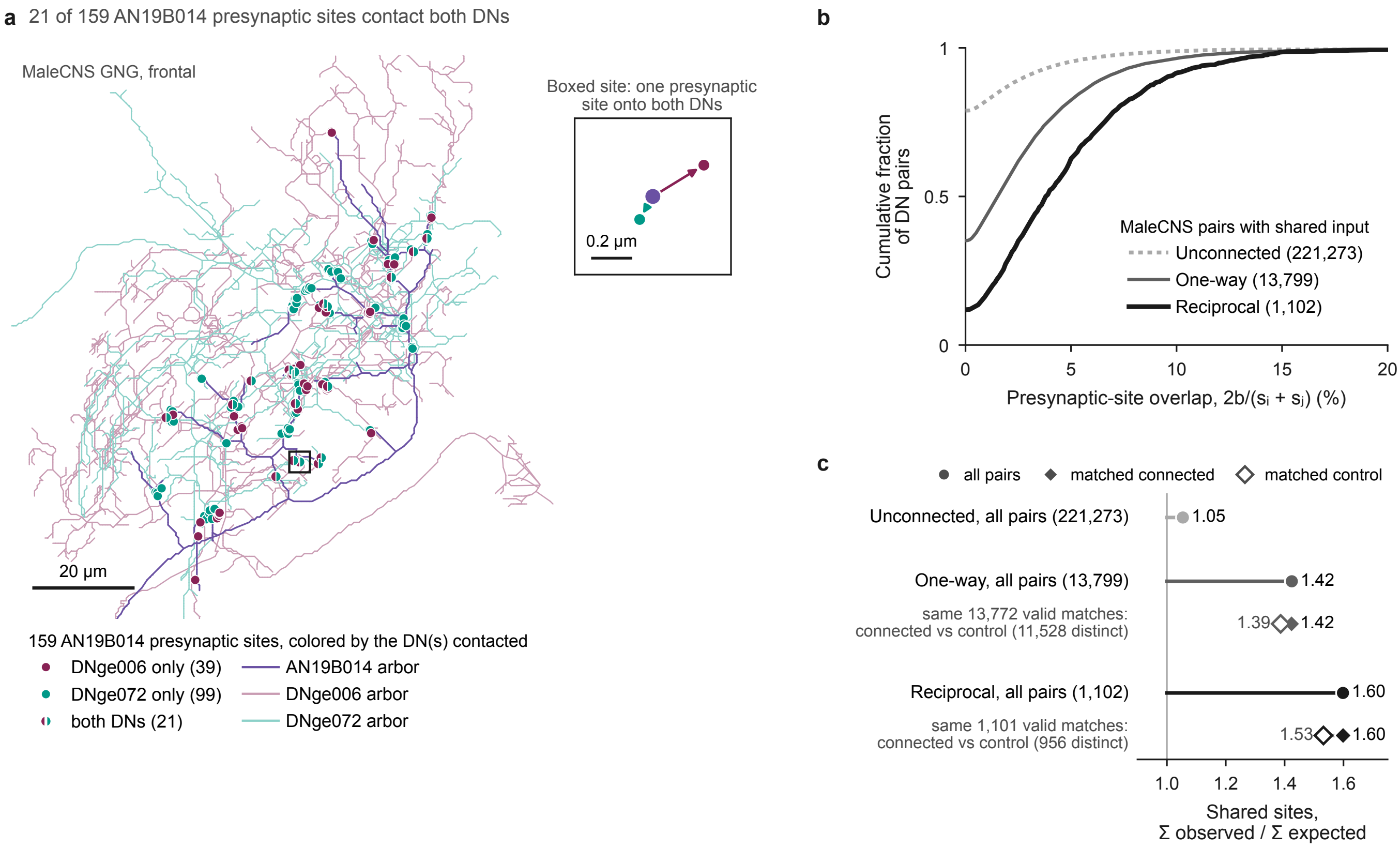

c, expected counts come from an allocation reference that preserves each upstream cell's contact counts onto cohort DNs and the contact slots at each site. Matched connected and control ratios use the same valid matches (both members with retained cord output); a reused control counts once per match. Sites, pairs and controls can be reused, so the ratios are descriptive.

**Supplementary Figure 3. Shared upstream neurons mainly contact DN pairs at separate presynaptic sites.** **a**, Frontal projection of the MaleCNS GNG with DNge006 (light wine), DNge072 (light teal) and AN19B014 (purple), and the 159 AN19B014 presynaptic sites with a postsynaptic partner on either DN, colored by the DN or DNs contacted. The inset enlarges the boxed site, the median sorted coordinate among the 21 shared sites; arrows run from the presynaptic site to its postsynaptic sites. **b**, Cumulative distributions of presynaptic-site overlap, $2b/(s_i + s_j)$, for all MaleCNS DN pairs with shared upstream sources, where $b$ counts sites of any shared upstream neuron that contact both DNs and $s_i, s_j$ count sites contacting each DN: unconnected (221,273 pairs; mean 0.77%), one-way (13,799; 2.57%) and reciprocal (1,102; 4.59%); the axis ends at 20%, beyond which lie fewer than 0.6% of the pairs in each class. For the single source in **a**, the overlap is 42/180 (23.3%). **c**, Aggregate observed/expected shared sites under a reference that preserves each upstream cell's contact counts onto cohort DNs and the contact slots at each site (vertical line, ratio 1). Circles, all pairs with shared input (number in parentheses). Diamonds compare connected pairs (filled) with their matched unconnected controls (open) over the same set of valid matches: 13,772 one-way matches (11,528 distinct controls) and 1,101 reciprocal matches (956). Controls were matched on voxel and regional overlap, shared-input cosine and count, input strengths, partner counts and side. A match is valid when both the connected pair and its control have retained identified cord output, which accounts for the reduction from 13,799 to 13,772 one-way and 1,102 to 1,101 reciprocal pairs. For each group, the ratio is the sum of observed shared sites divided by the sum of expected shared sites over the valid matches; a reused control contributes once per match. The matched connected ratios (1.42 and 1.60) are those of the full connected sets to two decimals. Sites, pairs and controls can be reused, so the ratios are descriptive.

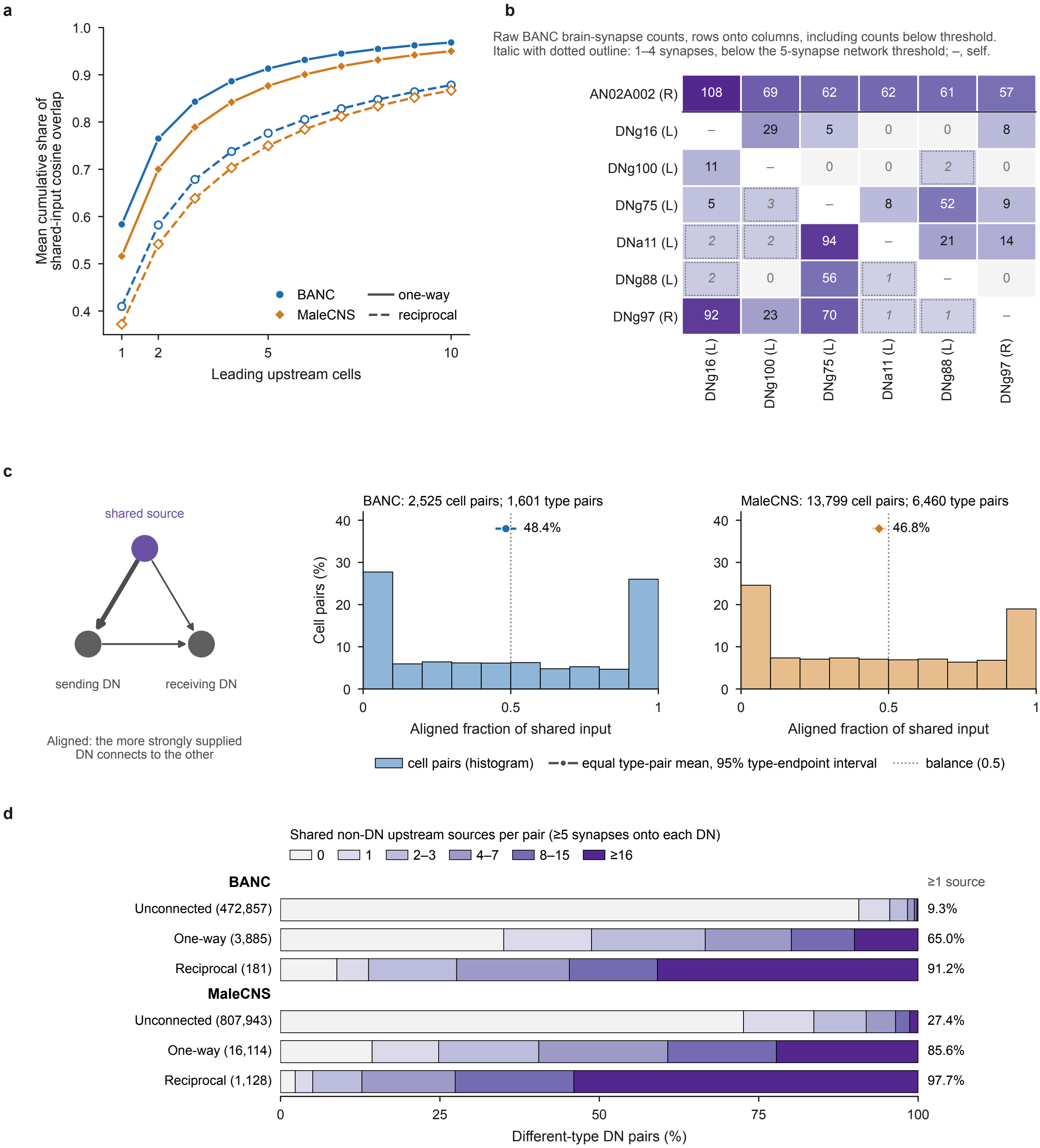

**Supplementary Figure 4. Individual upstream cells and connection direction.** **a**, Mean cumulative share of shared-input cosine overlap contributed by the leading upstream cells, for pairs with at least one shared upstream source: BANC 2,525 one-way and 165 reciprocal cell pairs; MaleCNS 13,799 and 1,102 (Supplementary Table 6). The one-way curves are also shown in main-text Fig. 3b. **b**, Brain synapses from one BANC AN02A002 cell to its six strongest different-type cohort-DN targets and among those DNs (rows onto columns; dashes omit self-connections; subthreshold counts included). L/R, annotated sides. **c**, For one-way connected pairs with shared input, the contribution-weighted fraction of shared-input overlap aligned with the connection, that is, from upstream cells that supply a larger fraction of all mapped non-DN input to the sending than to the receiving DN (pictogram; ties count one half). Bars, distribution over cell pairs; symbols and lines, equal type-pair mean with its 95% type-endpoint interval (1,601 BANC and 6,460 MaleCNS type pairs); dashed line, balance (Supplementary Table 7). **d**, Number of shared non-DN upstream sources (at least five synapses onto each DN) for all different-type DN pairs by connection class (at least five brain synapses in a direction): unconnected, 472,857 BANC and 807,943 MaleCNS pairs; one-way, 3,885 and 16,114; reciprocal, 181 and 1,128. Percentages give pairs with at least one shared source, as in Fig. 3a.

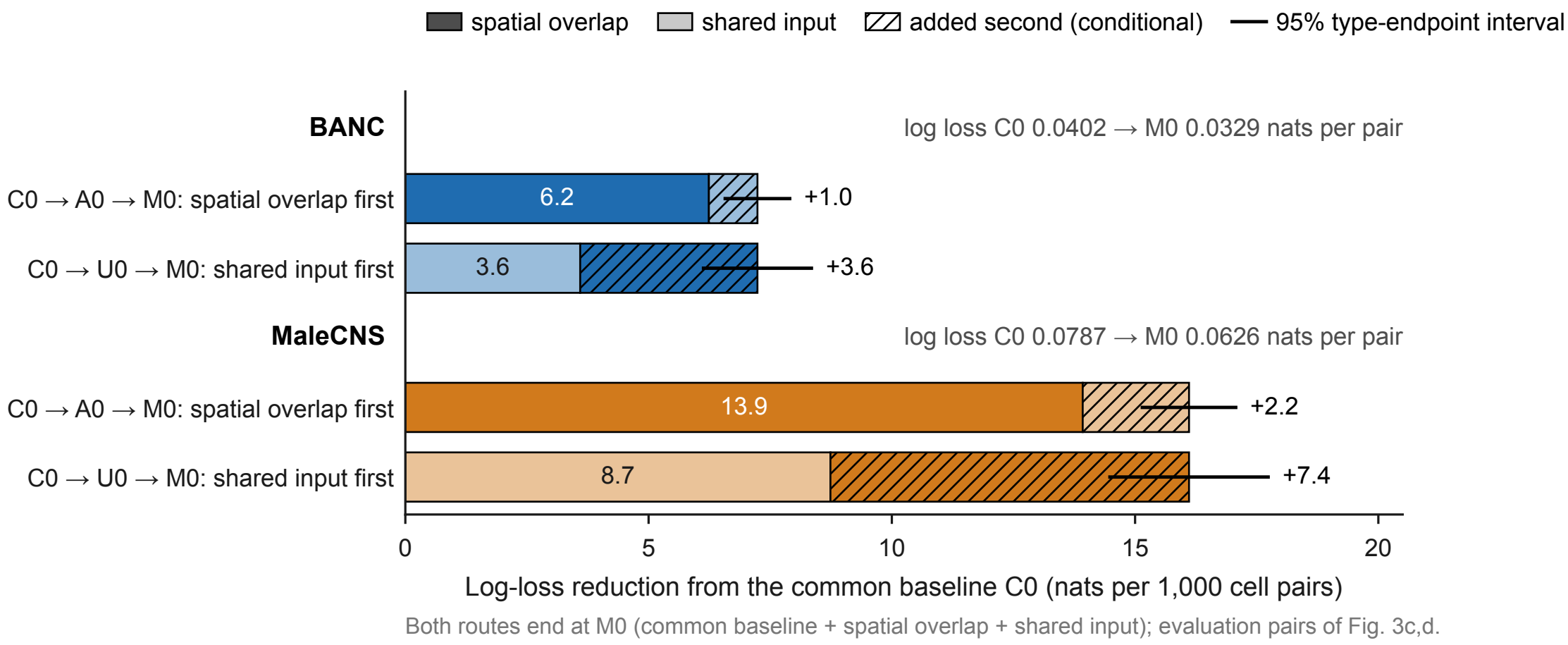

**Supplementary Figure 5. Conditional predictive contributions under two orders of feature addition.** Log-loss reduction from the common baseline when spatial overlap and shared input are added in either order, from C0 to M0, for the evaluation pairs of Fig. 3c,d. Solid segments add one description alone; hatched segments add the other conditionally (dark, spatial overlap; light, shared input); black lines, 95% type-endpoint intervals of the conditional increment. Numbers give reductions in nats per 1,000 cell pairs; the baseline and full-model log loss are shown for each specimen (Supplementary Tables 8–9).

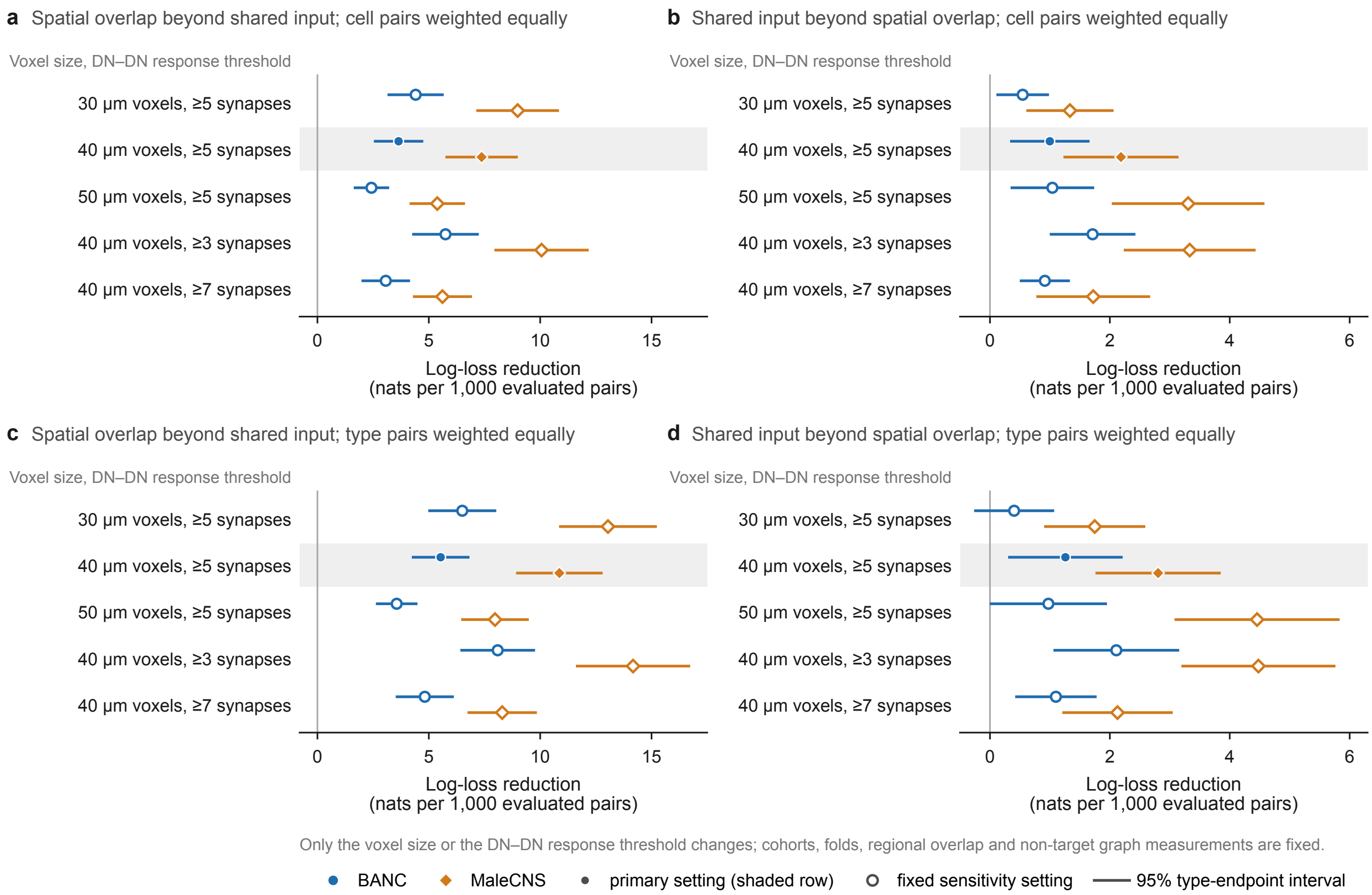

**Supplementary Figure 6. Sensitivity to voxel size and connection threshold.** **a**,**c**, Spatial overlap beyond shared input. **b**,**d**, Shared input beyond spatial overlap. Cell pairs (**a**,**b**) or type pairs (**c**,**d**) are weighted equally. Rows give voxel size and response threshold; the shaded row is the primary setting (filled symbols) and open symbols are the four fixed sensitivity settings. Log-loss reductions in nats per 1,000 pairs; bars, unadjusted 95% type-endpoint intervals. Blue circles, BANC; orange diamonds, MaleCNS. Cohorts, folds, regional overlap and non-target graph measurements are unchanged across settings. Response thresholds change only DN–DN labels, not source-edge retention or the three-site distance statistic (Supplementary Table 14).

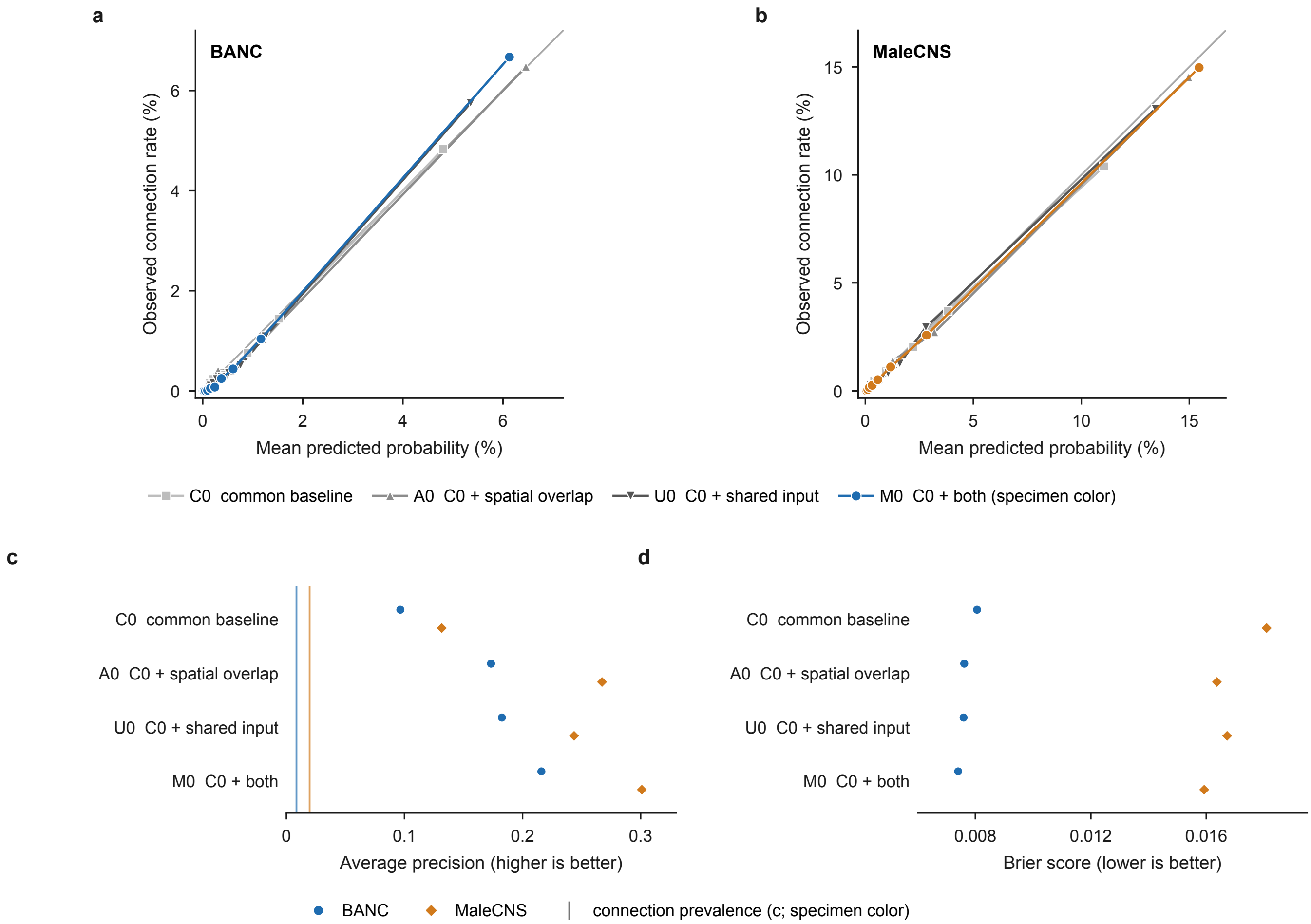

**Supplementary Figure 7. Calibration and ranking of the prediction models. a,b**, Mean predicted probability versus observed connection rate in ten prediction-quantile bins (ties kept together) for the common baseline C0 with coarse structure and breadth (squares) and after adding spatial overlap (triangles), shared input (inverted triangles) or both (circles, specimen color); grey line, equality. Bin membership differs between models. **c**, Average precision; vertical lines, connection prevalence. **d**, Brier score. Blue circles, BANC; orange diamonds, MaleCNS. Diagnostics use the saved predictions for 93,550 BANC and 162,638 MaleCNS cell pairs at natural prevalence, without recalibration (Supplementary Tables 15 and 16).

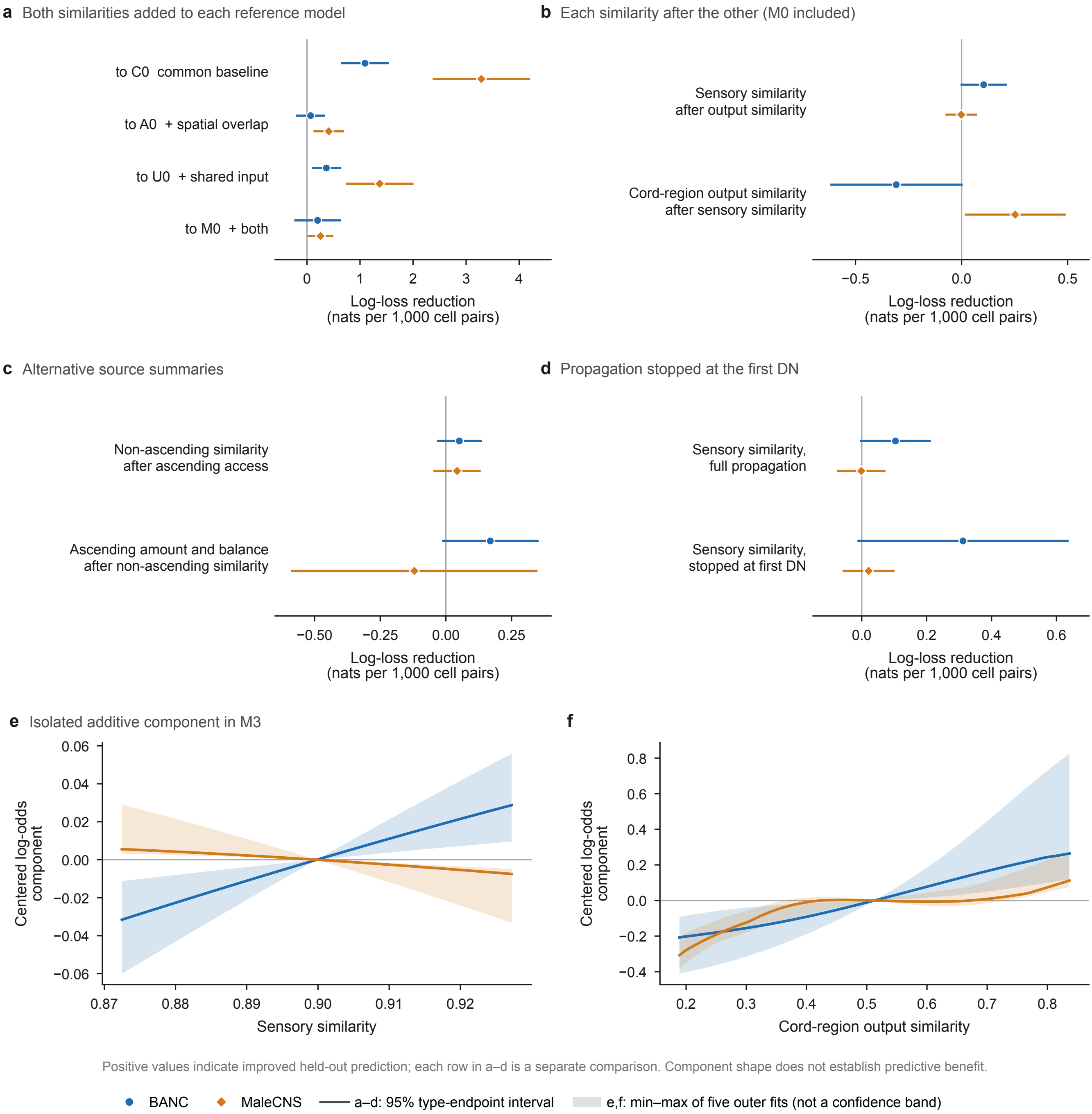

**Supplementary Figure 8. Coarse sensory and output similarity add limited information beyond local structure.** **a**, Held-out log-loss reduction after adding both similarities to the common baseline alone or with spatial overlap, shared input or both; each row is a separate comparison. **b**, Contribution of each similarity after the other, with both local descriptions included. **c**, Contributions of non-ascending source similarity after ascending access, and ascending amount and balance after non-ascending similarity. **d**, Sensory similarity after output similarity with propagation through the full graph or stopped at the first DN reached; both local descriptions are included. In **a**–**d**, reductions are in nats per 1,000 evaluated cell pairs and bars are unadjusted 95% type-endpoint intervals. Positive values indicate improved held-out prediction. **e**,**f**, Isolated additive components of sensory and output similarity in the model containing both local descriptions and both similarities, centered at the midpoint of the range supported in both specimens. Lines show the median of five outer fits and shading their minimum–maximum range, not a confidence interval. Component shape does not establish predictive benefit. Blue circles or lines denote BANC; orange diamonds or lines denote MaleCNS. The models are defined in Supplementary Note 3, with numerical contrasts in Supplementary Tables 9–10; source definitions and conditional-curve construction are in the Supplementary Methods.

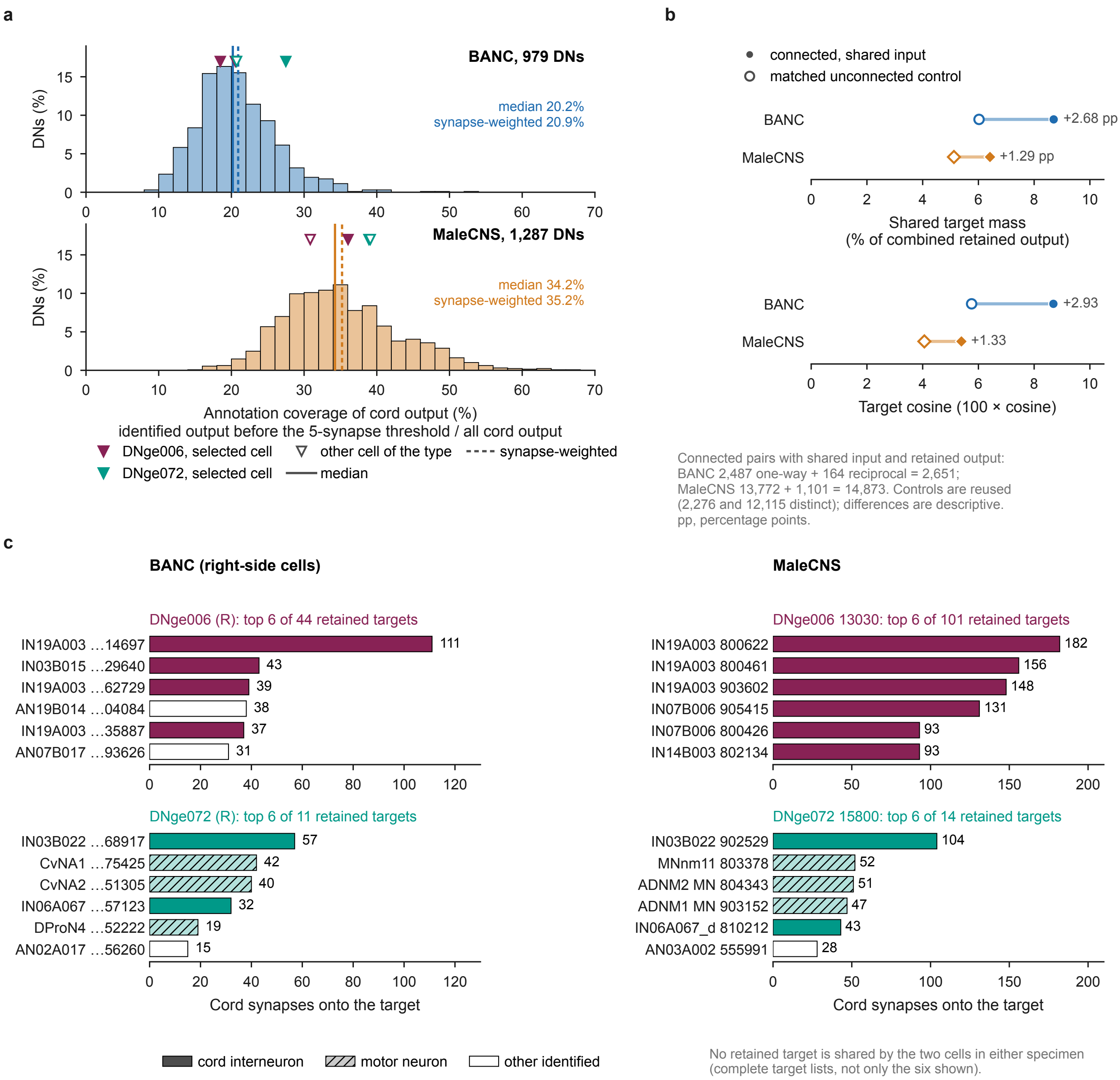

**Supplementary Figure 9. Output relationships and annotation coverage.** **a**, Distribution of prethreshold annotation coverage, the fraction of all named-cord output onto accepted identified neurons, for all 979 BANC and 1,287 MaleCNS cohort DNs. Solid and dashed lines show the cell median and synapse-weighted coverage. Filled and open triangles mark the selected DNge006/DNge072 cells and the other cell of each type, respectively. **b**, Connected pairs with shared input (filled) and their matched unconnected controls (open), requiring retained output from both members of both pairs: 2,651 BANC matches (2,487 one-way and 164 reciprocal) and 14,873 MaleCNS matches (13,772 one-way and 1,101 reciprocal). Shared-target mass is a percentage of combined retained output; the separate target-cosine axis is 100×cosine. Differences in mass are percentage points. Controls can be reused; these are descriptive comparisons (Supplementary Table 20). **c**, The six strongest retained identified targets of each selected cell, with type and short ID and target-class fill. Headers give each cell's full number of retained targets. The no-shared-target result uses the complete target sets, not just these six entries (main-text Fig. 4c; Supplementary Table 19).

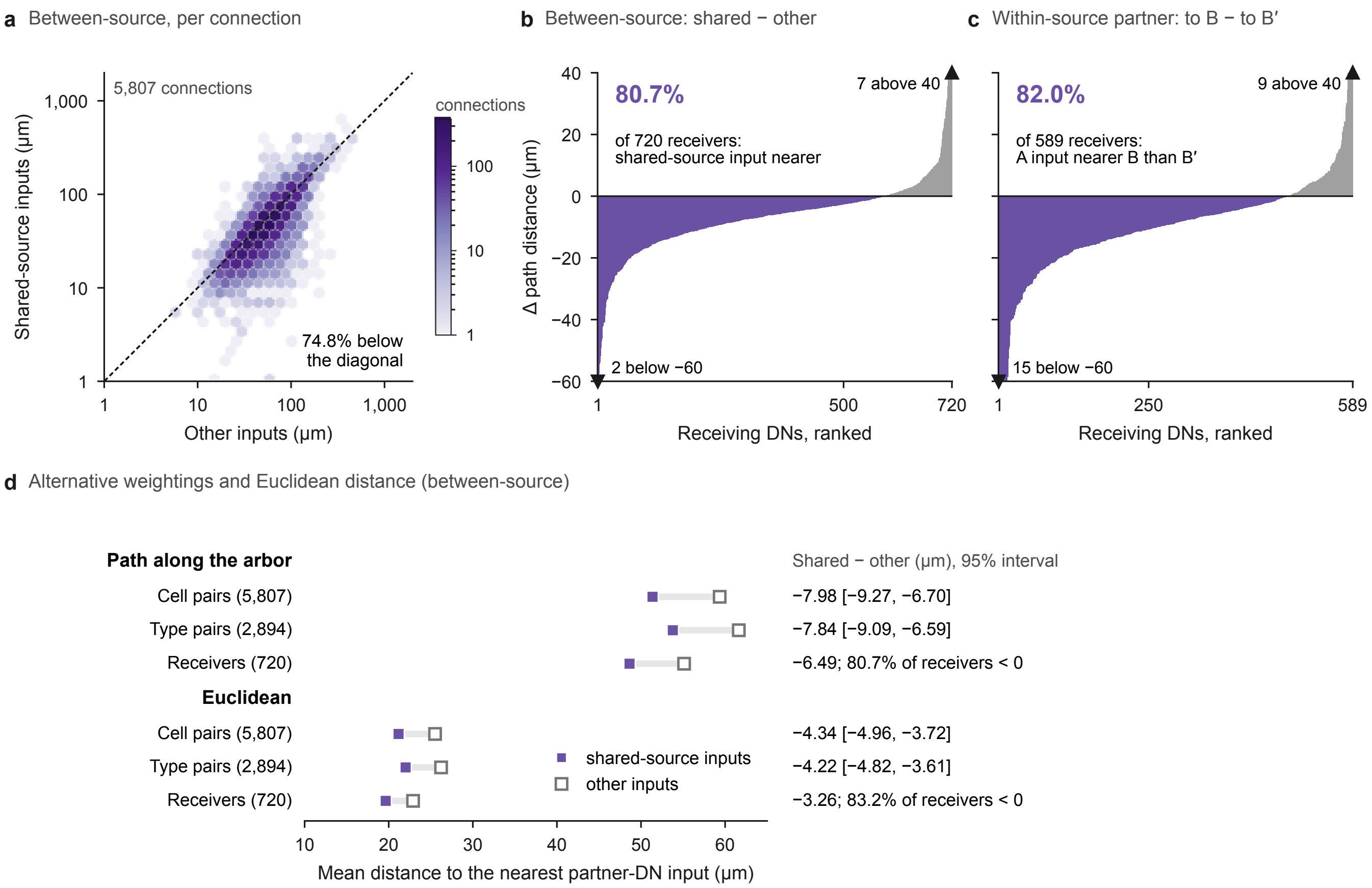

FlyWire/FAFB v783, an independent adult female brain: 1,210 typed DNs in 446 types. b,c, receiver-level values; purple, source-connected group nearer.
c, weighting source–receiver pairs equally gives −11.41 μm (19,246 pairs; descriptive). d, alternative summaries of the same retained records are not independent replications;
receiver averages are descriptive.

**Supplementary Figure 10. Independent FlyWire placement comparisons.** The endpoint cohort comprised 1,210 typed DNs in 446 types in an independently reconstructed adult female brain. **a**, Pair-level shared-source and other-input mean path distances for 5,807 one-way connections; hexagon color denotes connection count, and the dashed line equality. Shared-source inputs were nearer in 74.8% of pairs. **b**,**c**, Ranked receiver differences for the between-source (720 receivers) and within-source partner (589 receivers) comparisons. Purple indicates negative differences, grey nonnegative differences; triangles and counts mark observations outside the displayed range. Negative receiver fractions are 80.7% and 82.0%. **d**, Group means under cell-pair, type-pair and receiver weighting, using path and Euclidean distances on the same retained records. Numerical brackets are 95% type-endpoint intervals for the cell-pair and type-pair contrasts; receiver summaries are descriptive. Alternatives are summaries of the same observations, not independent replications. The primary within-source statistic weights receivers equally; equal source–receiver weighting gives $-11.41$ $\mu$m across 19,246 pairs. See Supplementary Tables 23–24.

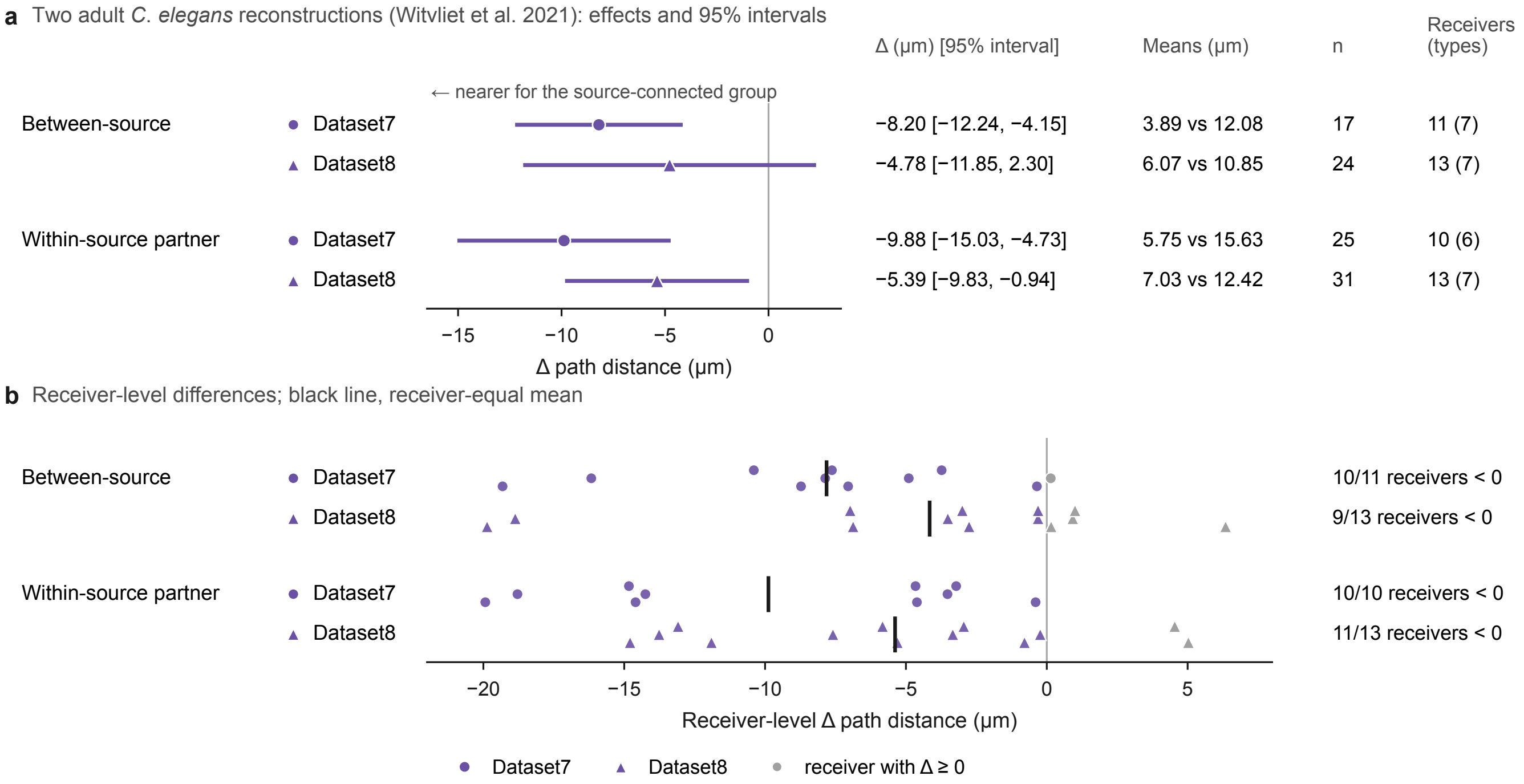

Population: 44 author-classified interneurons (24 classes) per animal; whole reconstructed brain arbor (no named regions); native postsynaptic attachments.
n: connections (between-source) or source–receiver pairs (within-source). Intervals are within-specimen asymptotic intervals (type-endpoint; crossed receiver-type and source clusters), with only 6 and 7 receiver-type clusters in the within-source comparison; finite-sample coverage is not established.
The between-source estimate weights connections equally, so it differs from the receiver-equal mean in b.

**Supplementary Figure 11. Exploratory placement comparisons in two adult C. elegans brains.** Each cohort contained 44 author-classified interneurons in 24 classes from the Witvliet reconstruction series (main-text reference 30). The spatial domain was the whole reconstructed brain arbor, and positions used native postsynaptic attachments. **a**, Differences, group means and 95% intervals for Dataset7 (circles) and Dataset8 (triangles). Between-source estimates weight directed connections equally and use type-endpoint intervals; within-source estimates weight receivers equally and use conditional intervals clustered by receiver type and source cell. The displayed $n$ counts directed connections or source–receiver pairs, respectively; receiver counts and type counts are listed separately. The Dataset8 between-source interval crosses zero. **b**, All receiver-level differences; black bars show receiver-equal means. Grey marks nonnegative values. Between-source means in **b** differ from the pair-equal estimates in **a**. Within-source comparisons involve only six and seven receiver-type clusters; finite-sample interval coverage is not established. Counts refer to observations within two animals, not independent biological replicates. See Supplementary Table 25.

## Supplementary Note 1. Distances along receiving neurites

Distances were measured on the official MaleCNS skeletons of 1,079 receiving DNs (main-text Methods). Supplementary Table 1 gives the population estimates under each weighting in cable and Euclidean distance (Supplementary Figure 1), and Supplementary Table 3 gives the distances for selected named receivers, including the GNG comparison of the DNge006-to-DNge072 example of main-text Fig. 2c,d (10.71 versus 18.96 $\mu$m) and the DNge019-to-DNg12_c case with the opposite ordering.

**Within-source partner comparison.** This comparison holds the non-DN source, receiver, actual source input records, exact side-specific endpoint region and connected skeleton component fixed. Sending partners with at least five source synapses are compared with partners with zero recorded source synapses in the unthresholded brain graph; counts of one to four are excluded. Every partner supplies at least three distinct physical reference sites. Analytic averaging over uniformly drawn three-site subsets standardizes the reference-set size of the distance statistic without changing actual synapse counts or introducing Monte Carlo variation. Methods gives the exact expectation and aggregation weights.

The analysis retains 922 of the 1,079 receivers in the between-source analysis; 157 have no source–region–component stratum meeting all criteria. The 68,609 eligible strata yield 49,742 source–receiver pairs from 8,453 sources. Receiver-equal means are 64.40 $\mu$m for partners also receiving source input and 73.54 $\mu$m for zero-link partners, a difference of $-9.14$ $\mu$m (conditional 95% interval $-11.21$ to $-7.07$ $\mu$m). Of 922 receiver averages, 749 (81.2%) are negative. Equal source–receiver weighting gives $-12.18$ $\mu$m; it is descriptive and is not the primary weighting. Supplementary Table 2 gives coverage and estimates. Partner-level calculations recorded source-edge counts, reference counts and expected distances, including the nearest-distance value without count standardization. Fine branch opportunity is not matched.

## Supplementary Note 2. Identified circuits and shared upstream cells

The cell-resolved analyses use the same cohorts as the prediction study. The source-sharing and source-contribution summaries included all different-type DN pairs, irrespective of fold assignment. Receiving-neurite and identified-target analyses used the eligible subsets listed in Supplementary Table 21. Every nonzero source contribution was ranked, including those for unconnected pairs and return counts below threshold.

In the named circuit, AN19B014 is the leading shared upstream cell of the selected DNge006 and DNge072 cells in both specimens, supplying all of the pair's shared-input cosine overlap in BANC and 81.7% in MaleCNS (Supplementary Table 4). MaleCNS annotations assign acetylcholine to AN19B014 and DNge006 and GABA to DNge072 (Supplementary Table 5); these type-level labels do not specify the sign of individual connections. Across the population, shared input was concentrated in few upstream cells, more strongly in one-way than in reciprocal pairs (Supplementary Figure 4a and Supplementary Table 6); Supplementary Figure 4b shows one broadly projecting BANC AN02A002 cell and the connections among its strongest DN targets. With type pairs weighted equally, alignment of shared input with connection direction was 48.4% in BANC and 46.8% in MaleCNS. The MaleCNS interval (45.1–48.5%) lay below 50%, indicating a small bias opposite to the stronger-to-weaker input orientation; neither specimen showed a majority in that orientation (Supplementary Figure 4c and Supplementary Table 7).

The full distributions of shared-source counts in unconnected, one-way and reciprocal pairs complement the presence/absence percentages in main-text Fig. 3a (Supplementary Figure 4d).

**Examples in population context.** The three example type pairs show how similar distributions of output across cord regions can accompany different combinations of shared brain territory, shared input and interconnection (Supplementary Figure 2). Supplementary Figure 2a places them among all 87,153 BANC and 111,628 MaleCNS unordered type pairs; in this descriptive map a type pair is connected if any cross-type cell pair meets the threshold, and type pairs outside the held-out evaluation are included. The examples were selected from BANC development data before their MaleCNS connections were inspected (main-text Methods).

DNge006 and DNge072 combined substantial spatial overlap and shared input with interconnection in both specimens (Supplementary Figure 2b,c). All four cross-type cell pairs in each specimen met the connection threshold. Mean voxel cosine overlap was 0.568 in BANC and 0.579 in MaleCNS; shared-input cosine was 0.114 and 0.144. In MaleCNS, 451 of 554 DN–DN synapse records lay in the GNG and another 100 in the left or right inferior posterior slope, locating the interconnection within the shared lower-brain territory. These are DN–DN synapses; the shared-input measurements summarize separate connections from upstream cells. The type-level totals include connections in both directions between cells other than the selected pair of main-text Fig. 2b, which had no recorded reverse connection.

DNp24 and DNp48 illustrate a different relationship (Supplementary Figure 2b,d). Their cord-output similarities were high, 0.942 in BANC and 0.816 in MaleCNS, and some BANC branches approached closely: the median sampled minimum separation was 0.444 $\mu$m. Yet their voxel overlap was lower (0.207 and 0.149), and shared-input overlap was zero in BANC and 0.0017 in MaleCNS. None of the four cross-type cell pairs in either specimen reached five synapses; MaleCNS contained one DNp48-to-DNp24 synapse record, below threshold. Similar output distributions and local proximity therefore coexisted with little shared input and no connection meeting the study threshold.

DNg97 and DNge046 show that a connected pair need not be high on both overlap measures (Supplementary Figure 2b,e). All six BANC and eight MaleCNS cross-type cell pairs were connected. Voxel overlap was high (0.581 and 0.556), whereas shared-input overlap was lower than for DNge006 and DNge072 (0.016 and 0.023). Of 688 MaleCNS interconnection records, 627 lay in the GNG. The two connected examples therefore combine spatial distribution and input convergence differently; complementarity across the population does not require both measures to be high for every connection. The measures have different definitions, so their magnitudes are not directly comparable.

Overlap values in Supplementary Table 12 average the cross-type cell pairs; the lower overlap of the unconnected example is descriptive, not matched on all spatial covariates. MaleCNS contact records may reuse a polyadic presynaptic site.

## Supplementary Note 3. Prediction models and conditional contributions

Supplementary Table 22 defines all feature blocks and model specifications. The core comparisons start from a common baseline (C0), add spatial overlap (A0) or shared input (U0), and combine both local descriptions (M0). Extensions assess coarse sensory and cord-region similarity, non-ascending versus ascending source descriptions, and first-arrival access. Each family has five outer fits. Mean loss uses the evaluated cell pairs at natural prevalence; type-pair weighting gives equal weight to each distinct type pair without refitting.

The two routes from the common baseline (C0) to the model with both local descriptions (M0) are shown in Supplementary Figure 5; main-text Fig. 3d shows their conditional increments. Supplementary Tables 8–10 list every model error and contrast.

**Local contributions within output groups.** DN types were grouped by their dominant observed cord compartment; these exploratory groups were not used for fitting. Adding both local descriptions reduced loss in all 14 BANC and all 24 MaleCNS group pairs shown (Supplementary Table 11). Individual groups share types and are not independent replications.

## Supplementary Note 4. Neighboring settings and prediction diagnostics

Four sensitivity settings were specified after the primary analysis and before they were fitted: 30- and 50-$\mu$m voxels at the five-synapse threshold, and three- and seven-synapse thresholds at 40 $\mu$m (Supplementary Figure 6). Only the two conditional contributions were evaluated. Scale changes alter voxel overlap while retaining regional overlap; threshold changes alter response labels while keeping non-target inputs fixed. Each setting changes one factor and reuses the same cohorts and complete-type folds; none replaces the primary specification. Every point estimate is positive, but the BANC shared-input contribution intervals include zero under equal type-pair weighting at both alternative spatial scales (Supplementary Table 14). These intervals are conditional on the graphs and fitted predictions.

The model with both local descriptions has a mean predicted connection probability of 0.008864 in BANC and 0.020752 in MaleCNS, compared with observed rates of 0.008530 and 0.019682, so overall predictions are modestly higher than observed (Supplementary Tables 15 and 16). Quantile calibration shows local departures around the identity line and does not establish perfect calibration (Supplementary Figure 7). The highest decile spans a broad probability range. Cell-pair and type-pair weights use the same predictions and are not independent replications.

## Supplementary Note 5. Coarse sensory and output similarity

DNs with similar sensory-access and cord-output profiles were more likely to connect, but much of this relationship overlapped with the local descriptions. Sensory access summarized multisynaptic paths from eight annotated source groups, including ascending neurons, after removal of DN–DN edges; cord output described the distribution of synapses across seven compartments (Supplementary Methods). Unlike shared upstream sources, a source profile can be similar even when different individual neurons supply the inputs. Sensory and output similarities were moderately related across type pairs (Spearman correlations 0.239 in BANC and 0.207 in MaleCNS).

Pairs above the development-set medians for both similarities connected more often than pairs below both medians: 1.857% versus 0.173% in BANC and 4.607% versus 0.280% in MaleCNS. Their excess connection rate over baseline predictions was substantially smaller once spatial overlap and shared input were included. Correspondingly, the benefit of adding both similarities fell from 1.094 to 0.200 nats per 1,000 pairs in BANC and from 3.287 to 0.256 in MaleCNS when both local descriptions were available (Supplementary Figure 8a). Adding either local description alone also reduced that benefit. The reductions were not strictly ordered: in BANC, the remaining benefit after spatial overlap alone was 0.069, smaller than after both local descriptions. These comparisons indicate overlapping predictive information across structural levels, without a simple additive partition.

With both local descriptions and output similarity included, sensory similarity reduced loss by 0.104 in BANC (95% interval, −0.005 to 0.213) and by −0.001 in MaleCNS (−0.076 to 0.073), giving no additional benefit supported in both specimens (Supplementary Figure 8b). The output contribution after sensory similarity also differed: it was negative in BANC (−0.308) and positive in MaleCNS (0.253), with the MaleCNS interval crossing zero under equal type-pair weighting (Supplementary Tables 9–10). The isolated similarity components are shown in Supplementary Figure 8e,f.

**Ascending sources and first arrival.** The eight-channel source composition includes ascending neurons, whose median share of propagated source mass is 0.270 in BANC and 0.239 in MaleCNS. We therefore removed only the ascending source column, leaving the graph unchanged, and separately modeled ascending access amount and balance. Non-ascending similarity after the ascending block reduces loss by 0.051 [−0.034, 0.136] in BANC and 0.042 [−0.048, 0.132] in MaleCNS; ascending access after non-ascending similarity contributes 0.169 [−0.015, 0.353] and −0.120 [−0.589, 0.349] (Supplementary Figure 8c). Neither contrast identifies a consistent additional source relationship across specimens. The ascending block measures the amount of a single channel and cannot distinguish different feedback pathways.

Deleting direct DN links still permits paths through a DN, an intermediate neuron and another DN. A first-arrival calculation stops propagation whenever any annotated DN is reached. Median retained source mass is 66.8% in BANC and 62.9% in MaleCNS, so subsequent DN visits affect the structural propagation measure. After output similarity and the full controls, first-arrival sensory similarity reduces loss by 0.312 [−0.013, 0.638] in BANC and 0.021 [−0.059, 0.101] in MaleCNS (Supplementary

Figure 8d). The BANC estimate remains positive after stopping indirect DN return, but its interval includes zero; neither calculation identifies a residual sensory contribution shared by both specimens. These mass ratios are properties of the discounted operator, not measured fractions of physiological feedback.

**Supporting direction and interaction models.** The exact variable sets are given in the model key before Supplementary Table 17. These supporting models retain local anatomical and connectivity covariates (block $L$ for presence and reciprocity; antisymmetric local predictors for direction) but omit the voxel/regional-overlap block $Q_A$ and the source-resolved shared-input block $Q_U$. For presence and reciprocity, the sensory and output blocks each contain similarity and breadth; direction uses signed profile and breadth contrasts as defined in the model key. The nonlinear interaction adds a tensor product of sensory and output spline bases after separate additive terms. The linear direction models use antisymmetric predictors, no intercept and scaling without centering, so exchanging endpoints complements the probability. Direction is evaluated only in singly connected pairs (780 BANC; 2,947 MaleCNS), and reciprocity conditional on connection, with 18 and 254 reciprocal pairs. These comparisons do not establish a common broad-to-narrow output hierarchy (Supplementary Table 17).

After coefficient inspection, an exploratory comparison tested narrower-to-broader sensory orientation with the local model. Observed fractions were 48.72% in BANC and 47.91% in MaleCNS, versus local-model expectations of 44.58% and 44.10%. These conditional residuals do not imply a majority direction. The comparison is post hoc; 94 MaleCNS output-entropy ties were omitted only from the separate output-breadth comparison.

## Supplementary Note 6. Identified cord targets

Prethreshold annotation coverage across all cohort DNs is shown in Supplementary Figure 9a; the denominator includes accepted and unresolved output targets.

Output analyses also retain counts over all raw target segments and at a one-synapse threshold for annotated neurons; unknown segments are not relabeled as neurons. Cell-class discrepancies between annotations, including motor versus efferent labels, are retained and do not establish functional equivalence across specimens. Direct DN-to-motor edges and two-edge routes through identified cord interneurons were tabulated separately: there were 3,899 BANC and 8,047 MaleCNS retained direct DN-to-motor edges, and 70,917 and 197,923 DN–motor-neuron pairs reachable through at least one identified interneuron. Reachability measures neither path gain nor recruited motor activity.

Supplementary Table 18 lists the output of every DNge006 and DNge072 cell summarized in main-text Fig. 4. For one-way pairs with shared input, the median share of output directed to shared identified targets was about 1% in both specimens (main-text Fig. 4c and Supplementary Table 19), and connected pairs had slightly greater overlap in retained identified output than matched unconnected controls (Supplementary Figure 9b and Supplementary Table 20).

## Supplementary Note 7. Independent placement comparisons in FlyWire

The female FlyWire/FAFB v783 brain provides a separate animal and reconstruction from MaleCNS. The endpoint cohort contained 1,210 DNs in 446 exact types, with matching skeletons. All 1,303 annotated DNs, including 93 without an exact type, were excluded from upstream sources. The comparison retained the five-contact threshold, the distinction between subthreshold source links in the between-source comparator and zero recorded links in the within-source comparator, and the expected nearest distance to three distinct reference sites. Brain-only data supplied no cord-output inclusion criterion or downstream replication.

Across 5,807 eligible one-way pairs, mean path distances were 51.37 $\mu$m for shared-source inputs and 59.36 $\mu$m for other inputs, a difference of $-7.98$ $\mu$m (95% type-endpoint interval, $-9.27$ to $-6.70$ $\mu$m). Of the pairs, 4,344 (74.8%) had a negative difference. Averaging within each receiver gave negative differences in 581 of 720 receivers (80.7%). Equal type-pair weighting and Euclidean distances on the same retained records gave the same direction (Supplementary Figure 10 and Table 23). Receiver-equal between-source summaries are descriptive and are distinct from the pair-equal primary estimate. The eligible records spanned 49 released endpoint regions; this count does not establish the direction in every region.

The within-source partner comparison retained 589 receivers in 293 exact types, 4,140 source cells and 19,246 source–receiver pairs, yielding 23,497 region–component strata. Receiver-equal means were 62.79 and 72.93 $\mu$m, a difference of $-10.15$ $\mu$m (conditional 95% interval, $-12.15$ to $-8.14$ $\mu$m); 483 of 589 receiver differences were negative (82.0%). Equal source–receiver weighting instead gave $-11.41$ $\mu$m (descriptive). Detailed coverage and alternative summaries are in Supplementary Table 24. The independently reconstructed fly brains support the same direction in both comparisons. Effect magnitudes are not directly equated across comparisons because their eligibility rules, reference statistics and weights differ.

## Supplementary Note 8. Exploratory extensions to other neuronal populations

**Adult C. elegans interneurons.** We analyzed both adult specimens (Dataset7 and Dataset8, 50 h) in the Witvliet developmental reconstruction series (main-text reference 30). Each had 44 author-classified interneurons in 24 classes in the typed endpoint cohort. All author-classified interneurons were excluded as sources, including any not eligible as endpoints. This is an analogous comparison in a different neuronal population, not an asserted homology to fly DNs or a reconstruction of an entire descending system. Native postsynaptic attachments and the whole reconstructed brain arbor supplied the spatial domain.

Dataset7 retained 17 between-source connections and 11 receivers. Its pair-equal difference was $-8.20$ $\mu$m (95% interval, $-12.24$ to $-4.15$ $\mu$m), with 10 of 11 receiver averages negative. Dataset8 retained 24 connections and 13 receivers, with difference $-4.78$ $\mu$m (95% interval, $-11.85$ to 2.30 $\mu$m); 9 of 13 receiver averages were negative. The Dataset8 interval includes zero and is retained in full.

The within-source differences were $-9.88$ $\mu$m in Dataset7 (95% interval, $-15.03$ to $-4.73$ $\mu$m; 25 source–receiver pairs and 10 receivers) and $-5.39$ $\mu$m in Dataset8 ($-9.83$ to $-0.94$ $\mu$m; 31 pairs and 13 receivers). Receiver differences were negative in 10 of 10 and 11 of 13 cells. There were only six and seven receiver-type clusters, respectively, so normal-interval finite-sample coverage is not established. All four estimates and all receiver values are shown in Supplementary Figure 11; Supplementary Table 25 gives the counts, means and intervals. These are within-animal conditional comparisons, not interval estimates from a population of independent animals. Previous clustering analyses included this developmental series (main-text reference 19); the present extension tests the specified source–partner relationship with fixed source positions and standardized reference-set size.

**Platynereis head-to-trunk projections.** The three-day larval publication archive (main-text reference 31) contained 138 author-annotated head-to-trunk projection cells. The typed endpoint cohort contained 99 cells in 35 types, and all 138 were excluded from upstream sources. Of 19 one-way candidate connections, four met the between-source criteria. All four shared one MN3 partner cell and contacted two MN1 and two MN2 receivers. Pair differences were +3.34, +18.51, $-4.60$ and +7.09 $\mu$m; the mean was +6.09 $\mu$m. Thus the retained pairs did not show a consistent nearer-shared-input ordering. The within-source comparison retained one receiver and two sources, with a descriptive difference of $-7.61$ $\mu$m. The between-source type dyads had no disjoint endpoints, and the within-source sample had one receiver-type cluster, making both interval estimates non-estimable. Supplementary Table 26 reports the actual directions and means without inferential bounds. These sparse observations are neither a positive replication nor evidence for an opposite population-wide relationship.

## Supplementary Methods

### Conventions

BANC, female brain-and-nerve-cord connectome; MaleCNS, male central nervous system connectome; DN, descending neuron; GNG, gnathal ganglion. Log-loss reductions are given in nats per 1,000 cell pairs unless stated otherwise, and positive values indicate better held-out prediction. The within-source partner comparison uses the crossed source/receiver-type interval defined below. Otherwise, intervals are unadjusted 95% type-endpoint intervals: they account for dependence among pairs that share a DN type within one connectome, not for variation between animals. Blue circles denote BANC and orange diamonds MaleCNS in population panels; wine and teal denote the first and second DN of a named pair, and purple denotes shared upstream input.

### Cohort flow and units of observation

Supplementary Table 21 distinguishes specimen, neuron, cell-pair, type-pair, source–receiver and repeated stratum counts. BANC cohort selection began with 986 DNs meeting the anatomical and annotated-target cord criteria. Seven had no external brain partners after removal of all DN–DN edges at the five-synapse threshold, leaving 979 DNs. All seven met the 50-contact cord criterion when all target segments were counted. The seven-region output profiles included all recorded target segments, but cohort membership remained defined by the annotated-target criterion. The BANC and MaleCNS cell-resolved analyses used these anatomically and cord-output-selected cohorts rather than all annotated DNs. FlyWire used its brain-only DN cohort without a cord-output filter; the other-species cohorts are defined below. A further 47 typed BANC DNs with valid brain skeletons and retained brain input failed the annotated-target cord criterion. Ten typed MaleCNS DNs with external brain records and finite geometry failed the named-cord count criterion; source access and skeleton availability were not evaluated further for those excluded cells.

### Record identity and receiving-skeleton exclusions

MaleCNS input records retain `body_pre`, `body_post`, the original integer pre/post coordinate fields and `primary_post`. Source records were not deduplicated. Physical reference sites for the within-source comparison were deduplicated on `x_post,y_post,z_post` within a fixed partner, receiver, region and component, without coordinate rounding; the uniform three-site expectation was applied to those sites. In the between-source nearest-distance calculation, repeated reference vertices have no effect on the minimum. Neither operation converts repeated source records into distinct physiological trials.

The 1,079 skeletons contained one component in 917 receivers, two in 114, three in 34, four in nine, five in two, and eight, eleven or twelve in one receiver each. They supplied 4,013,925 input records before reuse across partner comparisons. Across 26,295 tested pair–region rows, there were 38,507,366 input-record appearances, including pair reuse; 19,634 (0.0510%) were unreachable from the partner reference set. Seven regional rows and five directed pairs then failed eligibility. The count 19,634 is not a count of unique synapses. No components were joined, no SWC edges were resampled and no projection cutoff was applied. The median of receiver-specific median projection errors was 0.578 $\mu$m. Across receivers, the median, 95th percentile and maximum of the within-receiver 95th-percentile error were 1.327, 1.544 and 2.924 $\mu$m; the largest individual error was 8.286 $\mu$m.

Region labels retain the released side suffix during eligibility and distance calculations. In regional displays only, left/right labels are merged after distances are obtained; `CentralBrain-unspecified` is displayed as Unassigned. It is a retained central-brain label, not an imputed named region. The within-source analysis starts from the same 1,079 receivers but a different local qualification rule: at least three source records and three distinct partner sites within a component, plus both source-to-partner connection classes. It therefore contains 15,110 distinct partner–receiver edges, whereas the between-source calculation requires shared and other source groups within each fixed pair and retains 13,318 edges.

### Source annotation and cord-region mappings

The eight source columns preserve the canonical annotation mapping used to construct the graph. In BANC, sensory sources required `super_class=sensory`; `cell_function` beginning with visual mapped to visual, olfactory to olfactory, tactile/auditory/proprioception/unknown_mechanosensory to mechanosensory, and exact gustatory, hygrosensory or thermosensory labels to the corresponding groups. Visual-projection sources used that canonical class; ascending included ascending and sensory_ascending. In MaleCNS, sensory identity followed the canonical mapping of `cb_sensory` and `ol_sensory`, and modality used the released `class`; visual-projection and ascending columns used their canonical classes. Blank or unmatched labels did not become sensory seeds.

Cord regions were grouped before any identified-target filtering. BANC front, middle and hind leg comprised ProNM-T1/LNp_T1, MesoNM-T2/LNp_T2 and MetaNM-T3/LNp_T3, respectively; abdomen comprised ABDNM/ANm; upper tectulum comprised WTct, HTct, NTct and the HTct_UTct_T3, NTct_UTct_T1 and WTct_UTct_T2 labels; intermediate tectulum was IntTct; other named cord comprised AMNP, AMNp, LTct, mVAC, mVAC_T3 and Tct. MaleCNS leg groups used prefixes LegNp(T1), LegNp(T2) and LegNp(T3); abdomen was ANm; upper tectulum used WTct/HTct/NTct prefixes; intermediate was IntTct; other named cord comprised LTct and mVAC-prefixed regional labels. Nonmatching records were excluded from the seven-region profile. Counts used all recorded output to those regions, without a five-contact target threshold or target-annotation requirement. The BANC all-target cord counts and MaleCNS outgoing partner rows, rather than the Fig. 4 target matrix, supply these profiles.

### Nested fitting implementation and feature handling

Exact types are the `type` labels in the cohort tables. Left/right cells sharing a type label remain in the same group; no laterality suffix is added or removed for cross-validation. BANC five-fold GroupKFold assignments were defined on the initial 986-cell cohort and retained after the seven exclusions. MaleCNS uses five-fold GroupKFold on its 1,287-cell cohort. No folds were redrawn to balance connection outcomes. Inner GroupKFold uses three groups among outer-training cells; pair training excludes either validation type, and validation includes only pairs with both types in that inner fold. The 40,000-pair cap is sampled before inner splitting, with the same label-independent sample reused across model families.

Every scalar feature in the primary and source-variant model families listed in Supplementary Table 22 enters the scikit-learn `SplineTransformer` with `n_knots=4`, `degree=3`, `include_bias=False` and `extrapolation='constant'`. The four equally spaced knots include the observed training minimum and maximum. Binary and constant predictors use the same transformer, without a separate outcome-dependent selection step. Spline columns are standardized by training mean and standard deviation using StandardScaler; constant columns receive scale one. A separate intercept is appended. Knots and scaling are refitted within every inner partition and outer training set. Fixed graph construction, full-cohort centroid-spread normalization and external test-cell anatomy are unsupervised inputs; they are not estimated from the held-out response labels. The supporting direction, reciprocity and interaction models of Supplementary Table 17 are specified separately in the model key that precedes that table.

L-BFGS-B minimizes mean logistic loss plus a squared-coefficient penalty, leaving the intercept unpenalized, with maximum 500 iterations, `ftol=1e-10` and `gtol=1e-6`. An unsuccessful optimization raises an exception before a model is saved. For each outer fold, the selected penalty, inner-validation losses for the five candidate penalties, and train/test counts were recorded. The five candidate losses were stored in the order $(10^{-4}, 10^{-3}, 10^{-2}, 10^{-1}, 1)$, each pooled over validation pairs from the three inner folds. Saved model files contain coefficients, scalers, splines, feature names and training medians.

### Selection of named examples and display rules

The initial BANC candidate pool required both type names in MaleCNS, at least two BANC cells per type and an outer-fold index below three. Type-level source/output profiles were cell means. Candidates required the same dominant cord group (front leg, abdomen or upper tectulum) and both source and output similarities at or above their development-pool medians. Within each cord group, candidates were sorted by connected BANC cell-pair fraction (descending), total unthresholded DN–DN synapse count in both directions (descending) and then the two type names in lexical order (first type, then second type, ascending); the first row of each group was taken. This yielded DNg97–DNge046 (front leg; fraction 1.0, 226 synapses), DNge006–DNge072 (upper tectulum; 1.0, 205) and DNp24–DNp48 (abdomen). All four abdomen candidates had fraction 0 and no DN–DN synapses, so the abdomen example was decided by type name alone. Candidate rankings and selected cell pairs are included in the numerical outputs. The DNge019–DNg12_c example was the first row after sorting BANC one-way pairs with both types present in MaleCNS and a named leading source by shared-input cosine (descending). None of these selections used receiving-neurite distances. Within each example and specimen, cell pairs were restricted to one-way connected pairs when any existed and then sorted by shared-input cosine and spatial (voxel) cosine, both descending; the first pair was displayed. Examples illustrate these selection rules rather than an unbiased sample of circuits.

Anatomical displays use native-coordinate orthographic projections; display cropping does not clip the graphs used for distance calculation. For the MaleCNS external-site displays in Supplementary Figure 2, the original rows within each type are retained at stride $\max(1, \lfloor n/8000 \rfloor)$, where $n$ is its record count. DN–DN synaptic records are not subsampled. The shared-source input enlarged in Supplementary Figure 1d is the GNG record at the median path distance, supplied by AN07B052 (9.9 $\mu$m path and 7.0 $\mu$m straight-line distance). AN19B014 is separately ringed. The mean 10.7 $\mu$m in Fig. 2 pools all 12 shared sources. The 260 named-circuit records and the 554 type-pair DN–DN records include different cells and edges and are not interchangeable. Figure 3c bins voxel cosine in ten intervals with boundaries 0, 0.1, . . . , 1; value 1 enters the last bin. Nonzero shared-input strata are divided at the median among the evaluated cell pairs in each specimen (approximately 0.021 BANC and 0.0068 MaleCNS), not the full cohort or training set. Bins with fewer than 100 evaluated pairs are not displayed.

### Accepted targets, denominators and matching implementation

MaleCNS target acceptance required `status=Traced` and a nonempty superclass other than glia or not_a_neuron. BANC required `proofread=TRUE` or `roughly_proofread=TRUE`, excluding empty, glia, not_a_neuron and trachea classes. Motor endpoints were vnc_motor in MaleCNS or motor with `region=ventral_nerve_cord` in BANC. Intermediates were vnc_intrinsic or ventral_nerve_cord_intrinsic, respectively, under the same acceptance rules. Both edges of the two-step routes used the same named-cord regional scope and five-contact threshold. Unknown or unaccepted segments remain only in the all-contact denominator.

Matching used voxel cosine, regional cosine, shared-input cosine, log(1 + shared-source count), and sums/absolute differences of log(1 + all non-DN input count) and log(1 + retained upstream degree). Standard deviations were calculated over all different-type pairs within each specimen; a zero standard deviation was replaced by one. No covariate imputation was used. Translation by a mean would not alter distances; the implementation divides by standard deviation without centering. Euclidean nearest neighbors were found separately for the two same-side-status strata, using SciPy cKDTree with one neighbor. Exact-distance ties followed the nearest-neighbor selection of SciPy cKDTree, without an additional tie-breaking rule. Selected control identifiers and matching distances are saved with the numerical outputs. Controls can be reused across connected pairs and connection classes.

Output availability was not used to choose the nearest control. It was applied afterward, requiring retained identified output for both members of both pairs, with nonmissing outcome measurements. In MaleCNS this excluded 27 of 13,799 one-way matches: ten connected pairs and 23 controls lacked output, with six matches satisfying both exclusion reasons. One of 1,102 reciprocal matches lacked control output. The same output-availability screen was used for presynaptic-site ratios, even though presynaptic-site geometry itself does not require cord output. Retained one-way and reciprocal match sets reused 11,528 and 956 distinct controls. Exclusions, distances and covariate differences were summarized separately by specimen and connection class; standardized differences used the same all-pair standard deviations as matching. These summaries document the measured comparison without identifying a causal effect of interconnection.

### Neurotransmitter annotations

Official MaleCNS v1.0 body-neurotransmitter predictions were joined to cell annotations by exact body ID. Both tables were obtained from the public MaleCNS release identified in the main-text Data availability statement. Body prediction and score, type aggregate and score, consensus and ground-truth fields were retained. AN19B014, DNge006 and DNge072 each had two annotated cells. Scores are annotation-model outputs, not physiological confidence intervals; MaleCNS labels were not assigned to BANC cells. Supplementary Table 5 distinguishes body-level from type-level evidence.

### Regional summary weights

For Supplementary Figure 1b,c, left and right endpoint regions were merged and eligible pair–region contrasts were averaged equally. Regional groups with fewer than 100 comparisons were omitted. The Euclidean and path summaries use the same retained input records and weights. Main-text Fig. 2d is an empirical cumulative distribution over the 242 shared-source and 723 other non-DN GNG inputs of the selected receiver; 13 other-input distances exceed the displayed 60-$\mu$m range.

### Crossed-cluster interval for the within-source partner comparison

Let $m_j$ be the number of retained source cells for receiver $j$, $J$ the number of receivers, and $\delta_{uj}$ the source–receiver distance difference after averaging across region–component strata. The normalized weight is $a_{uj} = 1/(Jm_j)$, preserving equal weight for each receiver, and the residual score is $z_{uj} = a_{uj}(\delta_{uj} - \widehat{\mu})$. If $Z_t$, $Z_u$ and $Z_{tu}$ sum these scores by exact receiver type, source cell and their intersection, respectively, the signed variance is

$$\widehat{V} = \sum_t Z_t^2 + \sum_u Z_u^2 - \sum_{t,u} Z_{tu}^2.$$

The reported interval is $\widehat{\mu} \pm 1.96\sqrt{\widehat{V}}$. The variance estimate was positive in MaleCNS, FlyWire and both C. elegans adults; the sparse Platynereis estimate was not reported as an interval (see the structural conditions below). It accounts for these two cluster dimensions conditional on the observed graph and reference sets; it does not include every possible network dependence or variation between animals. Equal source–receiver weighting is reported separately and does not replace the primary receiver-equal estimate.

### Individual source contributions and directional input asymmetry

For retained non-DN brain synapse counts $w_{ui}$ and $w_{uj}$, the contribution of source $u$ to shared-input cosine is

$$c_u(i,j) = \frac{w_{ui}w_{uj}}{\sqrt{\sum_v w_{vi}^2}\sqrt{\sum_v w_{vj}^2}}, \qquad Q_U(i,j) = \sum_u c_u(i,j).$$

All nonzero terms were ranked for every different-type pair. Dividing by $Q_U$ gives fractions of cosine overlap for pairs with shared input. These fractions describe a source's contribution to overlap, not its fraction of all receiver synapses. Summaries included leading-cell fractions, the numbers of sources reaching 50% and 80% of overlap, inverse concentration and each source's DN-target breadth. Summed normalized contributions provide participation rankings, not degree-adjusted enrichment (Supplementary Note 2 and Figure 4a,b).

For a one-way connection $i \to j$, source-input fractions were $f_{ui} = w_{ui}/I_i$ and $f_{uj} = w_{uj}/I_j$, where $I_i$ included all mapped non-DN brain input before thresholding. The contribution-weighted mean of the indicator $f_{ui} > f_{uj}$, with ties assigned one half, quantified alignment from the more strongly to the less strongly supplied DN (Supplementary Figure 4c). Raw-count versions and signed fractional differences were retained separately. Both DN directions and return connections to upstream cells retained unthresholded counts. For leading sources of the selected examples and AN02A002/AN07B004 cells, immediate incoming and outgoing regional connections across brain and cord, with partner annotations, were extracted from the raw data.

### Fixed neighboring settings and prediction diagnostics

Four sensitivity settings were specified after the primary analysis and before they were fitted: 30- and 50-$\mu$m occupancy voxels at the five-synapse response threshold, and response thresholds of three and seven synapses at 40 $\mu$m. Each setting changes one factor. Scale changes re-bin the same BANC segment midpoints and MaleCNS external partner records at the same native-coordinate origin; regional overlap remains fixed. Threshold changes relabel presence from saved unthresholded DN–DN counts. Cohorts, non-target graph threshold, exclusion of the full DN block, source and output breadth, regional overlap, upstream overlap and coarse controls remain fixed. These are sensitivities to the spatial binning and response definition, not complete graph-threshold reconstructions or tests of simultaneous changes. Among the same 93,550 BANC pairs, the three-, five- and seven-synapse definitions yield 1,294, 798 and 589 positives; among 162,638 MaleCNS pairs they yield 4,669, 3,201 and 2,451. Loss increments compare models within a setting; response-threshold changes also change prevalence.

Only $A_0 = L + Q_A + H$, $U_0 = L + Q_U + H$ and $M_0 = L + Q_A + Q_U + H$ are compared, for spatial overlap after shared input and shared input after spatial overlap. The primary outer and inner type folds, training-sample rule, spline basis and penalty grid are retained. Scale changes reuse $U_0$, whose inputs and labels are unchanged; the other comparisons use 100 additional outer fits in total. The main models are the primary 40-$\mu$m, five-synapse fits. All four settings and both evaluation weights are reported, without selecting a new primary setting. These checks use the same specimens and held-out pairs as the primary analysis.

For the four primary descriptions $C_0$, $A_0$, $U_0$ and $M_0$, saved predictions provide log loss, Brier score, area under the receiver operating characteristic curve and average precision at natural prevalence. Type-pair-equal diagnostics assign each physical pair the reciprocal of its type pair's size. Calibration uses ten prediction-quantile bins per model and specimen, keeping tied probabilities together, and reports their sizes, positive counts, mean predictions and observed rates; fewer bins are allowed if ties collapse boundaries. No recalibration or classification-threshold optimization is performed. These are descriptive properties of the held-out predictions, not independent replications or between-animal estimates.

### Source decomposition and first arrival

Let $W_{ij}$ be the retained brain synapse count from presynaptic neuron $i$ to postsynaptic neuron $j$, after discarding edges with fewer than five synapses, self-edges and all annotated DN–DN edges. We normalized outgoing weights to form $P_{ij} = W_{ij}/\sum_k W_{ik}$; a row with no outgoing weight was set to zero. The propagation operator was $A = \alpha P^{\mathsf{T}}$, with discount $\alpha = 0.85$, so column $i$ sends source mass from neuron $i$ to its postsynaptic partners. The source matrix $B$ had one column for each annotated group: visual, visual projection, olfactory, mechanosensory, gustatory, hygrosensory, thermosensory and ascending. For group $g$ containing $n_g$ neurons, $B_{ig} = 1/n_g$ for its members and zero otherwise; an empty group contributed a zero column. Thus each nonempty group started with unit total mass, irrespective of its size.

The accumulated source access was

$$F = (I - A)^{-1}B = \sum_{\ell=0}^{\infty} A^{\ell}B.$$

Rows of $F$ correspond to neurons and columns to source groups. For each reached DN, its source composition was its row of $F$ divided by that row's sum. The implementation accumulated the series until the largest, across source columns, ratio of the absolute newly added mass to accumulated mass was below $10^{-11}$, with a maximum of 2,000 iterations. These are discounted structural-access profiles, not estimated firing rates or transition probabilities for physiological activity.

Non-ascending access is the first seven columns of $F$, renormalized per DN. Removing the ascending source leaves $P$ and every other source column unchanged. Eight BANC DNs and one MaleCNS DN have zero remaining access. They remain in the cohort; every non-ascending base contains the reached-endpoint count and a both-reached indicator. Similarity is encoded as zero if either endpoint is unreached, conditional on these availability controls, rather than interpreting an undefined composition as a probability vector.

The source-family base includes $L + Q_A + Q_U$, non-ascending and output entropies, output similarity and the availability indicators. Extensions add non-ascending similarity, an ascending block, or both. The ascending block consists of sums and absolute differences of ascending mass share and of $\log(1 + 10^4 F_{d,\text{asc}})$. It measures the amount and balance of one source channel, not similarity between separately identified ascending pathways. Non-ascending access is not labeled pure external sensation, because indirect processing and visual projection sources remain.

For first arrival, let $D$ be diagonal with value zero for every annotated DN and one for other neurons. Set $A_{\text{stop}} = AD$ and compute $F_{\text{stop}} = (I - A_{\text{stop}})^{-1}B$. Zeroing DN outgoing columns stops propagation after the first DN visit, without renormalizing any surviving transition. The first-arrival comparison replaces sensory entropy and similarity while retaining all other predictors and the output composition. It compares $L + Q_A + Q_U + H_{\text{stop}} + O$ with that base plus $S_{\text{stop}}$.

### Interval calculations and conditional curves

For dyadic uncertainty, let $e_d = \delta_d - \bar{\delta}$, let $S_t$ sum residuals over pairs incident on type $t$, and let $U_{tu}$ sum residuals within unordered type pair $(t, u)$. The signed variance is $(\sum_t S_t^2 - \sum_{t<u} U_{tu}^2)/N^2$. Type-pair weighting applies it after averaging within each unordered type pair. This accommodates shared endpoints conditional on these graphs; it does not estimate between-animal variation. Matching with reused controls introduces additional endpoint dependence, so matched comparisons are reported descriptively without this interval formula.

The signed variance is retained before taking a square root. A negative value within $32\epsilon(\sum_t S_t^2 + \sum_{t<u} U_{tu}^2)/N^2$ is labeled numerical roundoff; a more negative value invalidates the interval and yields missing bounds. No materially negative estimate is converted to a zero-width interval. All reported intervals had nonnegative variance estimates.

Conditional plots (Supplementary Figure 8e,f) show isolated additive spline components from $M_3$, centered at the midpoint of a common similarity range. For each similarity, we form quartiles of the other similarity and of its own entropy sum, take the within-stratum 5th–95th percentile ranges, and intersect these across both specimens. The resulting display ranges are 0.872464–0.927214 for sensory and 0.188992–0.836549 for output similarity. Other variables are held at training medians; because the model is additive, their constant contributions cancel when centering. Lines show the median of five outer fits and shading their minimum–maximum range, not a confidence band or causal effect. The overlap rule constrains two important covariates; it does not guarantee common support in every anatomical dimension.

### Presynaptic allocation reference

The allocation reference of Supplementary Figure 3c holds fixed each upstream cell's retained contact count onto cohort DNs, $N$, its contacts $n_i$ and $n_j$ with the two DNs, and each presynaptic site's number of contact slots $k$. The probability that both DNs occur at a site is $1 - p_0(n_i) - p_0(n_j) + p_0(n_i + n_j)$, where $p_0(n) = \prod_{t=0}^{k-1}(N - n - t)/(N - t)$, with impossible choices assigned probability zero. Summing across sites and shared upstream cells gives the reference count. The reference preserves contact counts but neither finer geometry nor the number of distinct target cells at a site. Observed/reference totals were 307,653/215,983 for one-way pairs, 105,558/66,044 for reciprocal pairs and 542,907/514,952 for unconnected pairs with shared input. Over the same valid matches, connected pairs and their matched controls gave 306,999/215,654 and 264,427/190,727 for one-way pairs (13,772 matches) and 105,481/65,977 and 83,389/54,435 for reciprocal pairs (1,101 matches); a reused control contributes once per match in which it appears. Pairs and sites are reused in these totals; the ratios are descriptive and are not independent-trial significance tests.

### Public inputs and dataset-specific placement adapters

**FlyWire.** The v783 connectivity archive supplied `flywire_synapses_783.feather`, `proofread_connections_783.feather` and `proofread_root_ids_783.npy`. The matching skeleton release supplied `sk_lod1_783_healed_ds2.parquet`. The official v2 annotation table was frozen at commit `327ae7bdcc18c7a6a5ac014a4bfa9c7560fe854e`. Exact root identifiers were retained as integers. The adapter extracted 1,889,753 nonself contact records from proofread neurons onto cohort DNs and reconciled their counts against the released graph. Individual coordinates and skeletons used native nanometres, scaled once by 0.001 to micrometres. Postsynaptic coordinates were assigned to the nearest receiving vertex using SciPy cKDTree, without an additional tie-breaking rule. The provider had already healed and downsampled the skeletons; the analysis added no edges or projection-distance cutoff. The maximum projection distance was 9.93 $\mu$m. Comparisons preserved released side-specific neuropil labels. Records with an unassigned label were retained as a separate domain but yielded no eligible between-source pair. No retained pair required a path across disconnected components. The between-source analysis first required at least three shared-source and three other input records within a region, then reapplied these counts after reachability filtering. The within-source analysis additionally required at least three source records, three distinct physical sites for each partner, and both partner classes in the same region and skeleton component. These rules are the same as in MaleCNS.

**C. elegans.** Source files were `Dataset7_skeletons.json`, `Dataset7_synapses.json` and their Dataset8 counterparts, together with the author's neuron classification, from commit `0646af9d25896ae660f97d462eab2d67282f5625` of `dwitvliet/nature2021`. Each native chemical active-zone record was expanded into its individual postsynaptic attachments. Different attachments onto the same cell remained different contacts; neither the number of EM sections nor synapse volume was used as multiplicity. The native `post_tid` was joined directly to the receiving skeleton. Neuron names remained available alongside reversible integer indices. Muscle, glial and unidentified targets or sources were not relabeled as neurons. The full author interneuron class was excluded from sources. Whole reconstructed brain arbors formed one spatial domain, with original parent edges and disconnected components preserved. Distances were in micrometres after multiplying native nanometres by 0.001. Between-source source classes used the five-contact graph; within-source partners required at least five source contacts or zero recorded contacts, with one-to-four-contact comparators excluded. The same three-record and distinct-three-reference-site rules and aggregation weights applied. Both adult specimens were included before their effect directions were inspected.

**Platynereis.** Public CATMAID project 30 (April 2024 publication archive), stack 21, supplied annotations and compact arbors. The head-to-trunk cohort was the intersection of the author's `episphere` and `head_trunk` annotations. Type names were taken from the publication repository at commit `ab75029040b8dba768d4c2cbdb26423aec59ae4a`. Sources required the `connectome_neuron` annotation and could not belong to the full 138-cell projection population. The 164 source segments lacking required neuronal annotation supplied 336 excluded records; annotation absence does not establish that they were non-neuronal. Chemical contacts used native connector-to-postsynaptic-treenode records. The receiving treenode, not the connector centroid, located each contact on an unchanged parent graph. Native nanometres were scaled by 0.001, and the whole arbor was one domain. The five-contact graph and three-site statistic were unchanged. Counts outside the acquired projection-target set were not used as evidence of zero source links.

**Inference and reproducibility.**

The additional datasets use the same type-endpoint and crossed-cluster variance formulas as the primary placement analysis. Before an interval is reported, the between-source type-dyad set must contain two pairs with no shared type endpoint; the within-source data must have at least two receiver-type and two source-cell clusters. The retained signed variance must also be finite and positive above numerical cancellation tolerance. Otherwise the effect remains descriptive and the interval is marked non-estimable. These are necessary structural conditions, not a claim of accurate normal coverage with few clusters. Thresholds, source classes and spatial domains were fixed before estimating the additional effects, without preregistration. Network prediction and cord-output analyses were not extended to these datasets.

## Supplementary Tables

**Supplementary Table 1. Distances of shared-source and other inputs to partner-DN inputs.** MaleCNS, micrometers (Fig. 1 and Supplementary Figure 1). Shared-source and other inputs are compared within the same receiver and endpoint brain region on identical retained records. Difference, shared minus other, with 95% type-endpoint intervals; receiver means are descriptive. $n$ counts directed cell pairs, unordered type pairs or receiving neurons according to the weighting row. Differences were calculated before rounding the two displayed means.

| Measure | Weight | $n$ | Shared | Other | Difference | 95% interval |
|---|---|---|---|---|---|---|
| Neurite path | Cell pair | 13318 | 60.77 | 70.18 | −9.42 | [−10.52, −8.31] |
| Neurite path | Type pair | 6260 | 64.22 | 73.76 | −9.54 | [−10.70, −8.39] |
| Neurite path | Receiver | 1079 | 54.51 | 62.77 | −8.26 | – |
| Euclidean | Cell pair | 13318 | 25.05 | 30.08 | −5.02 | [−5.55, −4.50] |
| Euclidean | Type pair | 6260 | 26.32 | 31.38 | −5.06 | [−5.57, −4.55] |
| Euclidean | Receiver | 1079 | 23.18 | 27.46 | −4.28 | – |

**Supplementary Table 2. Partner comparison holding source and receiver fixed.** Primary means weight sources equally within receivers and receivers equally. Shared-class partners receive at least five brain synapses from the source; zero-link partners receive none recorded. Distances are the exact expected nearest path distance to three distinct reference sites sampled without replacement. The interval is conditional on the graph and reference sets, clustered by receiver type and source cell (main-text Methods). The 1,140,253 partner-level rows and 3,461,863 combinations are reused comparisons, not independent samples. Edge eligibility differs from the between-source comparison.

| Quantity | Value |
|---|---|
| Receiving DNs in the between-source analysis | 1,079 |
| Eligible receivers | 922 |
| Receivers without an eligible stratum | 157 |
| Eligible receiver types | 415 |
| Unique non-DN source cells | 8,453 |
| Source–receiver pairs | 49,742 |
| Source–receiver–region–component strata | 68,609 |
| Partner-level source–receiver–region–component rows | 1,140,253 |
| Distinct partner–receiver edges | 15,110 |
| Cross-class partner combinations (reused) | 3,461,863 |
| Standardized reference sites per partner | 3 |
| Mean distance: shared-class partner | 64.40 $\mu$m |
| Mean distance: zero-link partner | 73.54 $\mu$m |
| Receiver-equal difference | −9.14 $\mu$m |
| Conditional 95% interval | [−11.21, −7.07] $\mu$m |
| Receivers with negative mean | 749/922 (81.2%) |
| Source–receiver pairs with negative difference | 75.1% |
| Equal source–receiver difference (descriptive) | −12.18 $\mu$m |

**Supplementary Table 3. Cable distances in selected named receivers.** MaleCNS path lengths (micrometers) within the same postsynaptic region; $n$, input records. No selected comparison had unreachable projected vertices.

| Directed pair | Region | Shared $n$ | Other $n$ | Shared distance | Other distance |
|---|---|---|---|---|---|
| DNge006 to DNge072 | IPS(L) | 20 | 81 | 26.57 | 35.42 |
| DNge006 to DNge072 | GNG | 242 | 723 | 10.71 | 18.96 |
| DNge006 to DNge072 | Unassigned | 9 | 15 | 2.60 | 17.83 |
| DNge046 to DNg97 | SAD | 41 | 729 | 63.92 | 82.63 |
| DNge046 to DNg97 | GNG | 290 | 3805 | 38.89 | 74.39 |
| DNge019 to DNg12_c | GNG | 163 | 144 | 42.12 | 37.79 |

**Supplementary Table 4. Leading shared upstream cell of each example cell pair.** Synapse counts onto the first and second DN follow the order of the pair name; overlap share is the upstream cell's share of the pair's shared-input cosine overlap. The DNge006/DNge072 rows are the selected cells of Fig. 2. Fractions describe cosine overlap, not all input synapses.

| Specimen | DN pair | Upstream type | To first DN | To second DN | Overlap share |
|---|---|---|---|---|---|
| BANC | DNge006 / DNge072 | AN19B014 | 41 | 26 | 1.000 |
| BANC | DNg97 / DNge046 | AN02A002 | 57 | 20 | 0.497 |
| BANC | DNge019 / DNg12_c | AN19B015 | 36 | 25 | 0.551 |
| MaleCNS | DNge006 / DNge072 | AN19B014 | 60 | 120 | 0.817 |
| MaleCNS | DNg97 / DNge046 | AN02A002 | 148 | 41 | 0.555 |
| MaleCNS | DNp24 / DNp48 | OA-VPM4 | 9 | 6 | 0.545 |
| MaleCNS | DNge019 / DNg12_c | GNG404 | 27 | 69 | 0.352 |

**Supplementary Table 5. Neurotransmitter annotations of the named-circuit cells in MaleCNS.** Official MaleCNS v1.0 annotations. ACh, acetylcholine; GABA, $\gamma$-aminobutyric acid. Scores are annotation-model scores, not measures of physiological effect; the type score pools the two cells of a type and is not an independent prediction. Labels are not transferred to BANC.

| Cell | Body ID | Consensus | Body score (%) | Type score (%) |
|---|---|---|---|---|
| DNge006 R | 10668 | ACh | 95.86 | 95.81 |
| DNge006 L | 13030 | ACh | 95.77 | 95.81 |
| DNge072 R | 14887 | GABA | 86.54 | 86.16 |
| DNge072 L | 15800 | GABA | 85.77 | 86.16 |
| AN19B014 L | 21743 | ACh | 96.82 | 96.44 |
| AN19B014 R | 1050379364 | ACh | 96.05 | 96.44 |

**Supplementary Table 6. Concentration of shared input among upstream cells.** Cell pairs with at least one shared upstream source; the one-way rows correspond to main-text Fig. 3b. Median cells, median number of shared upstream cells; top 1 median, median share of shared-input cosine overlap from the leading cell; top 5 mean, mean share from the five leading cells; median $n_{80}$, median minimum number of leading cells reaching 80% of the overlap. For fewer than five shared sources, top-five share includes every shared source and equals one.

| Specimen | Connection | Pairs | Median cells | Top 1 median | Top 5 mean | Median $n_{80}$ |
|---|---|---|---|---|---|---|
| BANC | Unconnected | 43961 | 1 | 1.000 | 0.985 | 1 |
| BANC | One-way | 2525 | 4 | 0.544 | 0.913 | 2 |
| BANC | Reciprocal | 165 | 12 | 0.369 | 0.776 | 5 |
| MaleCNS | Unconnected | 221273 | 2 | 0.769 | 0.970 | 2 |
| MaleCNS | One-way | 13799 | 7 | 0.464 | 0.877 | 3 |
| MaleCNS | Reciprocal | 1102 | 18 | 0.318 | 0.750 | 6 |

**Supplementary Table 7. Shared-input asymmetry relative to connection direction.** One-way connected pairs with shared input (Supplementary Figure 4c). Orientation estimates give the contribution-weighted fraction of shared input aligned with the connection (0.5, balance), defined by input fractions or synapse counts; signed input fraction is the contribution-weighted difference in input fraction between the sending and receiving DN (0, balance). Intervals are 95% type-endpoint intervals.

| Specimen | Measure | Weight | Estimate | 95% interval | Pairs |
|---|---|---|---|---|---|
| BANC | Fraction-based orientation | Cell pair | 0.4851 | [0.4524, 0.5179] | 2525 |
| BANC | Fraction-based orientation | Type pair | 0.4835 | [0.4521, 0.5149] | 1601 |
| BANC | Count-based orientation | Cell pair | 0.4662 | [0.4349, 0.4975] | 2525 |
| BANC | Count-based orientation | Type pair | 0.4616 | [0.4343, 0.4889] | 1601 |
| BANC | Signed input fraction | Cell pair | −0.0007 | [−0.0045, 0.0030] | 2525 |
| BANC | Signed input fraction | Type pair | −0.0010 | [−0.0042, 0.0021] | 1601 |
| MaleCNS | Fraction-based orientation | Cell pair | 0.4701 | [0.4503, 0.4899] | 13799 |
| MaleCNS | Fraction-based orientation | Type pair | 0.4679 | [0.4506, 0.4852] | 6460 |

| Specimen | Measure | Weight | Estimate | 95% interval | Pairs |
|---|---|---|---|---|---|
| MaleCNS | Count-based orientation | Cell pair | 0.4596 | [0.4437, 0.4755] | 13799 |
| MaleCNS | Count-based orientation | Type pair | 0.4629 | [0.4487, 0.4770] | 6460 |
| MaleCNS | Signed input fraction | Cell pair | −0.0017 | [−0.0033, −0.0002] | 13799 |
| MaleCNS | Signed input fraction | Type pair | −0.0017 | [−0.0030, −0.0004] | 6460 |

**Supplementary Table 8. Held-out errors of all prediction models.** Model codes and feature blocks are defined in Supplementary Table 22. Evaluation includes 93,550 BANC cell pairs in 17,265 type pairs and 162,638 MaleCNS cell pairs in 22,137 type pairs. Loss is natural-log binary cross-entropy with cell pairs or type pairs weighted equally.

| Specimen | Model | Cell-pair loss | Type-pair loss |
|---|---|---|---|
| BANC | C0 | 0.040169 | 0.060261 |
| BANC | A0 | 0.033930 | 0.051075 |
| BANC | U0 | 0.036575 | 0.055348 |
| BANC | M0 | 0.032932 | 0.049817 |
| BANC | C1 | 0.039075 | 0.058613 |
| BANC | M1 | 0.032424 | 0.049275 |
| BANC | M2 | 0.032836 | 0.049788 |
| BANC | M3 | 0.032732 | 0.049698 |
| BANC | A1 | 0.033861 | 0.050863 |
| BANC | U1 | 0.036208 | 0.054729 |
| BANC | E0 | 0.033124 | 0.050180 |
| BANC | E1 | 0.033074 | 0.050175 |
| BANC | E2 | 0.032956 | 0.049939 |
| BANC | E3 | 0.032905 | 0.049934 |
| BANC | T0 | 0.033073 | 0.050062 |
| BANC | T1 | 0.032760 | 0.049739 |
| MaleCNS | C0 | 0.078685 | 0.110965 |
| MaleCNS | A0 | 0.064762 | 0.090793 |
| MaleCNS | U0 | 0.069948 | 0.098849 |
| MaleCNS | M0 | 0.062577 | 0.087988 |
| MaleCNS | C1 | 0.075398 | 0.106434 |
| MaleCNS | M1 | 0.062573 | 0.087938 |
| MaleCNS | M2 | 0.062319 | 0.087755 |
| MaleCNS | M3 | 0.062321 | 0.087714 |
| MaleCNS | A1 | 0.064351 | 0.090292 |
| MaleCNS | U1 | 0.068577 | 0.097068 |
| MaleCNS | E0 | 0.062373 | 0.088071 |
| MaleCNS | E1 | 0.062198 | 0.087939 |
| MaleCNS | E2 | 0.062360 | 0.088220 |
| MaleCNS | E3 | 0.062318 | 0.088249 |
| MaleCNS | T0 | 0.062148 | 0.087847 |
| MaleCNS | T1 | 0.062127 | 0.087786 |

**Supplementary Table 9. Model contrasts with cell pairs weighted equally.** $\Delta$ is the reduction in held-out log loss from the first to the second model in nats per pair; estimates and interval bounds are multiplied by 1,000 for display, and positive values favor the second model. Intervals are 95% type-endpoint intervals. M3 to E1 and M3 to T1 change the access definition and are not nested additions. All pooled signed variances were nonnegative.

| Specimen | Comparison | $\Delta \times 1{,}000$ | 95% interval |
|---|---|---|---|
| BANC | A0 to M0 | 0.998 | [0.336, 1.661] |
| BANC | U0 to M0 | 3.644 | [2.535, 4.753] |
| BANC | C0 to A0 | 6.239 | [4.421, 8.058] |
| BANC | C0 to U0 | 3.594 | [2.514, 4.674] |
| BANC | C0 to M0 | 7.238 | [5.434, 9.042] |
| BANC | C0 to C1 | 1.094 | [0.640, 1.548] |

| Specimen | Comparison | $\Delta \times 1{,}000$ | 95% interval |
|---|---|---|---|
| BANC | M0 to M1 | 0.508 | [0.028, 0.987] |
| BANC | M0 to M2 | 0.096 | [−0.292, 0.484] |
| BANC | M0 to M3 | 0.200 | [−0.238, 0.638] |
| BANC | M2 to M3 | 0.104 | [−0.005, 0.213] |
| BANC | M1 to M3 | −0.308 | [−0.621, 0.005] |
| BANC | C1 to M3 | 6.344 | [4.586, 8.101] |
| BANC | A0 to A1 | 0.069 | [−0.204, 0.341] |
| BANC | U0 to U1 | 0.367 | [0.088, 0.646] |
| BANC | E0 to E1 | 0.050 | [−0.035, 0.135] |
| BANC | E0 to E2 | 0.168 | [−0.017, 0.353] |
| BANC | E0 to E3 | 0.219 | [0.013, 0.425] |
| BANC | E2 to E3 | 0.051 | [−0.034, 0.136] |
| BANC | E1 to E3 | 0.169 | [−0.015, 0.353] |
| BANC | T0 to T1 | 0.312 | [−0.013, 0.638] |
| BANC | M3 to E1 | −0.342 | [−0.757, 0.073] |
| BANC | M3 to T1 | −0.029 | [−0.208, 0.150] |
| MaleCNS | A0 to M0 | 2.185 | [1.226, 3.145] |
| MaleCNS | U0 to M0 | 7.371 | [5.745, 8.997] |
| MaleCNS | C0 to A0 | 13.923 | [11.041, 16.806] |
| MaleCNS | C0 to U0 | 8.738 | [6.197, 11.279] |
| MaleCNS | C0 to M0 | 16.109 | [12.697, 19.521] |
| MaleCNS | C0 to C1 | 3.287 | [2.370, 4.205] |
| MaleCNS | M0 to M1 | 0.003 | [−0.071, 0.077] |
| MaleCNS | M0 to M2 | 0.257 | [0.019, 0.496] |
| MaleCNS | M0 to M3 | 0.256 | [0.013, 0.499] |
| MaleCNS | M2 to M3 | −0.001 | [−0.076, 0.073] |
| MaleCNS | M1 to M3 | 0.253 | [0.015, 0.491] |
| MaleCNS | C1 to M3 | 13.077 | [10.012, 16.142] |
| MaleCNS | A0 to A1 | 0.411 | [0.122, 0.699] |
| MaleCNS | U0 to U1 | 1.371 | [0.734, 2.008] |
| MaleCNS | E0 to E1 | 0.176 | [−0.034, 0.385] |
| MaleCNS | E0 to E2 | 0.013 | [−0.443, 0.469] |
| MaleCNS | E0 to E3 | 0.055 | [−0.401, 0.512] |
| MaleCNS | E2 to E3 | 0.042 | [−0.048, 0.132] |
| MaleCNS | E1 to E3 | −0.120 | [−0.589, 0.349] |
| MaleCNS | T0 to T1 | 0.021 | [−0.059, 0.101] |
| MaleCNS | M3 to E1 | 0.123 | [−0.255, 0.501] |
| MaleCNS | M3 to T1 | 0.193 | [−0.288, 0.675] |

**Supplementary Table 10. Model contrasts with type pairs weighted equally.** As Supplementary Table 9, with each unordered type pair weighted equally, using the same saved predictions without refitting.

| Specimen | Comparison | $\Delta \times 1{,}000$ | 95% interval |
|---|---|---|---|
| BANC | A0 to M0 | 1.258 | [0.303, 2.214] |
| BANC | U0 to M0 | 5.531 | [4.237, 6.826] |
| BANC | C0 to A0 | 9.186 | [7.166, 11.206] |
| BANC | C0 to U0 | 4.913 | [3.421, 6.405] |
| BANC | C0 to M0 | 10.444 | [8.340, 12.549] |
| BANC | C0 to C1 | 1.649 | [0.934, 2.363] |
| BANC | M0 to M1 | 0.542 | [0.085, 0.999] |
| BANC | M0 to M2 | 0.028 | [−0.495, 0.552] |
| BANC | M0 to M3 | 0.119 | [−0.434, 0.672] |
| BANC | M2 to M3 | 0.091 | [−0.050, 0.231] |
| BANC | M1 to M3 | −0.423 | [−0.788, −0.058] |
| BANC | C1 to M3 | 8.915 | [6.873, 10.957] |
| BANC | A0 to A1 | 0.212 | [−0.194, 0.618] |
| BANC | U0 to U1 | 0.618 | [0.210, 1.027] |
| BANC | E0 to E1 | 0.005 | [−0.151, 0.160] |

| Specimen | Comparison | $\Delta \times 1{,}000$ | 95% interval |
|---|---|---|---|
| BANC | E0 to E2 | 0.241 | [−0.106, 0.588] |
| BANC | E0 to E3 | 0.246 | [−0.111, 0.603] |
| BANC | E2 to E3 | 0.004 | [−0.150, 0.159] |
| BANC | E1 to E3 | 0.241 | [−0.105, 0.587] |
| BANC | T0 to T1 | 0.323 | [−0.043, 0.689] |
| BANC | M3 to E1 | −0.478 | [−0.912, −0.043] |
| BANC | M3 to T1 | −0.042 | [−0.276, 0.193] |
| MaleCNS | A0 to M0 | 2.805 | [1.760, 3.850] |
| MaleCNS | U0 to M0 | 10.861 | [8.919, 12.803] |
| MaleCNS | C0 to A0 | 20.172 | [17.165, 23.180] |
| MaleCNS | C0 to U0 | 12.116 | [9.629, 14.604] |
| MaleCNS | C0 to M0 | 22.977 | [19.551, 26.404] |
| MaleCNS | C0 to C1 | 4.532 | [3.306, 5.757] |
| MaleCNS | M0 to M1 | 0.050 | [−0.050, 0.150] |
| MaleCNS | M0 to M2 | 0.233 | [−0.070, 0.536] |
| MaleCNS | M0 to M3 | 0.273 | [−0.044, 0.590] |
| MaleCNS | M2 to M3 | 0.041 | [−0.062, 0.143] |
| MaleCNS | M1 to M3 | 0.223 | [−0.080, 0.527] |
| MaleCNS | C1 to M3 | 18.719 | [15.699, 21.739] |
| MaleCNS | A0 to A1 | 0.500 | [0.098, 0.903] |
| MaleCNS | U0 to U1 | 1.781 | [0.989, 2.573] |
| MaleCNS | E0 to E1 | 0.132 | [−0.276, 0.540] |
| MaleCNS | E0 to E2 | −0.149 | [−0.728, 0.431] |
| MaleCNS | E0 to E3 | −0.178 | [−0.799, 0.443] |
| MaleCNS | E2 to E3 | −0.029 | [−0.271, 0.212] |
| MaleCNS | E1 to E3 | −0.310 | [−0.903, 0.283] |
| MaleCNS | T0 to T1 | 0.061 | [−0.056, 0.179] |
| MaleCNS | M3 to E1 | −0.225 | [−0.562, 0.113] |
| MaleCNS | M3 to T1 | −0.071 | [−0.490, 0.347] |

**Supplementary Table 11. Local contributions within output groups.** DN types are grouped by their dominant observed cord compartment: FL, ML and HL, front, middle and hind leg; AB, abdomen; UT and IT, upper and intermediate tectulum; OT, other named cord regions. Groups were not used for fitting; group pairs with at least 20 evaluated type pairs and ten connected cell pairs are shown. Values are type-pair-equal log-loss reductions (nats per 1,000 pairs) for both local descriptions added to the baseline (Both), spatial overlap beyond shared input (Spatial|shared) and shared input beyond spatial overlap (Shared|spatial). Connected counts are cell pairs. Groups share types and are not independent replications.

| Specimen | Groups | Type pairs | Connected | Both | Spatial\|shared | Shared\|spatial |
|---|---|---|---|---|---|---|
| BANC | FL/FL | 1322 | 162 | 18.720 | 12.319 | −0.736 |
| BANC | FL/ML | 341 | 35 | 27.352 | 13.642 | 4.970 |
| BANC | FL/HL | 498 | 65 | 17.840 | 11.472 | 0.801 |
| BANC | FL/UT | 2389 | 80 | 6.871 | 2.970 | 0.634 |
| BANC | FL/IT | 2262 | 102 | 10.157 | 4.910 | 1.884 |
| BANC | FL/OT | 686 | 21 | 11.525 | 5.500 | 1.533 |
| BANC | ML/IT | 329 | 12 | 19.395 | 5.745 | 7.123 |
| BANC | HL/HL | 46 | 10 | 36.819 | 7.715 | 13.291 |
| BANC | HL/IT | 396 | 20 | 9.763 | 11.832 | −5.318 |
| BANC | AB/IT | 570 | 10 | 3.304 | 2.157 | 0.859 |
| BANC | UT/UT | 1158 | 56 | 8.992 | 4.071 | 2.923 |
| BANC | UT/IT | 2222 | 95 | 8.496 | 4.031 | 1.977 |
| BANC | IT/IT | 1005 | 70 | 19.179 | 8.651 | 1.943 |
| BANC | IT/OT | 554 | 13 | 13.989 | 7.377 | 1.391 |
| MaleCNS | FL/FL | 1267 | 499 | 51.812 | 19.778 | 11.116 |
| MaleCNS | FL/ML | 196 | 77 | 51.619 | 9.923 | 10.553 |
| MaleCNS | FL/HL | 1732 | 555 | 39.471 | 23.762 | 1.688 |
| MaleCNS | FL/AB | 1028 | 123 | 28.513 | 11.242 | 5.822 |
| MaleCNS | FL/UT | 2310 | 230 | 15.022 | 5.532 | 1.944 |
| MaleCNS | FL/IT | 1541 | 90 | 14.529 | 7.167 | 1.409 |
| MaleCNS | FL/OT | 1331 | 183 | 29.616 | 11.230 | 5.102 |

| Specimen | Groups | Type pairs | Connected | Both | Spatial\|shared | Shared\|spatial |
|---|---|---|---|---|---|---|
| MaleCNS | ML/HL | 139 | 38 | 35.150 | 11.268 | 3.268 |
| MaleCNS | ML/OT | 120 | 16 | 43.775 | 20.039 | 2.776 |
| MaleCNS | HL/HL | 592 | 178 | 40.596 | 19.670 | 5.790 |
| MaleCNS | HL/AB | 693 | 70 | 17.160 | 9.492 | 3.423 |
| MaleCNS | HL/UT | 1560 | 62 | 11.626 | 5.919 | 1.282 |
| MaleCNS | HL/IT | 987 | 47 | 19.273 | 8.206 | 1.649 |
| MaleCNS | HL/OT | 905 | 175 | 32.587 | 20.754 | 1.157 |
| MaleCNS | AB/AB | 193 | 45 | 19.878 | 10.519 | −2.238 |
| MaleCNS | AB/UT | 893 | 18 | 12.837 | 7.394 | 0.631 |
| MaleCNS | AB/IT | 617 | 20 | 21.330 | 9.763 | 0.633 |
| MaleCNS | AB/OT | 522 | 63 | 10.538 | 2.252 | 3.694 |
| MaleCNS | UT/UT | 1006 | 225 | 22.481 | 10.492 | 2.028 |
| MaleCNS | UT/IT | 1319 | 171 | 9.708 | 5.461 | 1.980 |
| MaleCNS | UT/OT | 1260 | 32 | 11.064 | 8.287 | −0.393 |
| MaleCNS | IT/IT | 427 | 131 | 13.037 | −1.150 | 7.102 |
| MaleCNS | IT/OT | 756 | 49 | 26.240 | 13.109 | 2.373 |
| MaleCNS | OT/OT | 361 | 94 | 36.455 | 22.944 | 1.693 |

**Supplementary Table 12. Overlap measures of the example type pairs.** Values average four cell pairs per specimen for DNge006–DNge072 and DNp24–DNp48, and six BANC or eight MaleCNS pairs for DNg97–DNge046 in Supplementary Figure 2. Voxel cos., 40-$\mu$m voxel cosine; ROI cos., brain-region cosine; input cos., input-weighted shared-input cosine; input Jac., presence-based upstream-source Jaccard index.

| Specimen | Type pair | Voxel cos. | ROI cos. | Input cos. | Input Jac. |
|---|---|---|---|---|---|
| BANC | DNp24/DNp48 | 0.207 | 0.463 | 0.0000 | 0.0000 |
| BANC | DNg97/DNge046 | 0.581 | 0.745 | 0.0163 | 0.0149 |
| BANC | DNge006/DNge072 | 0.568 | 0.575 | 0.1143 | 0.0568 |
| MaleCNS | DNp24/DNp48 | 0.149 | 0.234 | 0.0017 | 0.0039 |
| MaleCNS | DNg97/DNge046 | 0.556 | 0.841 | 0.0234 | 0.0248 |
| MaleCNS | DNge006/DNge072 | 0.579 | 0.983 | 0.1442 | 0.0607 |

**Supplementary Table 13. Cable proximity of the example type pairs in BANC.** Each brain segment was divided into ⌈length/2 $\mu$m⌉ equal subsegments and sampled at their midpoints; each sample carries its subsegment length. Median min., median over cell pairs of the minimum separation; A and B fraction within 5, mean fraction of the first and second type's cable within 5 $\mu$m of the other type.

| BANC type pair | Median min. ($\mu$m) | A fraction within 5 | B fraction within 5 |
|---|---|---|---|
| DNg97/DNge046 | 0.388 | 0.224 | 0.319 |
| DNge006/DNge072 | 0.420 | 0.364 | 0.330 |
| DNp24/DNp48 | 0.444 | 0.241 | 0.094 |

**Supplementary Table 14. Conditional contributions under fixed neighboring settings.** Setting gives voxel size ($\mu$m)/response threshold (synapses); 40/5 is the primary specification (Supplementary Figure 6). Gains are log-loss reductions in nats per 1,000 evaluated pairs with 95% type-endpoint intervals; positive values mean lower error.

| Specimen | Setting | Added description | Weight | Gain [95% interval] |
|---|---|---|---|---|
| BANC | 40/5 | Spatial overlap beyond shared input | Cell pair | 3.644 [2.535, 4.753] |
| BANC | 40/5 | Spatial overlap beyond shared input | Type pair | 5.531 [4.237, 6.826] |
| BANC | 40/5 | Shared input beyond spatial overlap | Cell pair | 0.998 [0.336, 1.661] |
| BANC | 40/5 | Shared input beyond spatial overlap | Type pair | 1.258 [0.303, 2.214] |
| BANC | 30/5 | Spatial overlap beyond shared input | Cell pair | 4.409 [3.148, 5.671] |
| BANC | 30/5 | Spatial overlap beyond shared input | Type pair | 6.502 [4.976, 8.029] |
| BANC | 30/5 | Shared input beyond spatial overlap | Cell pair | 0.546 [0.107, 0.984] |
| BANC | 30/5 | Shared input beyond spatial overlap | Type pair | 0.403 [−0.264, 1.071] |
| BANC | 50/5 | Spatial overlap beyond shared input | Cell pair | 2.425 [1.630, 3.220] |
| BANC | 50/5 | Spatial overlap beyond shared input | Type pair | 3.558 [2.625, 4.491] |
| BANC | 50/5 | Shared input beyond spatial overlap | Cell pair | 1.041 [0.344, 1.738] |

| Specimen | Setting | Added description | Weight | Gain [95% interval] |
|---|---|---|---|---|
| BANC | 50/5 | Shared input beyond spatial overlap | Type pair | 0.975 [−0.003, 1.952] |
| BANC | 40/3 | Spatial overlap beyond shared input | Cell pair | 5.749 [4.253, 7.245] |
| BANC | 40/3 | Spatial overlap beyond shared input | Type pair | 8.092 [6.415, 9.770] |
| BANC | 40/3 | Shared input beyond spatial overlap | Cell pair | 1.712 [0.997, 2.426] |
| BANC | 40/3 | Shared input beyond spatial overlap | Type pair | 2.108 [1.059, 3.157] |
| BANC | 40/7 | Spatial overlap beyond shared input | Cell pair | 3.068 [1.978, 4.158] |
| BANC | 40/7 | Spatial overlap beyond shared input | Type pair | 4.820 [3.514, 6.126] |
| BANC | 40/7 | Shared input beyond spatial overlap | Cell pair | 0.915 [0.497, 1.334] |
| BANC | 40/7 | Shared input beyond spatial overlap | Type pair | 1.100 [0.421, 1.779] |
| MaleCNS | 40/5 | Spatial overlap beyond shared input | Cell pair | 7.371 [5.745, 8.997] |
| MaleCNS | 40/5 | Spatial overlap beyond shared input | Type pair | 10.861 [8.919, 12.803] |
| MaleCNS | 40/5 | Shared input beyond spatial overlap | Cell pair | 2.185 [1.226, 3.145] |
| MaleCNS | 40/5 | Shared input beyond spatial overlap | Type pair | 2.805 [1.760, 3.850] |
| MaleCNS | 30/5 | Spatial overlap beyond shared input | Cell pair | 8.989 [7.132, 10.846] |
| MaleCNS | 30/5 | Spatial overlap beyond shared input | Type pair | 13.043 [10.849, 15.237] |
| MaleCNS | 30/5 | Shared input beyond spatial overlap | Cell pair | 1.334 [0.606, 2.062] |
| MaleCNS | 30/5 | Shared input beyond spatial overlap | Type pair | 1.748 [0.904, 2.591] |
| MaleCNS | 50/5 | Spatial overlap beyond shared input | Cell pair | 5.381 [4.137, 6.626] |
| MaleCNS | 50/5 | Spatial overlap beyond shared input | Type pair | 7.972 [6.454, 9.489] |
| MaleCNS | 50/5 | Shared input beyond spatial overlap | Cell pair | 3.306 [2.033, 4.579] |
| MaleCNS | 50/5 | Shared input beyond spatial overlap | Type pair | 4.456 [3.079, 5.833] |
| MaleCNS | 40/3 | Spatial overlap beyond shared input | Cell pair | 10.061 [7.945, 12.177] |
| MaleCNS | 40/3 | Spatial overlap beyond shared input | Type pair | 14.169 [11.605, 16.734] |
| MaleCNS | 40/3 | Shared input beyond spatial overlap | Cell pair | 3.332 [2.232, 4.431] |
| MaleCNS | 40/3 | Shared input beyond spatial overlap | Type pair | 4.478 [3.193, 5.763] |
| MaleCNS | 40/7 | Spatial overlap beyond shared input | Cell pair | 5.613 [4.287, 6.938] |
| MaleCNS | 40/7 | Spatial overlap beyond shared input | Type pair | 8.295 [6.737, 9.852] |
| MaleCNS | 40/7 | Shared input beyond spatial overlap | Cell pair | 1.723 [0.773, 2.672] |
| MaleCNS | 40/7 | Shared input beyond spatial overlap | Type pair | 2.128 [1.206, 3.049] |

**Supplementary Table 15. Prediction errors and ranking at natural prevalence.** Cell, each cell pair weighted equally; Type, each unordered type pair weighted equally. Prev., connection prevalence, the no-information reference for average precision (AP); AUROC, area under the receiver operating characteristic curve. Prevalence differs between weightings because weights change, not because pairs are resampled. Metrics use the saved predictions without refitting or recalibration (Supplementary Figure 7).

| Specimen | Model | Weight | Prev. | Log loss | Brier | AUROC | AP |
|---|---|---|---|---|---|---|---|
| BANC | C0 | Cell | 0.00853 | 0.040169 | 0.008058 | 0.8501 | 0.0965 |
| BANC | C0 | Type | 0.01357 | 0.060261 | 0.012765 | 0.8287 | 0.1015 |
| BANC | A0 | Cell | 0.00853 | 0.033930 | 0.007614 | 0.9263 | 0.1733 |
| BANC | A0 | Type | 0.01357 | 0.051075 | 0.012010 | 0.9117 | 0.1839 |
| BANC | U0 | Cell | 0.00853 | 0.036575 | 0.007592 | 0.8836 | 0.1825 |
| BANC | U0 | Type | 0.01357 | 0.055348 | 0.012041 | 0.8599 | 0.1847 |
| BANC | M0 | Cell | 0.00853 | 0.032932 | 0.007404 | 0.9350 | 0.2160 |
| BANC | M0 | Type | 0.01357 | 0.049817 | 0.011702 | 0.9175 | 0.2220 |
| MaleCNS | C0 | Cell | 0.01968 | 0.078685 | 0.018091 | 0.8473 | 0.1316 |
| MaleCNS | C0 | Type | 0.02955 | 0.110965 | 0.026837 | 0.8247 | 0.1464 |
| MaleCNS | A0 | Cell | 0.01968 | 0.064762 | 0.016369 | 0.9236 | 0.2673 |
| MaleCNS | A0 | Type | 0.02955 | 0.090793 | 0.024072 | 0.9132 | 0.2872 |
| MaleCNS | U0 | Cell | 0.01968 | 0.069948 | 0.016722 | 0.8896 | 0.2437 |
| MaleCNS | U0 | Type | 0.02955 | 0.098849 | 0.024769 | 0.8759 | 0.2610 |
| MaleCNS | M0 | Cell | 0.01968 | 0.062577 | 0.015927 | 0.9301 | 0.3010 |
| MaleCNS | M0 | Type | 0.02955 | 0.087988 | 0.023446 | 0.9204 | 0.3222 |

**Supplementary Table 16. Overall observed and predicted connection rates.** Cell pairs weighted equally. O/E, observed/expected ratio; a value of one indicates agreement in the overall rate only, not calibration within subgroups.

| Specimen | Model | Observed rate | Mean prediction | O/E |
|---|---|---|---|---|
| BANC | C0 | 0.008530 | 0.008972 | 0.9507 |
| BANC | A0 | 0.008530 | 0.008916 | 0.9567 |
| BANC | U0 | 0.008530 | 0.009033 | 0.9443 |
| BANC | M0 | 0.008530 | 0.008864 | 0.9624 |
| MaleCNS | C0 | 0.019682 | 0.020863 | 0.9434 |
| MaleCNS | A0 | 0.019682 | 0.020749 | 0.9486 |
| MaleCNS | U0 | 0.019682 | 0.020853 | 0.9438 |
| MaleCNS | M0 | 0.019682 | 0.020752 | 0.9484 |

**Model key for Supplementary Table 17.** The model labels in this table refer to the supporting-model specifications and are distinct from the C0–M3 models used in the primary connectivity analysis (Supplementary Table 22).

*Reciprocity and presence tasks (symmetric predictors). local* contains the 16 columns of block $L$: `brain_distance`; `x_overlap`, `y_overlap` and `z_overlap`; sums and absolute differences of external in-degree, out-degree, in-strength and out-strength; `same_side`; `direct_partner_similarity`; `fine_partner_similarity`; and `onehop_similarity`. Unlike C0 ($L+H$), it omits the breadth block $H$. *sensory* adds `sensory_similarity` with the sum and absolute difference of sensory entropy; *output* adds `output_similarity` with the sum and absolute difference of output entropy; *joint* contains both additions (22 columns, the column set of C1, that is, C0 with $S$ and $O$). None of these models contains the spatial-overlap block $Q_A$ or the shared-input block $Q_U$, so *joint* is not M3. In the linear reciprocity task, *interaction* adds the product of sensory and output similarity (23 columns). In the presence task, every *joint* column enters the cubic spline basis described in the Supplementary Methods (five columns per feature), and *interaction* adds the tensor product of the sensory- and output-similarity spline bases (25 columns).

*Direction task (antisymmetric predictors, endpoint $i$ minus endpoint $j$). local* contains the differences of external in-degree, out-degree, in-strength and out-strength, of brain-centroid $x$, $y$ and $z$, of log-transformed $x$, $y$ and $z$ extents, of direct- and fine-partner entropy, and of the eight one-hop source channels (20 columns). *sensory* adds the differences of sensory entropy, sensory containment and the eight sensory channels; *output* adds the differences of output entropy, output containment and the seven cord-region channels; *joint* contains both additions (39 columns); *interaction* adds the sensory-entropy difference multiplied by output similarity and the output-entropy difference multiplied by sensory similarity (41 columns). The containment of endpoint $i$ is its profile mass in channels where the other endpoint has at least 5% of its mass.

*Fitting and samples.* Direction and reciprocity models are linear penalized logistic models without splines, with penalties selected from $\{0.01, 0.1, 1, 10\}$. Direction has no intercept and scales columns by their standard deviation without centering, so exchanging endpoints complements the probability; reciprocity centers and scales columns and adds an unpenalized intercept. The presence models center and scale their spline columns, add an unpenalized intercept and select penalties from $\{10^{-4}, 10^{-3}, 10^{-2}, 0.1, 1\}$. Direction uses singly connected pairs (780 BANC; 2,947 MaleCNS), with the outcome $i \to j$ for $i < j$; reciprocity uses pairs connected in at least one direction (798; 3,201; 18 and 254 reciprocal); the presence interaction uses all evaluated pairs (93,550; 162,638). All tasks use the five type-grouped outer folds, evaluate pairs with both endpoints in the held-out fold and select penalties by three-fold inner GroupKFold. $\Delta$ loss is reference minus extended mean log loss in nats per pair.

**Supplementary Table 17. Direction, reciprocity and interaction models with coarse relationship blocks.** These supporting models retain local anatomical and connectivity covariates (block $L$ for presence and reciprocity; antisymmetric local predictors for direction) but omit the voxel/regional-overlap block $Q_A$ and the source-resolved shared-input block $Q_U$. For presence and reciprocity, the sensory and output blocks each contain similarity and breadth; direction uses signed profile and breadth contrasts (Supplementary Note 5). Model labels are defined in the model key that precedes this table; they are supporting-model specifications, distinct from the C0–M3 models of the primary analysis. Δ loss is unscaled (nats per pair); intervals use signed variance estimates from the saved predictions.

| Specimen | Task | Comparison | Δ loss | 95% interval |
|---|---|---|---|---|
| BANC | Interaction | joint to interaction | 0.00006 | [−0.00003, 0.00014] |
| BANC | Direction | local to sensory | 0.00513 | [−0.00810, 0.01837] |
| BANC | Direction | local to output | 0.01233 | [−0.01724, 0.04190] |
| BANC | Direction | local to joint | 0.00707 | [−0.02657, 0.04070] |
| BANC | Direction | sensory to joint | 0.00193 | [−0.02660, 0.03047] |
| BANC | Direction | output to joint | −0.00526 | [−0.01950, 0.00898] |
| BANC | Direction | joint to interaction | −0.00035 | [−0.00159, 0.00090] |
| BANC | Reciprocity | local to sensory | 0.00013 | [−0.00090, 0.00116] |
| BANC | Reciprocity | local to output | 0.00100 | [−0.00136, 0.00336] |
| BANC | Reciprocity | local to joint | 0.00102 | [−0.00104, 0.00308] |
| BANC | Reciprocity | sensory to joint | 0.00089 | [−0.00131, 0.00308] |
| BANC | Reciprocity | output to joint | 0.00002 | [−0.00107, 0.00111] |
| BANC | Reciprocity | joint to interaction | −0.00215 | [−0.00363, −0.00067] |
| MaleCNS | Interaction | joint to interaction | −0.00002 | [−0.00018, 0.00014] |
| MaleCNS | Direction | local to sensory | 0.00815 | [−0.00010, 0.01640] |
| MaleCNS | Direction | local to output | 0.01374 | [0.00254, 0.02494] |
| MaleCNS | Direction | local to joint | 0.01533 | [0.00317, 0.02749] |
| MaleCNS | Direction | sensory to joint | 0.00718 | [−0.00223, 0.01658] |
| MaleCNS | Direction | output to joint | 0.00159 | [−0.00358, 0.00677] |
| MaleCNS | Direction | joint to interaction | 0.00012 | [−0.00052, 0.00077] |
| MaleCNS | Reciprocity | local to sensory | 0.00258 | [−0.00035, 0.00552] |
| MaleCNS | Reciprocity | local to output | −0.00110 | [−0.00412, 0.00191] |
| MaleCNS | Reciprocity | local to joint | 0.00124 | [−0.00173, 0.00421] |
| MaleCNS | Reciprocity | sensory to joint | −0.00134 | [−0.00432, 0.00164] |
| MaleCNS | Reciprocity | output to joint | 0.00235 | [−0.00042, 0.00511] |
| MaleCNS | Reciprocity | joint to interaction | −0.00032 | [−0.00069, 0.00004] |

**Supplementary Table 18. Identified cord output of every DNge006 and DNge072 cell.** Two cells per type and specimen (Fig. 4). Identified is the accepted-neuron synapse count before thresholding; retained applies the five-synapse target-edge threshold; all includes unaccepted target segments in the named cord regions. Coverage is identified/all. The following block reports target composition and endpoint counts.

| Specimen | Cell | Native ID | Identified | Retained | All | Coverage (%) |
|---|---|---|---|---|---|---|
| BANC | DNge072 L | 720575941452165801 | 242 | 201 | 1168 | 20.7 |
| BANC | DNge072 R | 720575941514949881 | 301 | 240 | 1095 | 27.5 |
| BANC | DNge006 R | 720575941526955266 | 1027 | 792 | 5559 | 18.5 |
| BANC | DNge006 L | 720575941632059372 | 1189 | 957 | 5786 | 20.5 |
| MaleCNS | DNge006 R | 10668 | 2456 | 2218 | 7962 | 30.8 |
| MaleCNS | DNge006 L | 13030 | 2897 | 2581 | 8037 | 36.0 |
| MaleCNS | DNge072 R | 14887 | 540 | 498 | 1381 | 39.1 |
| MaleCNS | DNge072 L | 15800 | 444 | 399 | 1141 | 38.9 |

| Specimen | Cell | IN (%) | Motor (%) | Retained MN/all (%) | Direct MN | Via IN MN |
|---|---|---|---|---|---|---|
| BANC | DNge072 L | 32.8 | 52.2 | 9.0 | 4 | 17 |
| BANC | DNge072 R | 42.9 | 46.7 | 10.2 | 5 | 12 |
| BANC | DNge006 R | 71.1 | 14.6 | 2.1 | 9 | 124 |
| BANC | DNge006 L | 65.9 | 18.0 | 3.0 | 9 | 150 |
| MaleCNS | DNge006 R | 76.7 | 14.0 | 3.9 | 12 | 267 |
| MaleCNS | DNge006 L | 79.1 | 11.8 | 3.8 | 10 | 257 |

| Specimen | Cell | IN (%) | Motor (%) | Retained MN/all (%) | Direct MN | Via IN MN |
|---|---|---|---|---|---|---|
| MaleCNS | DNge072 R | 46.6 | 44.0 | 15.9 | 5 | 57 |
| MaleCNS | DNge072 L | 44.9 | 46.1 | 16.1 | 6 | 31 |

IN and Motor are shares of retained identified output. Retained MN/all is retained motor-neuron synapses (target edges with at least five synapses) divided by all recorded cord output, the All column above. Direct MN and Via IN MN count distinct motor neurons directly or through one accepted cord interneuron, respectively. The two sets can overlap and are not added.

**Supplementary Table 19. Annotation coverage and alternative output-overlap summaries.** Output overlap is summarized for one-way pairs with shared input and retained output from both DNs (main-text Fig. 4c). Coverage uses all 979 BANC and 1,287 MaleCNS DNs before the edge threshold. Pair summaries use the same 2,516 and 13,789 one-way pairs with shared input and nonzero retained identified output at five synapses; the one-synapse and all-segment alternatives do not add pairs. Zero-vector pairs are excluded by that common availability screen. All-segment summaries include unidentified fragments and are not interpreted as identified neuronal targets.

| Specimen | Measure | Value |
|---|---|---|
| BANC | Annotated coverage, synapse weighted | 0.2093 |
| BANC | Median DN annotation coverage | 0.2019 |
| BANC | One-way pairs with shared input and output | 2516 |
| BANC | Pairs without shared target at ≥5 | 1109 |
| BANC | Median shared-target output mass | 0.0102 |
| BANC | Median annotated-target cosine, ≥5 | 0.0043 |
| BANC | Median annotated-target cosine, ≥1 | 0.0191 |
| BANC | Median all-segment cosine, ≥1 | 0.0136 |
| MaleCNS | Annotated coverage, synapse weighted | 0.3522 |
| MaleCNS | Median DN annotation coverage | 0.3425 |
| MaleCNS | One-way pairs with shared input and output | 13789 |
| MaleCNS | Pairs without shared target at ≥5 | 5031 |
| MaleCNS | Median shared-target output mass | 0.0107 |
| MaleCNS | Median annotated-target cosine, ≥5 | 0.0044 |
| MaleCNS | Median annotated-target cosine, ≥1 | 0.0117 |
| MaleCNS | Median all-segment cosine, ≥1 | 0.0099 |

**Supplementary Table 20. Connected pairs minus matched unconnected controls.** Mean differences for connected pairs with shared input (Supplementary Figure 9b). Controls were matched on voxel and regional overlap, shared-input cosine and count, input strengths, partner counts and side, and can be reused, so differences are descriptive. The analysis outputs include matched-pair identifiers, connection classes, inclusion flags and control reuse. Both pairs require retained cord output after matching. A mass difference of 0.02678 means 2.678 percentage points of combined retained output, not a 2.678% relative increase.

| Specimen | Measure | Mean difference | Pairs |
|---|---|---|---|
| BANC | Target cosine | 0.02934 | 2651 |
| BANC | Shared target mass | 0.02678 | 2651 |
| MaleCNS | Target cosine | 0.01329 | 14873 |
| MaleCNS | Shared target mass | 0.01293 | 14873 |
| MaleCNS | Presynaptic-site overlap | 0.00527 | 14873 |

**Supplementary Table 21. Cohort flow and analysis units.** Counts describe analysis units within the listed reconstructed specimens, not numbers of independent animals. BANC, MaleCNS and FlyWire are distinct flies; Dataset7 and Dataset8 are separate adult worms; Platynereis is one larva.

| Specimen | Stage | Retained units | Rule |
|---|---|---|---|
| BANC | Initial eligible cohort | 986 DNs | Exact-ID brain skeleton; typed; brain input; initial annotated-target cord readout ≥ 50. |
| BANC | Final network cohort | 979 DNs; 418 types | Seven initial cells lack external brain input after removal of DN–DN edges. |

| Specimen | Stage | Retained units | Rule |
|---|---|---|---|
| BANC | All different-type pairs | 476,923 cell pairs | Unordered pairs of cells with different exact types. |
| BANC | Prediction evaluation | 93,550 cell pairs | Both endpoint types are in the same held-out fold; 17,265 type pairs. |
| MaleCNS | Final network cohort | 1,287 DNs; 473 types | Typed; external brain records and source access; $\geq 50$ named-cord contacts. |
| MaleCNS | All different-type pairs | 825,185 cell pairs | Same pair definition as BANC. |
| MaleCNS | Prediction evaluation | 162,638 cell pairs | Both endpoint types held out; 22,137 type pairs. |
| MaleCNS | One-way connections | 16,114 directed pairs | One direction has $\geq 5$ brain synapses and the reverse has $< 5$. |
| MaleCNS | Initial position comparison | 13,323 directed pairs | At least three shared and three other source records in a region. |
| MaleCNS | Between-source path comparison | 13,318 directed pairs | Reachability and regional count requirements reapplied; 1,079 receivers; 6,260 type pairs. |
| MaleCNS | Within-source receiver pool | 1,079 receivers | Receivers from the between-source analysis. |
| MaleCNS | Within-source retained receivers | 922 receivers; 415 types | 157 lack a region–component stratum with both eligible partner classes. |
| MaleCNS | Within-source units | 49,742 source–receiver pairs | 8,453 sources; 68,609 region–component strata; sources/receivers reused. |
| MaleCNS | Within-source edge coverage | 15,110 partner–receiver edges | Different qualification rules from the between-source comparison. |
| BANC | Retained identified cord output | 972 DNs | Seven final-cohort DNs have no accepted target edge with $\geq 5$ contacts. |
| MaleCNS | Retained identified cord output | 1,283 DNs | Four final-cohort DNs have no accepted target edge with $\geq 5$ contacts. |
| FlyWire | Typed endpoint cohort | 1,210 DNs; 446 types | Brain-only; skeleton-covered; all 1,303 annotated DNs excluded as sources. |
| FlyWire | Between-source | 5,807 pairs; 720 receivers | 2,894 unordered type pairs; same five-contact and region rules. |
| FlyWire | Within-source receivers | 589 receivers; 293 types | 131 of the 720 receiver pool lack eligible strata. |
| FlyWire | Within-source units | 19,246 source–receiver pairs | 4,140 sources; 23,497 region–component strata; 7,448 partner–receiver edges. |
| C. elegans D7 | Typed endpoint cohort | 44 interneurons; 24 classes | All author-classified interneurons excluded as sources. |
| C. elegans D7 | Between-source | 17 pairs; 11 receivers | Whole reconstructed brain arbor; seven receiver types. |
| C. elegans D7 | Within-source | 25 source–receiver pairs; 10 receivers | 22 sources; six receiver types; native attachments. |
| C. elegans D8 | Typed endpoint cohort | 44 interneurons; 24 classes | Same cohort definition as Dataset7. |
| C. elegans D8 | Between-source | 24 pairs; 13 receivers | Whole reconstructed brain arbor; seven receiver types. |
| C. elegans D8 | Within-source | 31 source–receiver pairs; 13 receivers | 22 sources; seven receiver types; native attachments. |
| Platynereis | Projection cohort | 138 head-to-trunk cells | All 138 excluded as sources; 99 typed endpoints in 35 types. |
| Platynereis | Between-source | 4 pairs; 4 receivers | From 19 one-way candidates; one common partner cell; two type pairs. |
| Platynereis | Within-source | 2 source–receiver pairs; 1 receiver | Two sources; one receiver type; interval nonestimable. |

**Supplementary Table 22. Feature definitions and model specifications.**

J(p, q) = 1 - JSD(p, q)/log 2, using natural logs and 0 log 0 = 0; no square root is taken. Profiles are row-normalized; zero rows remain zero. One-hop profiles are zero in 121/979 BANC and 17/1,287 MaleCNS DNs. The implemented expression encodes one zero endpoint as 0.5 and two as 1. These zeros mark absent direct seed input, not probability distributions. Missing values are not imputed. All scalar features use the training-fitted spline and scaling pipeline in Supplementary Methods.

| Feature family | Block | Definition and source |
|---|---|---|
| brain_distance | L | $\log(1 + \|c_i - c_j\|/s)$, where $s = \|\mathrm{SD}(c)\|$. BANC centroid: segment-midpoint cable-weighted mean; MaleCNS: record-weighted mean of DN-owned external brain coordinates. Spread is calculated within the final cohort, without responses. |
| x_overlap, y_overlap, z_overlap | L | For axis $a$, $\max(0, \min(h_{ia}, h_{ja}) - \max(l_{ia}, l_{ja}))/\min(w_{ia}, w_{ja})$. Coordinates are divided by $s$; $w = \max(h - l, 10^{-6})$. BANC bounds use brain vertices; MaleCNS uses external DN-owned coordinates. |
| external in/out degree sum and absolute_difference | L | For each direction, $z_i = \log(1 + k_i)$ for the number of nonzero input-graph edges; use $z_i + z_j$ and $\|z_i - z_j\|$. Input graph has edge threshold five and no DN–DN block. |
| external in/out strength sum and absolute_difference | L | As above with retained synapse count replacing degree. These are sums/differences of transformed values, not transforms of pairwise sums. |
| same_side | L | One only when both side strings are equal and either left or right. Center, unknown and mismatched sides encode zero. |
| direct_partner_similarity | L | $J(p_i, p_j)$ for normalized incoming synapse counts pooled by canonical_class; every released mapped class, including blank categories, has its own coordinate. |
| fine_partner_similarity | L | $J(p_i, p_j)$ for normalized incoming counts pooled by cell_class. This retains classes, not individual source IDs. |
| onehop_similarity | L | $J(p_i, p_j)$ for row-normalized $(AB)_i$ in the eight source columns. $A = 0.85P^{\intercal}$ therefore also weights a source by its outgoing count and its seed-group normalization. |
| sensory_entropy sum and difference | H | $h(p) = -\sum_g p_g \log p_g$ for the eight-channel resolvent profile; use $h_i + h_j$ and $\|h_i - h_j\|$. Zero terms contribute zero. |
| output_entropy sum and difference | H | Same entropy summaries for normalized counts across seven named cord regions, without a per-target threshold or identified-target requirement. |
| occupancy_cosine | QA | $\sum_v a_{iv} a_{jv}/(\|a_i\|\|a_j\|)$. $a$ is BANC cable length by segment midpoint, or MaleCNS external contact-row count at the DN-owned endpoint, in 40-$\mu$m native-origin voxels. |
| occupancy_jaccard | QA | $\|V_i \cap V_j\|/\|V_i \cup V_j\|$ for nonempty voxels. Occupancy counts are unthresholded. Zero denominator gives zero; overlaps are clipped to $[0, 1]$ for floating-point error. |
| roi_cosine, roi_jaccard | QA | Cosine of external brain synapse-count profiles and Jaccard of nonzero side-specific region sets. Records with exactly one DN endpoint contribute at the DN-owned endpoint; no per-edge threshold. |
| upstream_cosine | QU | Cosine of incoming synapse-count vectors indexed by individual non-DN source ID after the five-contact edge threshold. Different classes of source can be shared; identity is not reduced to a class label. |
| upstream_jaccard | QU | Jaccard of individual upstream-ID sets with retained input edges. Zero denominator gives zero, as for spatial overlaps. |
| sensory_similarity | S | $J(p_i, p_j)$ for the eight-group resolvent profile. A nonempty seed group has total initial mass one. Final-cohort cells have nonzero total access. |
| output_similarity | O | $J(o_i, o_j)$ for row-normalized unthresholded cord-region counts. Every cohort cell has at least 50 contacts in those groups. |
| nonascending profiles and availability | E models | Renormalize the first seven resolvent columns. Eight BANC and one MaleCNS cell have zero remaining access; similarity is encoded zero if either is unreached. Reached-endpoint count and both-reached indicator enter every E base. |
| ascending amount and balance | E2, E3 | Sums and absolute differences of ascending mass fraction and $\log(1 + 10^4 F_{i,\mathrm{asc}})$. These describe one source channel, not different ascending pathways. |
| first_arrival_similarity and breadth | T models | Replace source profiles by row-normalized $(I - AD)^{-1}B$, with $D_{ii} = 0$ for every annotated DN and one otherwise. This stops after the first DN visit without renormalizing surviving transitions. |
| C0, A0, U0, M0 | Core models | C0=$L + H$; A0=C0+$Q_A$; U0=C0+$Q_U$; M0=C0+$Q_A + Q_U$. The conditional contrasts are A0 to M0 and U0 to M0. |

| Feature family | Block | Definition and source |
|---|---|---|
| C1, A1, U1; M1, M2, M3 | Extensions | C1/A1/U1 add $S + O$ to their corresponding bases. M1=M0+$S$; M2=M0+$O$; M3=M0+$S + O$. All use the same evaluation pairs. |
| E0–E3 and T0–T1 | Source variants | E0 includes all local blocks, non-ascending/output breadth, $O$ and availability. E1 adds non-ascending similarity; E2 ascending amount/balance; E3 both. T0 uses first-arrival/output breadth, all local blocks and $O$; T1 adds first-arrival similarity. |

**Supplementary Table 23. FlyWire between-source distance comparisons.**

| Metric | Weighting | $n$ | Shared | Other | $\Delta$ | 95% interval |
|---|---|---|---|---|---|---|
| Path | cell pair | 5,807 | 51.37 | 59.36 | −7.98 | [−9.27, −6.70] |
| Path | type pair | 2,894 | 53.78 | 61.63 | −7.84 | [−9.09, −6.59] |
| Path | receiver | 720 | 48.64 | 55.12 | −6.49 | Descriptive |
| Euclidean | cell pair | 5,807 | 21.17 | 25.51 | −4.34 | [−4.96, −3.72] |
| Euclidean | type pair | 2,894 | 22.01 | 26.23 | −4.22 | [−4.82, −3.61] |
| Euclidean | receiver | 720 | 19.64 | 22.90 | −3.26 | Descriptive |

Distances and differences are in $\mu$m. Negative differences mean nearer shared-source inputs. The primary estimate weights cell pairs equally. Type pairs are unordered exact-type pairs; receiver means average their eligible incoming partner connections. Path and Euclidean comparisons use the same retained records. Receiver fractions with a negative difference are 581/720 (80.7%) for path distance and 599/720 (83.2%) for Euclidean distance. Of 5,807 cell pairs, 4,344 have negative path differences. Intervals account for type endpoints within the FlyWire specimen. Associated figures: main Fig. 1b and Supplementary Figure 10. Values are rounded to two decimal places where applicable; differences were calculated before rounding.

**Supplementary Table 24. FlyWire within-source partner comparisons.**

| Quantity | Value |
|---|---|
| Receiver pool | 720 |
| Retained receivers | 589 |
| Receiver exact types | 293 |
| Source cells | 4,140 |
| Source–receiver pairs | 19,246 |
| Region–component strata | 23,497 |
| Distinct partner–receiver edges | 7,448 |
| Distinct reference sites per draw | 3 |
| Mean distance, source-connected partners ($\mu$m) | 62.79 |
| Mean distance, zero-recorded-input partners ($\mu$m) | 72.93 |
| Receiver-equal difference ($\mu$m) | −10.15 |
| Conditional standard error ($\mu$m) | 1.02 |
| Conditional 95% interval ($\mu$m) | [−12.15, −8.14] |
| Receivers with negative difference | 483/589 (82.0%) |
| Source–receiver-equal difference ($\mu$m) | −11.41 |

The primary estimate weights each receiver equally after averaging its retained sources and eligible region–component strata. Partner groups receive at least five source contacts or zero recorded contacts; each partner supplies at least three distinct receiving reference sites. The interval uses crossed receiver-type and source-cell clusters conditional on the observed graph. Equal source–receiver weighting is descriptive. See main Fig. 1b and Supplementary Figure 10. Values are rounded to two decimal places where applicable; differences were calculated before rounding.

**Supplementary Table 25. Placement comparisons in two adult C. elegans reconstructions.**

| Adult | Comparison | Group means | Δ | 95% interval | $n$ / receivers |
|---|---|---|---|---|---|
| D7 | Between sources | 3.89 / 12.08 | −8.20 | [−12.24, −4.15] | 17 / 11 |
| D7 | Within-source partners | 5.75 / 15.63 | −9.88 | [−15.03, −4.73] | 25 / 10 |
| D8 | Between sources | 6.07 / 10.85 | −4.78 | [−11.85, 2.30] | 24 / 13 |
| D8 | Within-source partners | 7.03 / 12.42 | −5.39 | [−9.83, −0.94] | 31 / 13 |

All distances are in $\mu$m. Each adult cohort has 44 interneurons in 24 classes. For between-source comparisons, group means are shared versus other input distances and $n$ is the number of directed connections. For within-source comparisons, means are source-connected versus zero-recorded-input partner distances and $n$ counts source–receiver pairs. Both within-source samples contain 22 sources; Dataset7 has six receiver-type clusters and Dataset8 seven. Between-source receivers represent seven types in each adult. Negative receiver fractions are 10/11 and 9/13 (between-source), and 10/10 and 11/13 (within-source). Dataset8's between-source interval includes zero. The spatial domain is the whole reconstructed brain arbor; intervals are within-specimen asymptotic intervals, with finite-sample coverage unestablished. See Supplementary Figure 11. Values are rounded to two decimal places where applicable; differences were calculated before rounding.

**Supplementary Table 26. Descriptive Platynereis placement comparisons.**

| Partner / summary | Receiver / sample | Shared | Other | $\Delta$ | Records |
|---|---|---|---|---|---|
| 3337770 (MN3) | 3319051 (MN1) | 23.68 | 20.34 | +3.34 | 62/40 |
| 3337770 (MN3) | 3319201 (MN1) | 36.96 | 18.45 | +18.51 | 81/35 |
| 3337770 (MN3) | 3319419 (MN2) | 38.93 | 43.53 | −4.60 | 42/37 |
| 3337770 (MN3) | 3319963 (MN2) | 29.71 | 22.62 | +7.09 | 16/40 |
| Between-source mean | 4 receivers | 32.32 | 26.23 | +6.09 | – |
| Within-source mean | 1 receiver; 2 sources | 31.00 | 38.61 | −7.61 | – |

Distances are in $\mu$m. Individual rows describe between-source comparisons and give shared/other input-record counts. The final row instead compares partners receiving source input with partners receiving none recorded, using the expected nearest distance to three distinct sites and receiver-equal weighting. From 138 head-to-trunk cells, 99 typed cells in 35 types formed the endpoint cohort; 19 one-way candidate connections yielded the four retained pairs. Three pair differences are positive and one negative. All four pairs share one MN3 partner cell. Both between-source and within-source cluster intervals are non-estimable: no between-source type dyads have disjoint endpoints, and there is only one within-source receiver type. No zero-width interval or inferred precision is assigned. The whole arbor forms one spatial domain. See Supplementary Note 8. Values are rounded to two decimal places where applicable; differences were calculated before rounding.